\documentclass[12pt]{cernart}

\usepackage{graphicx}
\usepackage{amsmath}
\usepackage{amssymb}
\usepackage{times}
\usepackage{cite}
\usepackage{multirow}
\usepackage[table]{xcolor}
\usepackage{appendix}

\begin{document}
\setcounter{topnumber}{3}
\renewcommand{\topfraction}{0.999}
\renewcommand{\bottomfraction}{0.99}
\renewcommand{\textfraction}{0.0}
\setcounter{totalnumber}{6}
\renewcommand{\thefootnote}{\alph{footnote}}

\begin{titlepage}
\flushright{\includegraphics[width=2cm]{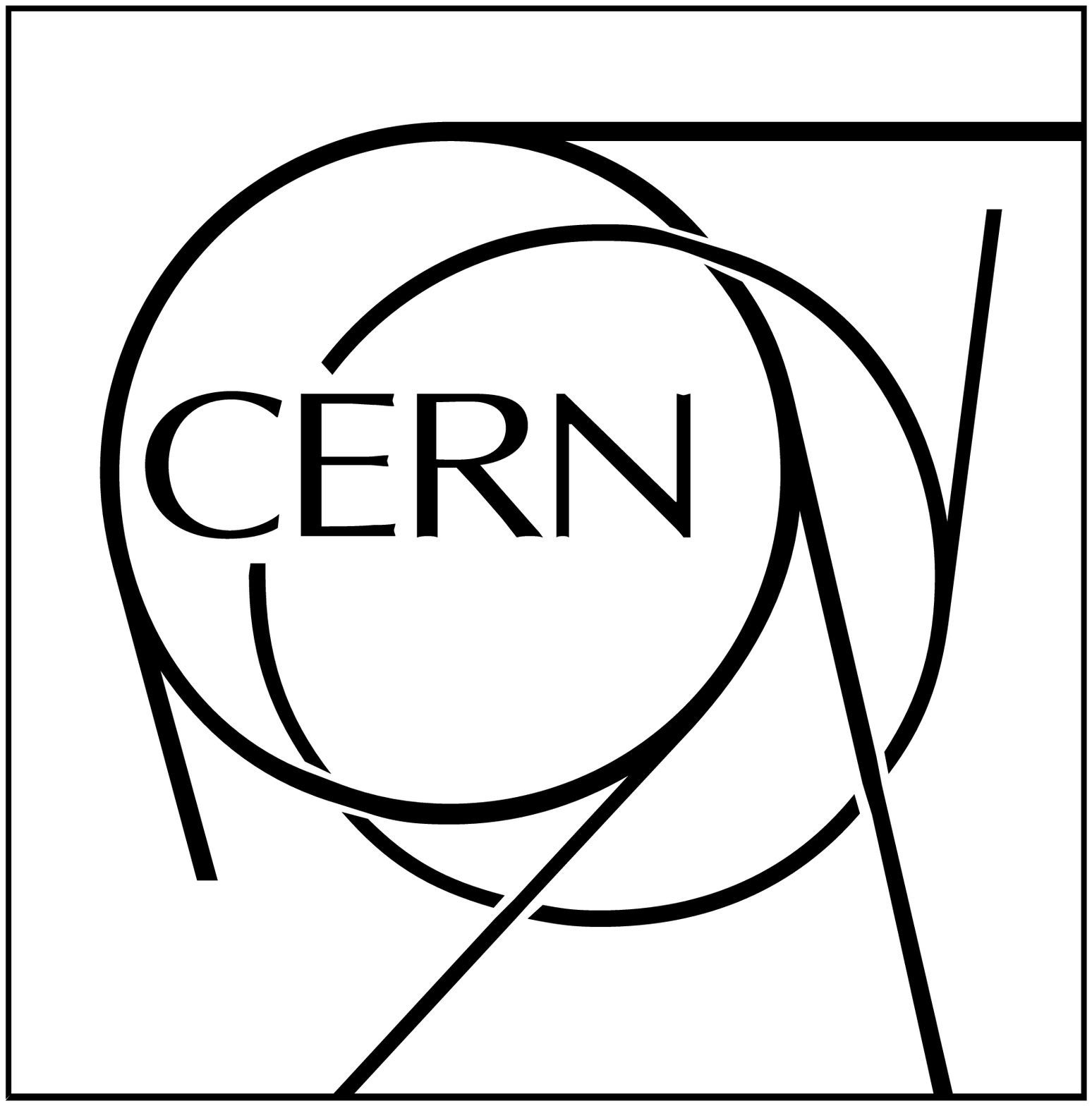}}
\docnum{CERN-PH-EP-2012-154}
\date{24 October 2012}
%
\title{\large{A survey of backward proton and pion production in p+C interactions
              at beam momenta from 1 to 400~GeV/c}}
\begin{Authlist}
\vspace{2mm}
\noindent
O.~Chvala$^{4,6}$, H.~G.~Fischer$^{3,}$\footnote{e-mail: Hans.Gerhard.Fischer@cern.ch}, M.~Makariev$^{5}$, A.~Rybicki$^{2}$, D.~Varga$^{1}$, S.~Wenig$^{3}$

\vspace*{2mm} 

%
$^{1}$E\"otv\"os Lor\'and University, Budapest, Hungary \\
$^{2}$H. Niewodnicza\'nski Institute of Nuclear Physics, Polish Academy of Sciences, Cracow, Poland \\
$^{3}$CERN, Geneva, Switzerland\\
$^{4}$Charles University, Faculty of Mathematics and Physics, Institute of
             Particle and Nuclear Physics, Prague, Czech Republic \\
$^{5}$Institute for Nuclear Research and Nuclear Energy, BAS, Sofia, Bulgaria\\
$^{6}$now at University of Tennessee, Knoxville, TN, USA\\
\end{Authlist}
%
\begin{center}
{\small{\it to be published in EPJC }}
\end{center}
\vspace*{10mm} 

\begin{abstract}
\vspace{-3mm}

New data on proton and pion production in p+C interactions
from the CERN PS and SPS accelerators are used in conjunction
with other available data sets to perform a comprehensive survey of 
backward hadronic cross sections. This survey covers the complete
backward hemisphere in the range of lab angles from 10 to 180
degrees, from 0.2 to 1.4~GeV/c in lab momentum and from 1 to 400~GeV/c 
in projectile momentum. Using the constraints of continuity
and smoothness of the angular, momentum and energy dependences a
consistent description of the inclusive cross sections is established 
which allows the control of the internal consistency of the nineteen available
data sets.

\end{abstract}
 
\clearpage
\end{titlepage}

%
%
\section{Introduction} 
\vspace{3mm}
\label{sec:intro}

An impressive amount of data on backward hadron production
in p+C interactions has been collected over the past four
decades. A literature survey reveals no less than 19 experiments
which have contributed a total amount of more than 3500
data points covering wide areas in projectile momentum, lab
angle and lab momentum.

Looking at the physics motivation and at the distribution in time of these efforts, 
two distinct classes of experimental approaches become evident. 15 experiments cluster in 
a first period during the two decades between 1970 and 1990. All
these measurements have been motivated by the nuclear part
of proton-nucleus collisions, in particular by the width of the
momentum distributions in the nuclear rest system which reach 
far beyond the narrow limits expected from nuclear binding alone.
These studies have ceased in the late 1980's with the advent
of relativistic heavy ion collisions and their promise of "new"
phenomena beyond the realm of classic nuclear physics.

A second class of very recent measurements has appeared and is
being pursued after the turn of the century, with publications
starting about 2008. Here the motivation is totally different.
It is driven by the necessity of obtaining hadronic reference data 
for the study of systematic effects in cosmic ray and neutrino physics, in
particular concerning atmospheric and long base line experiments
as well as eventual novel neutrino factories. The main aim of
these studies is the comparison to and the improvement of hadronic production models -- 
models which are to be considered as multi-parameter descriptions of the non-calculable sector of
the strong interaction, with very limited predictive power.

This new and exclusive aim has led to the strange situation that 
if all recent publications contain detailed comparisons to available production models, 
no comparison to existing data is attempted. It remains therefore unclear how 
these new results compare to the wealth of already available data and 
whether they in fact may over-ride and replace the existing results.

In this environment the studies conducted since 15 years by the NA49 experiment 
at the CERN SPS have a completely different aim. Here it
is attempted to trace a model-independent way from the basic
hadron-nucleon interaction via hadron-nucleus to nucleus-nucleus
collisions. This aim needs precision data from a large variety of
projectile and target combinations as well as a maximum phase space
coverage. As the acceptance of the NA49 detector is limited to
lab angles below 45 degrees, it is indicated to use existing backward data 
in the SPS energy range in order to extend the acceptance coverage for 
the asymmetric proton-nucleus interactions. This requires
a careful study of the dependence on cms energy and of the
reliability of the results to be used.

In the course of this work it appeared useful and even mandatory
to provide a survey of all available data over the full scale
of interaction energies, the more so as no overview of the
experimental situation is available to date. This means that the
present study deals with projectile momenta from 1 to 400~GeV/c,
for a lab angle range from 10 to 180~degrees, and for lab momenta
from 0.2 to 1.2~GeV/c.

%
%
\section{Variables and kinematics} 
\vspace{3mm}
\label{sec:var_kin}

Most available data have been obtained as a function of the lab
momentum $p_{\textrm{lab}}$ (or kinetic energy $T_{\textrm{lab}}$) at constant lab angle
$\Theta_{\textrm{lab}}$. In this publication all given yields are transformed to
the double differential invariant cross section

\begin{equation}
  \label{eq:ddcs}
  f(p_{\textrm{lab}},\Theta_{\textrm{lab}}) = \frac{E}{p_{\textrm{lab}}^2} \frac{d^2\sigma}{dp_{\textrm{lab}}d\Omega} . 
\end{equation}

In this context the term "backward" needs a precise definition. One
possibility would be to define as "backward" the region of lab angles
$\Theta_{\textrm{lab}} >$~90~degrees. The present paper uses instead a definition which
refers to the cms frame with the basic variables Feynman $x_F$ and
transverse momentum $p_T$, defining as "backward" the particle yields at $x_F <$~0.
This allows a clear separation of the projectile
fragmentation region at positive $x_F$ with a limited feed-over into negative 
$x_F$ and the target fragmentation region at negative $x_F$ with a limited
feed-over into positive $x_F$.
At the same time the notion of "kinematic limit" in participant fragmentation is clearly brought out 
at $x_F$~=~$\pm$~1 and the contributions from intranuclear cascading may be clearly
visualized and eventually separated.

\begin{figure}[b]
  \begin{center}
  	\includegraphics[width=13.5cm]{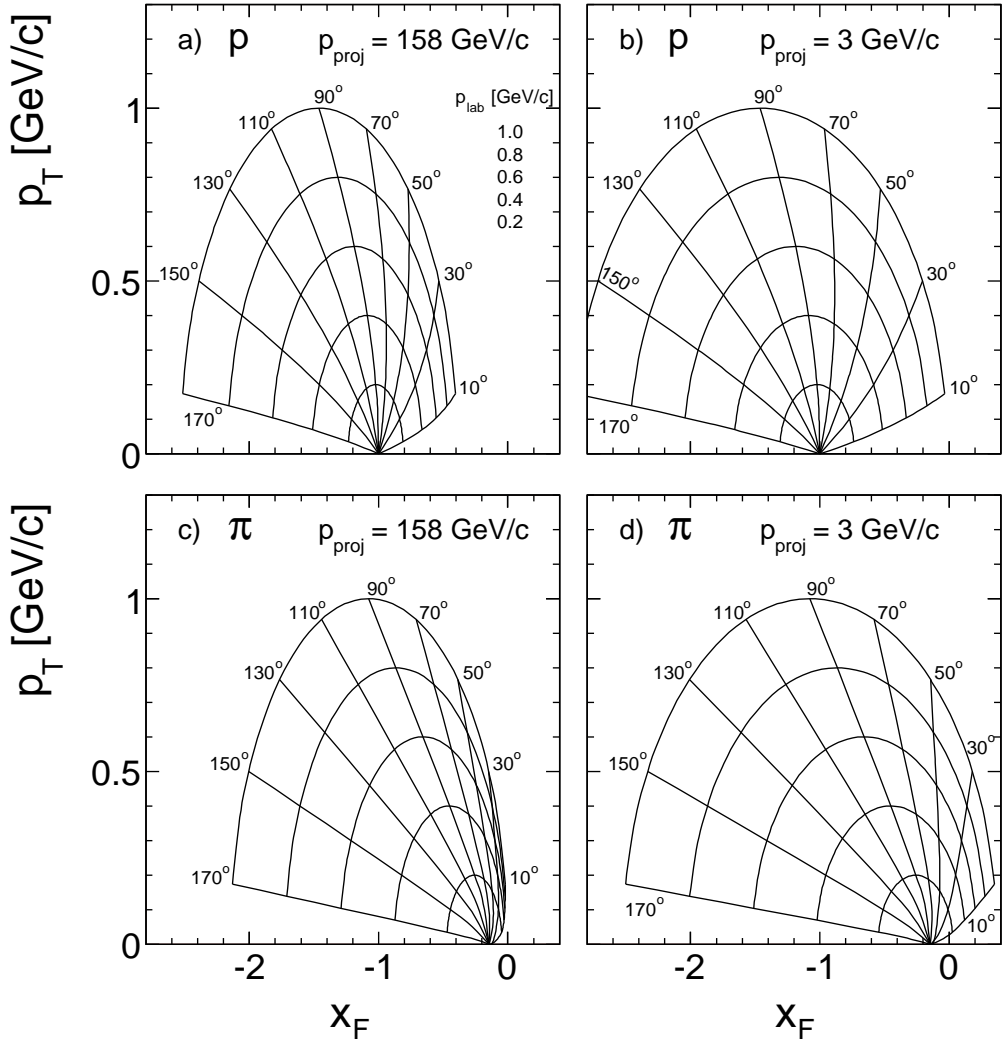}
 	    \caption{Lines of constant $p_{\textrm{lab}}$ and $\Theta_{\textrm{lab}}$ in the
               cms frame spanned by Feynman $x_F$ and $p_T$ for protons and pions
               at two different projectile momenta, a) protons at 158~GeV/c,
               b) protons at 3~GeV/c, c) pions at 158~GeV/c and d) pions at 3~GeV/c}
  	 \label{fig:kinematics}
  \end{center}
\end{figure}
 
The correlation between the two pairs of variables is presented
in Fig.~\ref{fig:kinematics} which shows lines of constant $p_{\textrm{lab}}$ and $\Theta_{\textrm{lab}}$ in the
coordinate frame of $x_F$ and $p_T$ for protons and pions for the two values of
projectile momentum at 158 and 3~GeV/c which are representative of the 
typical range of interaction energies discussed in this paper.

Several comments are due in this context. The definition of Feynman
$x_F$ has been modified from the standard one,

\begin{equation}
	x_F = \frac{p_l}{p_{\textrm{max}}} = \frac{p_l}{\sqrt{s}/2}
	\label{eq:xf}
\end{equation}              
to

\begin{equation}
	x_F = \frac{p_l}{\sqrt{s/4 - m_p^2}} 
\end{equation}
with $m_p$ the proton mass. This takes care of baryon number
conservation and regularises the kinematic borders at low interaction
energies. The $s$ dependence in Fig.~\ref{fig:kinematics} is small to negligible for
lab angles above about 50~degrees both for pions and protons but
becomes noticeable at small $\Theta_{\textrm{lab}}$. If at SPS energy the
full range of lab momenta up to 1.4~GeV/c and angles above 10~degrees 
is confined to the backward region both for protons and
pions, the coverage for pions extends to positive $x_F$ at
low lab angles and low beam momenta.

Another remark concerns the overlap between target fragmentation
and nuclear cascading. For protons, at all lab angles above about 70~degrees 
the kinematic limit for fragmentation of a target nucleon
at rest in the lab system is exceeded. For pions on the other
hand this is not the case as their $x_F$ value for $p_{\textrm{lab}}$~=~0 is at

\begin{equation}
	|x_F| = \frac{m_{\pi}}{m_p} = 0.148
\end{equation}                               

This means that over the full range of lab angles and up to large
$p_{\textrm{lab}}$ values the contribution from target participants mixes with
the nuclear component. The separation of the two processes therefore
becomes an important task, see Sect.~\ref{sec:separation} of this paper.

A last remark is due to the limits of experimental coverage. All
existing experiments run out of statistics at cross section levels
of about 10~$\mu$b, that is about 4 orders of magnitude below the maximum
yields. As visible from the momentum ranges indicated in Tables 1 and 2,
this corresponds to a typical upper momentum cut-off in the region of
1~GeV/c.

%
%
\section{The Experimental Situation}
\vspace{3mm}
\label{sec:exp_sit}

The backward phase space coverage in p+C interactions is surprisingly
complete if compared with the forward direction and even with
the available data in the elementary p+p collisions. This is apparent 
from the list of experiments given in Tables~\ref{tab:prot_data} and \ref{tab:pion_data} with their ranges in beam
momentum, lab angle, and lab momentum. Although some effort has been
spent to pick up all published results, this list is not claimed
to be exhaustive as some results given as "private communication",
in conference proceedings or unpublished internal reports might
have escaped attention.

\begin{table}[h]
\renewcommand{\tabcolsep}{0.2pc} 
\renewcommand{\arraystretch}{1.25}
\scriptsize
\begin{center}
\begin{tabular}{|cccccccc|}
\hline
interaction           &    Experiment            & projectile momentum &  lab angle coverage                &  $p_{\textrm{lab}}$ coverage   &  number of     &   \multicolumn{2}{c|}{errors [\%]} \\
                      &                          &       (GeV/c)       &           (degrees)                    &      (GeV/c)                   &  data points   &  $\langle \sigma_{stat} \rangle$ & $\langle \sigma_{syst} \rangle$ \\ \hline
\multirow{6}{6mm}{p+C}& Bayukov \cite{bayukov}   &        400          &  70, 90, 118, 137, 160                 &      0.4--1.3                &       35       &  6 & 20         \\
                      & NA49 \cite{pc_proton}    &        158          &     10, 20, 30, 40                     &      0.3--1.6                &       40       &  7 &  5     \\
                      & Belyaev \cite{belyaev_prot} & 17, 23, 28, 34 ,41, 49, 56  &    159                         &      0.3--1.2             &      125       &  5 & 15   \\
                      & HARP-CDP  \cite{harp-cdp}    &    3, 5, 8, 12, 15  &  25, 35, 45, 55, 67, 82, 97, 112       &      0.45--1.5               &      202       &  4 &  6      \\
                      & Burgov \cite{burgov}     &    2.2, 6.0, 8.5    &            162                         &      0.35--0.85              &       36       & 15 &  5     \\
                      & Bayukov \cite{bayukov1}  &    1.87, 4.5, 6.57  &            137                         &      0.3--1.1                &       55       & 10 & 20       \\
                      & Geaga  \cite{geaga}      &  1.8, 2.9, 5.8      &   180                                  &      0.3--1.0                &       50       & 17 & 15       \\
                      & Frankel \cite{frankel}   &         1.22        &            180                         &      0.45--0.8               &        6       &  7 &    \\
                      & Komarov \cite{komarov}   &         1.27        &  105, 115, 122, 130, 140, 150, 160     &      0.34--0.54          &  $\scriptstyle \sim$200  & 8  &  15  \\ \hline
n+C                   & Franz \cite{franz}        &   0.84, 0.99, 1.15  & 51, 61, 73, 81, 98, 120, 140, 149, 160 &       0.3--0.8               &      553       & 5  & 10 \\ \hline
\end{tabular}
\end{center}
\caption{Data sets for proton production in p+C and n+C
         collisions from seven experiments giving the ranges
         covered in projectile momentum, lab angle, and lab
         momentum, the number of measured data points and errors}
\label{tab:prot_data}
\end{table}

\begin{table}[h]
\renewcommand{\tabcolsep}{0.2pc} 
\renewcommand{\arraystretch}{1.25}
\scriptsize
\begin{center}
\begin{tabular}{|ccccccc|}
\hline
Experiment               & projectile momentum &  lab angle coverage                            &  $p_{\textrm{lab}}$ coverage                &  number of    &   \multicolumn{2}{c|}{errors [\%]} \\
                         &      (GeV/c)        &           (degrees)                            &      (GeV/c)                              &  data points  &  $\langle \sigma_{stat} \rangle$ & $\langle \sigma_{syst} \rangle$  \\ \hline
Nikiforov \cite{niki}    &        400          &    70, 90, 118, 137, 160                       &       0.2--1.3                           &   59          &  12  &  \\
NA49 \cite{pc_pion}    &        158          & 5, 10, 15, 20, 25, 30, 35, 40, 45              &       0.1--1.2                           &  174          &   5  &  4  \\
Belyaev \cite{belyaev_pion}& 17, 22, 28, 34, 41, 47, 57  &    159                               &       0.25--1.0                           &      218      &  4 & 15   \\
Abgrall  \cite{abgrall}    &    31               &  0.6--22.3                                    &       0.2--18                           &    624        &   6 & 7   \\
HARP-CDP \cite{harp-cdp}     &   3, 5, 8, 12, 15   & 25, 35, 45, 55, 67, 82, 97, 112                &       0.2--1.6                           &  829          &   6  &  8 \\
HARP \cite{harp}      &   3, 5, 8, 12       & 25, 37, 48, 61, 72, 83, 95, 106,117            &       0.125--0.75                         &  605          &  \multicolumn{2}{c|}{12}   \\
Burgov \cite{burgov1}    &    2.2, 6.0, 8.5    &         162                                    &       0.25--0.6                          &   29          &   \multicolumn{2}{c|}{20}  \\
Baldin \cite{baldin}     &       6.0, 8.4      &         180                                    &       0.2--1.25                          &   45          &   \multicolumn{2}{c|}{10}  \\
Cochran \cite{cochran}   &         1.38        & 15, 20, 30, 45, 60, 75, 90, 105, 120, 135, 150 &       0.1--0.7                           &  199          &   3   &  12  \\
Crawford \cite{crawford} &         1.20        &          22.5, 45, 60, 90, 135                 &       0.1--0.4                           &   50          &   8   &   7  \\
\hline
\end{tabular}
\end{center}
\caption{Data sets for pion production in p+C
         collisions from seven experiments giving the ranges
         covered in projectile momentum, lab angle, and lab
         momentum, the number of measured data points and errors}
\label{tab:pion_data}
\end{table}

For secondary protons, Table~\ref{tab:prot_data}, the important amount of 
low energy n+C data by Franz et al. \cite{franz} has been added to the survey as
the isospin factors for the transformation into p+C results
have been studied and determined with some precision, see Sect.~\ref{sec:prot}.

For secondary pions, Table~\ref{tab:pion_data}, the situation is somewhat complicated
by the fact that two independent sets of results have been published 
by the HARP-CDP \cite{harp-cdp} and the HARP \cite{harp} groups, based
on identical input data obtained with the same detector.
An attempt to clarify this partially contradictory situation 
is presented in Sect.~\ref{sec:harp} of this paper.

Unfortunately, no commonly agreed scale in the three basic variables
$\Theta_{\textrm{lab}}$, $p_{\textrm{lab}}$ and $p_{\textrm{beam}}$ of the double-differential cross sections 
has been defined by the different collaborations providing the
data contained in Tables~\ref{tab:prot_data} and \ref{tab:pion_data}. This leads to the fact that not
a single couple out of the more than 3500 data points contained in
these Tables may be directly compared. The application of an
interpolation scheme as described in Sect.~\ref{sec:comp} is therefore an
absolute necessity. Ideally the thus obtained interpolated cross
sections would form an internally consistent sample of results
which would be coherent within the given experimental errors.
As will become apparent in the following data comparison, this
assumption is surprisingly well fulfilled for the majority of
the experiments. Only four of the 20 quoted groups of results
fall significantly out of this comparison; those will be discussed
in Sect.~\ref{sec:dev_exp} of this paper. In this sense the overall survey
of the backward proton and pion production results in a powerful
constraint for the comparison with any new data sample.

%
%
\section{Data comparison}
\vspace{3mm}
\label{sec:comp}

As stated above the main problem in bringing the wealth of available
data into a consistent picture is given by the generally disparate
position in phase space and interaction energy of the different
experiments. The triplet of lab variables given by the beam momentum
$p_{\textrm{beam}}$, the lab momentum $p_{\textrm{lab}}$ and the lab angle $\Theta_{\textrm{lab}}$ has been
used for the establishment of the following interpolation scheme.
In addition and of course, the statistical and systematic errors 
have to be taken into account in the data comparison.

%
%
\subsection{Errors}
\vspace{3mm}
\label{sec:errors}

The last columns of Tables~\ref{tab:prot_data} and \ref{tab:pion_data} contain some information about 
the statistical and systematic errors of the different experiments.
The given numbers are to be regarded as mean values excluding some
upward tails as they are inevitable at the limits of the covered
phase space in particular for the statistical uncertainties. In
some cases only rudimentary information about the systematic errors
is available or the systematic and statistical errors are even
combined into one quantity. In the latter cases these values are
given in between the respective columns of Table~\ref{tab:pion_data}.

Inspection of these approximate error levels reveals a rather broad
band of uncertainties ranging from about 4\% to about 20\%, the
latter limit being generally defined by overall normalization
errors. The presence of extensive data sets well below the 10\% range
of both statistical and systematic errors gives  however some
hope that a resulting overall consistency on this level might
become attainable by the extensive use of data interpolation.

The term "interpolation" is to be regarded in this context as
a smooth interconnection of the data points in any of the three
phase space variables defined above. This interconnection is 
generally done by eyeball fits which offer, within the error
limits shown above, sufficient accuracy. If the distributions
in $\Theta_{\textrm{lab}}$ and interaction energy are anyway not describable
by straight-forward arithmetic parametrization, the $p_{\textrm{lab}}$
dependences are, as discussed in Sect.~\ref{sec:pldep} below, in a majority 
of cases approximately exponential. In these cases exponential
fits have been used if applicable.

As additional constraint physics asks of course for smoothness
and continuity in all three variables simultaneously. Therefore
the resulting overall data interpolation has to attempt 
a three-dimensional consistency.

If the data interpolation helps, by the inter-correlation of data
points, to reduce the local statistical fluctuations, it does
of course not reduce the systematic uncertainties. It is rather
on the level of systematic deviations that the consistency of
different experimental results is to be judged. It will become
apparent from the detailed discussion described below that
the majority of the quoted experiments allows for the establishment
of a surprisingly consistent overall description in all three
variables.

%
%
\subsection{Dependence on cms energy $s$}
\vspace{3mm}
\label{sec:sdep}

As the data discussed here span an extremely wide range of cms energy
from close to production threshold to the upper range of Fermilab
energies, a suitable compression of the energy scale has been
introduced in order to be able to present the results in a close-to-equidistant 
fashion against energy. The form chosen here is the
variable $1/\sqrt{s}$. This choice is suggested by the considerable
amount of work invested in studying the approach of hadronic cross
sections to the scaling limit at high energy in the 1970's \cite{whitmore}.
In fact the Regge parametrization suggested a smooth dependence
of the cross sections as $s^{-\alpha}$, with $\alpha$~=~0.25--0.5 depending
on the choice of trajectories involved. Such behaviour was indeed
found experimentally. In the present study the cross sections
turn out to have only a mild $1/\sqrt{s}$ dependence for $\sqrt{s} \gtrsim$~5~GeV,
a dependence which is however different for pions and protons.
This dependence is strongly modified below $\sqrt{s} \sim$~2.5~GeV due to
threshold effects.

%
%
\subsection{Angular dependence}
\vspace{3mm}
\label{sec:thdep}

A convenient and often used scale for the lab angle dependence
is given by $cos(\Theta_{\textrm{lab}})$. This scale has the advantage of 
producing shapes that are again to zero order exponential.
Of course, continuity through $\Theta_{\textrm{lab}}$~=~180~degrees imposes an
approach to 180~degrees with tangent zero. As the data samples
are generally not measured at common values of $\Theta_{\textrm{lab}}$, a
fixed grid of angles has been defined based on the $\Theta_{\textrm{lab}}$
values of the HARP-CDP experiment \cite{harp-cdp} dominating the range
from 25 to 112~degrees. Measured  values down to 10~degrees and
in the higher angular range at 137, 160, and 180~degrees have been added. 
Measurements not corresponding to these grid values are
interpolated using the $cos(\Theta_{\textrm{lab}})$ distributions specified
below.

%
%
\subsection{Lab momentum dependence}
\vspace{3mm}
\label{sec:pldep}

All data discussed here have been transformed into invariant
cross sections (\ref{eq:ddcs}). This facilitates the presentation in different 
coordinate systems and eliminates the trivial approach
of the phase space element to zero with decreasing momentum.
In addition, most of the invariant $p_{\textrm{lab}}$ distributions are close 
to exponential within the measured $p_{\textrm{lab}}$ range. There are
notable deviations mostly at low momentum and in the lower (higher)
range of lab angles for pions and protons, respectively, as well
as in the approach to threshold. In these cases an eyeball fit
has been used which can be reliably performed within the error
margins indicated above.

\begin{figure}[h]
  \begin{center}
  	\includegraphics[width=12cm]{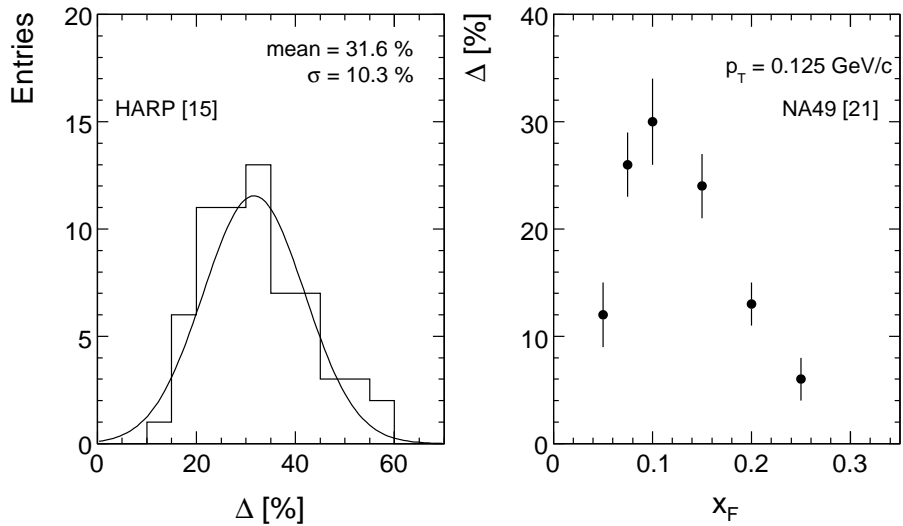}
 	 	\caption{a) Distribution of the deviation of the data points
                 at $p_{\textrm{lab}}$~=~0.125~GeV/c from the exponential fits for $\pi^+$ and $\pi^-$
                 at all angles and beam momenta, b) Deviation of $\pi^+$ cross sections
                 at $p_T$~=~0.125~GeV/c from exponential fits to the higher $p_T$ region
                 in p+p interactions as a function of $x_F$}
  	 \label{fig:first_point}
  \end{center}
\end{figure}

At low lab momenta physics requires a deviation from the exponential shape 
as the invariant cross sections must approach $p_{\textrm{lab}}$~=~0 with tangent zero. 
This limit appears in general at $p_{\textrm{lab}} <$~0.2~GeV/c
for pions and $p_{\textrm{lab}} <$~0.5~GeV/c for protons. The data presented here
fall practically all above these momentum limits. Only the HARP
experiment \cite{harp} gives results at $p_{\textrm{lab}}$~=~0.125~GeV/c for pions where
indeed a substantial deviation from the exponential shape is
visible. This is shown in Fig.~\ref{fig:first_point} where the deviation from exponential
fits at this $p_{\textrm{lab}}$ is given in percent for all angles and beam
momenta together with similar deviations observed in p+p interactions \cite{pp_pion}.

A number of examples of momentum distributions for protons and pions
is given in the following Figs.~\ref{fig:pldist_harpcdp} and \ref{fig:plabex} which show the
invariant cross sections as a function of $p_{\textrm{lab}}$ and the corresponding
exponential fits

\begin{equation}
	\label{eq:slope}
	f(p_{\textrm{lab}},\Theta_{\textrm{lab}},p_{\textrm{beam}}) = A(\Theta_{\textrm{lab}},p_{\textrm{beam}})*exp(-p_{\textrm{lab}}/B(\Theta_{\textrm{lab}},p_{\textrm{beam}}))
\end{equation}
which are, whenever necessary, supplemented by hand interpolations into the non-exponential regions.

A first group of distributions in the medium angular range at
45 and 97~degrees is presented in Fig.~\ref{fig:pldist_harpcdp} for the HARP-CDP data
concerning protons and pions, including exponential fits.

\begin{figure}[h]
  \begin{center}
  	\includegraphics[width=14cm]{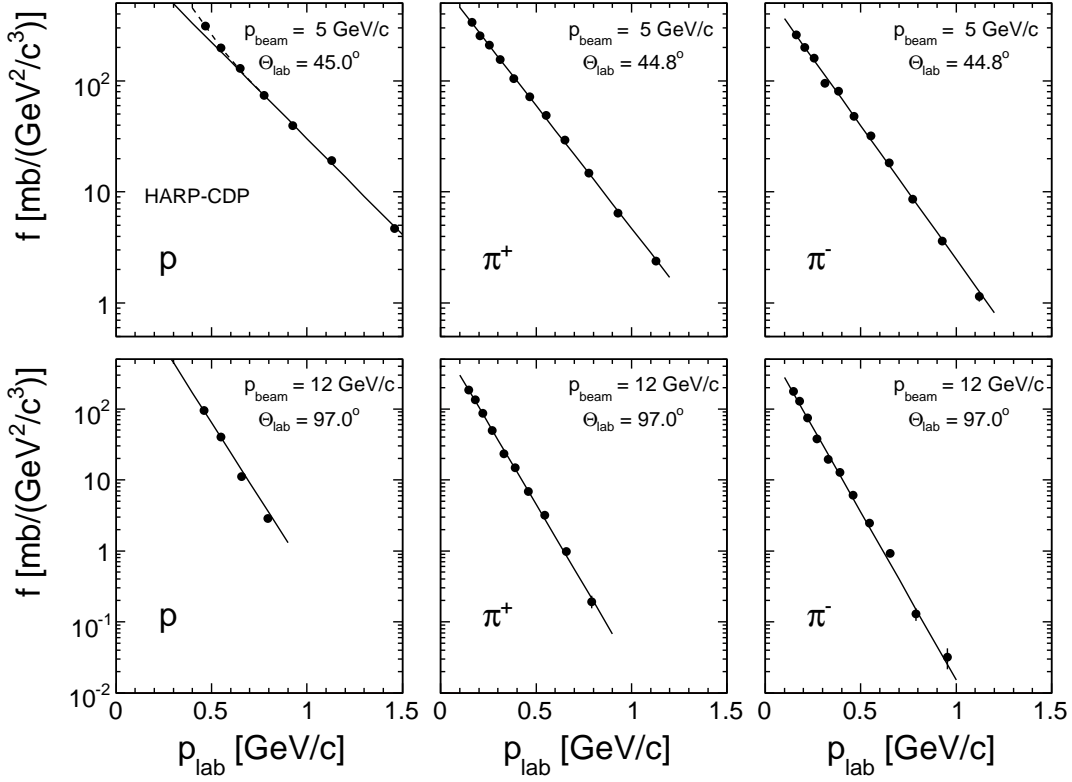}
 	 	\caption{Invariant cross sections for protons, $\pi^+$ and $\pi^-$
                 as a function of $p_{\textrm{lab}}$ at $\Theta_{\textrm{lab}}$~=~45 and 97~degrees. 
                 Full lines: exponential fits. Broken line:
                 hand-interpolation into the non-exponential region}
  	 \label{fig:pldist_harpcdp}
  \end{center}
\end{figure}

\begin{figure}[h]
  \begin{center}
  	\includegraphics[width=14cm]{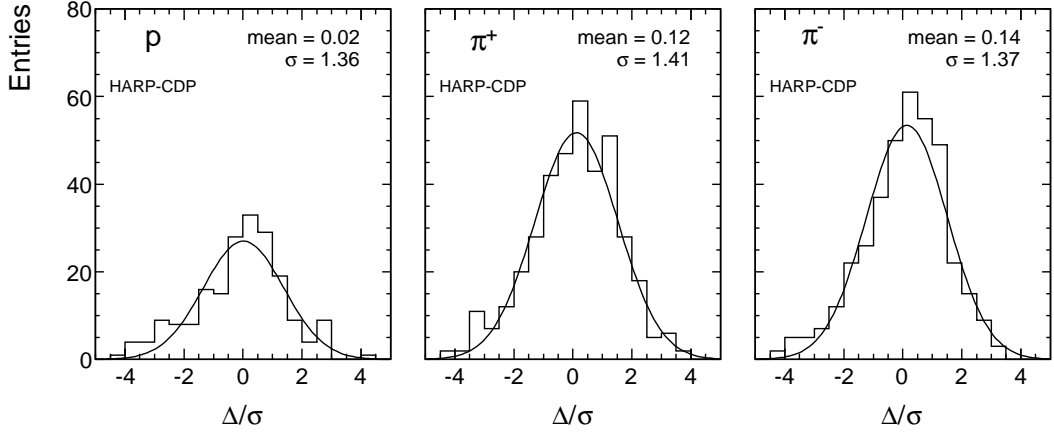}
 	 	\caption{Normalized residual distributions for protons and
                 pions for the complete set of beam momenta and angles of the
                 HARP-CDP data with the exception of a few points at low angles
                 and momenta which clearly exhibit non-exponential behaviour}
  	 \label{fig:diff_dist}
  \end{center}
\end{figure}

Evidently the exponential shape is within errors in general a good approximation 
to the momentum dependence. More quantitative
information is contained in the normalized residual distributions
of the data points,

\begin{equation}
   r_{\textrm{norm}} = \Delta/\sigma     
\end{equation}
where $\Delta$ is the difference between data and fit and sigma the
statistical error of the given data point. Should the fit describe
the physics and should systematic effects be negligible, the
distribution of $r_{\textrm{norm}}$ is expected to be Gaussian with rms equal
to unity. The $r_{\textrm{norm}}$ distributions are given for the totality
of the HARP-CDP data in Fig.~\ref{fig:diff_dist}.

These distributions are well described by centred Gaussians. The
resulting rms values are however somewhat bigger than one signalling
systematic experimental effects or a deviation of physics from
the simple exponential parametrization. In view of the statistical
errors of 4\% to 6\% given by HARP-CDP (Tables~\ref{tab:prot_data} and \ref{tab:pion_data}) these deviations
are on the level of a few percent which is below the error margin
to be anyway expected from the present general data survey.

Further examples of $p_{\textrm{lab}}$ distributions from other experiments are
given in Fig.~\ref{fig:plabex} for a selection of particle type, beam momenta and angles.

\begin{figure}[h]
  \begin{center}
  	\includegraphics[width=14cm]{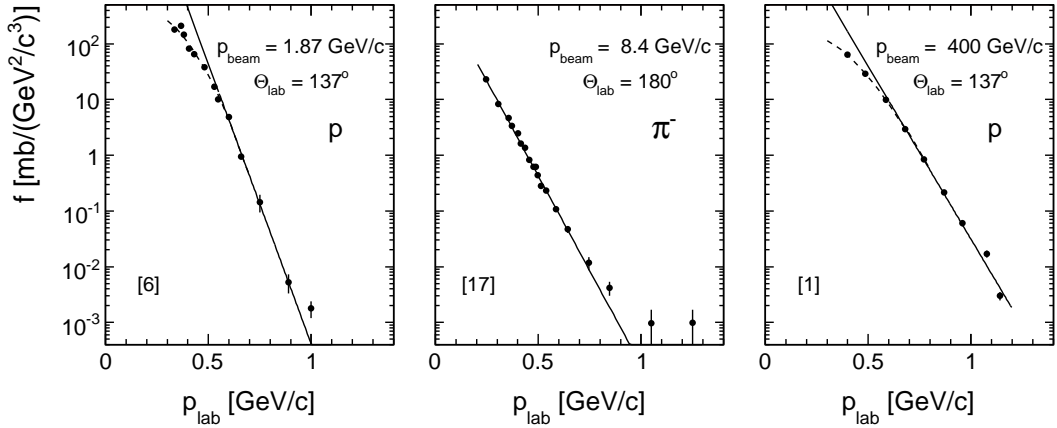}
 	 	\caption{Several examples of invariant cross sections as
                 a function of $p_{\textrm{lab}}$ for a variety of particle type, lab angle
                 and beam momenta including exponential fits (full lines)
                 and, when necessary, eyeball fits into the non-exponential
                 regions of $p_{\textrm{lab}}$ (broken lines).}
  	 \label{fig:plabex}
  \end{center}
\end{figure}

\begin{figure}[h]
  \begin{center}
  	\includegraphics[width=16.cm]{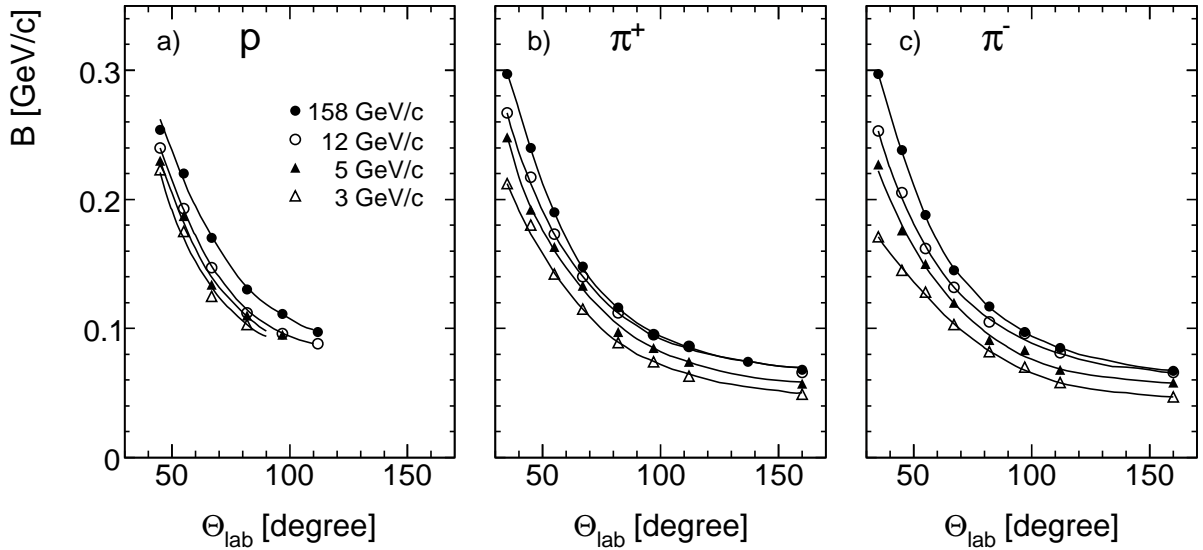}
 	  \caption{Inverse slopes $B(\Theta_{\textrm{lab}},p_{\textrm{beam}})$ as a function
              of $\Theta_{\textrm{lab}}$ for four beam momenta from 3 to 158~GeV/c,
              a) for protons, b) for $\pi^+$ and c) for $\pi^-$. The full lines
              are drawn to guide the eye}
  	 \label{fig:inverse}
  \end{center}
\end{figure}

Again the basically exponential shape of these distributions is
evident. Characterizing the exponential fits by their inverse slopes 
$B(\Theta_{\textrm{lab}},p_{\textrm{beam}})$ a smooth and distinct dependence 
on lab angle and beam momentum becomes visible as shown in Fig.~\ref{fig:inverse}.

Compared to the strong dependence of B on $\Theta_{\textrm{lab}}$ which ranges from
0.3 to 0.05~GeV/c, the only modest dependence on $p_{\textrm{beam}}$ of $\sim$0.03~GeV/c 
for beam momenta from 3 to 158~GeV/c is noticeable.

Following the above data parametrization a generalized
grid of $p_{\textrm{lab}}$ values between 0.2 and 1.2~GeV/c, in steps of 0.1~GeV/c,
may now be established. Concerning the lower and upper limits of
this grid, an extrapolation beyond the limits given by the
experimental values has been performed in some cases. This
extrapolation does not exceed the bin width of the respective
data lists and is therefore defendable in view of the generally
smooth, gentle and well-defined $p_{\textrm{lab}}$ dependences.

%
%
\subsection{Physics constraints}
\vspace{3mm}
\label{sec:ph}

In the absence of theoretical predictability in the soft sector
of the strong interaction, any attempt at bringing a multitude of experimental results 
into a common and consistent picture has to
rely on a minimal set of model-independent physics constraints.
In fact a "democratic" averaging of eventually contradictory data
sets would only add confusion instead of clarity.

%
%
\subsubsection{Continuity}
\vspace{3mm}
\label{sec:ph1}

Two examples of the continuity constraint have already been mentioned
above: invariant $p_{\textrm{lab}}$ distributions have to approach zero momentum
horizontally that is with tangent zero. The same is true for
angular distributions in their approach to 180~degrees.

%
%
\subsubsection{Smoothness}
\vspace{3mm}
\label{sec:ph2}

It is a matter of experimental experience in the realm of soft
hadronic interactions that in general distributions in any kind
of kinematic variable tend to be "smooth" in the sense of absence
of abrupt local upwards or downwards variations. The widespread
use of simple algebraic parametrizations has its origin in this
fact, specifically in the absence of local maxima and minima,
with the eventual exception of threshold behaviour of which
some examples will become visible below.

%
%
\subsubsection{Charge conservation and isospin symmetry}
\vspace{3mm}
\label{sec:ph3}

Charge conservation has of course to be fulfilled by any type
of experimental result. This means for instance that for the interaction 
of a positively charged projectile (proton) with an isoscalar nucleus (Carbon) 
the $\pi^+/\pi^-$ ratio has to be
greater or equal to unity over the full phase space invoking
isospin symmetry (and of course the experience from a wide
range of experimental results). The presence of data with
$\pi^+/\pi^- <$~1 therefore immediately indicates experimental
problems. The inspection of $\pi^+/\pi^-$ ratios has the further
advantage that a large part of the systematic uncertainties,
notably the overall normalization errors, cancel in this
ratio.

%
%
\subsubsection{Isospin rotation of secondary baryons and projectile}
\vspace{3mm}
\label{sec:ph4}

It has been shown that in proton induced nuclear collisions the
yields of the secondary protons and neutrons are related
by a constant factor of about 2 which is in turn related to the ratio 
of the basic nucleon-nucleon interaction \cite{anderson}. 
Similarly, when rotating the projectile isospin from proton to neutron, 
it has been predicted that the yield ratio of secondary protons from
proton and neutron projectiles

\begin{equation}
   R^{\textrm{p/n}} = \frac{f(p+C \rightarrow p')}{f(n+C \rightarrow p)}
\end{equation}
should be constant and equal to 2.5 for light nuclei \cite{boal}. The extensive and precise 
low-energy data set of Franz et al. \cite{franz} from n+C interactions has therefore been included 
in the present survey. These data present a welcome extension of the $1/\sqrt{s}$ scale into 
the region 0.47 to 0.49 which is not covered for most of the angular range with proton projectiles. 
As shown below, these data fit indeed very well, after re-normalization, into the general 
$1/\sqrt{s}$ dependence of secondary protons where the low energy data 
by Frankel et al. \cite{frankel} and Komarov et al. \cite{komarov} 
at angles between 112 and 180~degrees provide an independent control of the normalization.

%
%
\subsubsection{Establishing a consistent set of data}
\vspace{3mm}
\label{sec:ph5}

With these constraints in mind, and having established the
parametrization and interpolation of the $p_{\textrm{lab}}$ distributions as
discussed above, one may now proceed to the attempt at sorting
the 19 available experiments into a consistent global data
set. It would of course be rather surprising if all experiments
would fit into this global picture within their respective error
limits. In fact it turns out that this procedure establishes
a very strong constraint for possible deviations, as a large
majority of results is creating a perfectly consistent picture
both for protons and for pions. Only four of the 19 data sets cannot be brought 
into consistency with all other experiments without gravely affecting and contradicting 
the above constraints. These data are not included in the following global
interpolation scheme. They will be discussed separately in Sect.~\ref{sec:dev_exp} below.

%
%
\section{The proton data}
\vspace{3mm}
\label{sec:prot}

%
%
\subsection{$\mathbf {1/\sqrt{s}}$ dependence}
\vspace{3mm}
\label{sec:prot_sdep}

\begin{figure}[h]
  \begin{minipage}[t]{0.5\linewidth} 
  \begin{center}
  	\includegraphics[width=7.1cm]{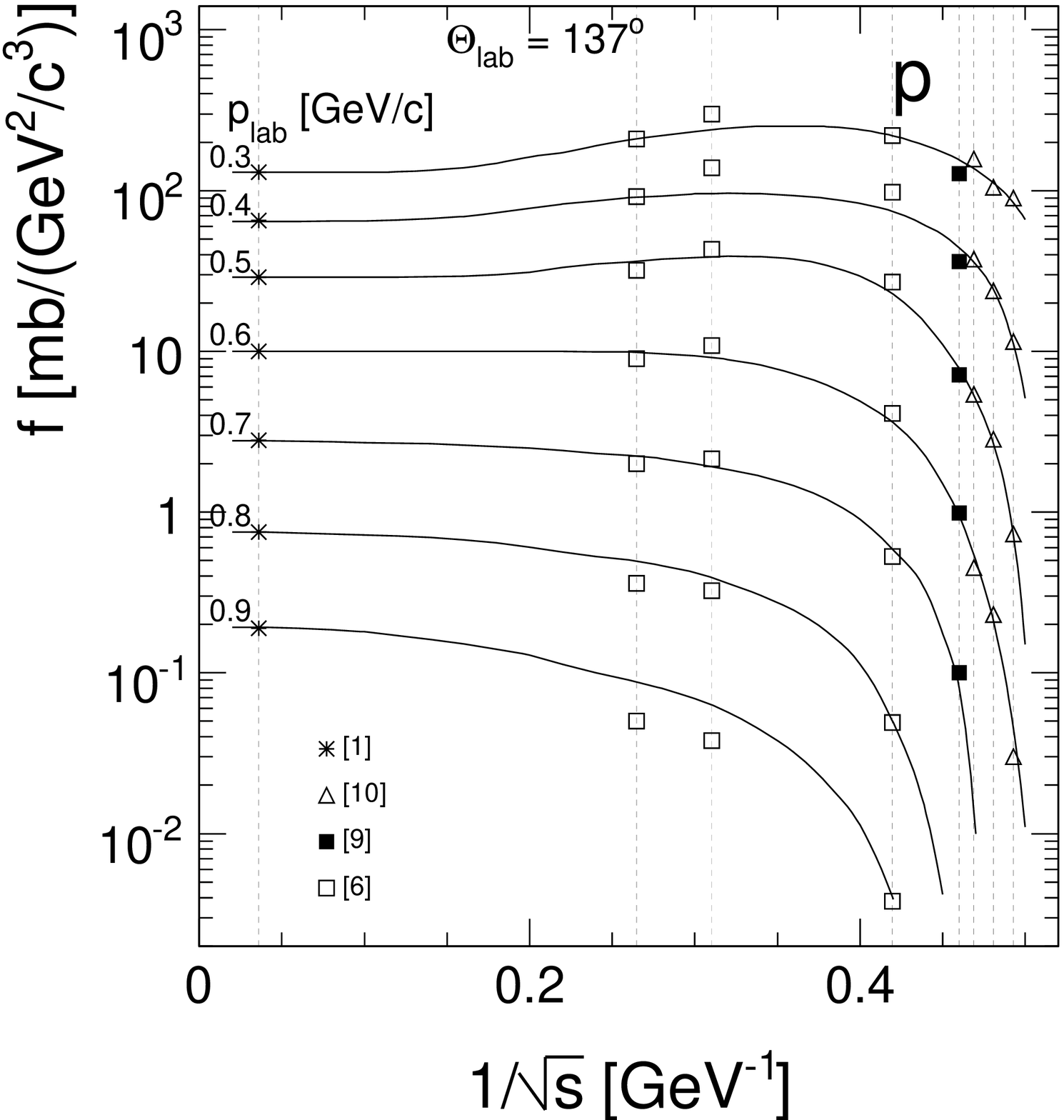}
  \end{center}
  \end{minipage}
  \begin{minipage}[t]{0.5\linewidth} 
  \begin{center}
  	\includegraphics[width=7.1cm]{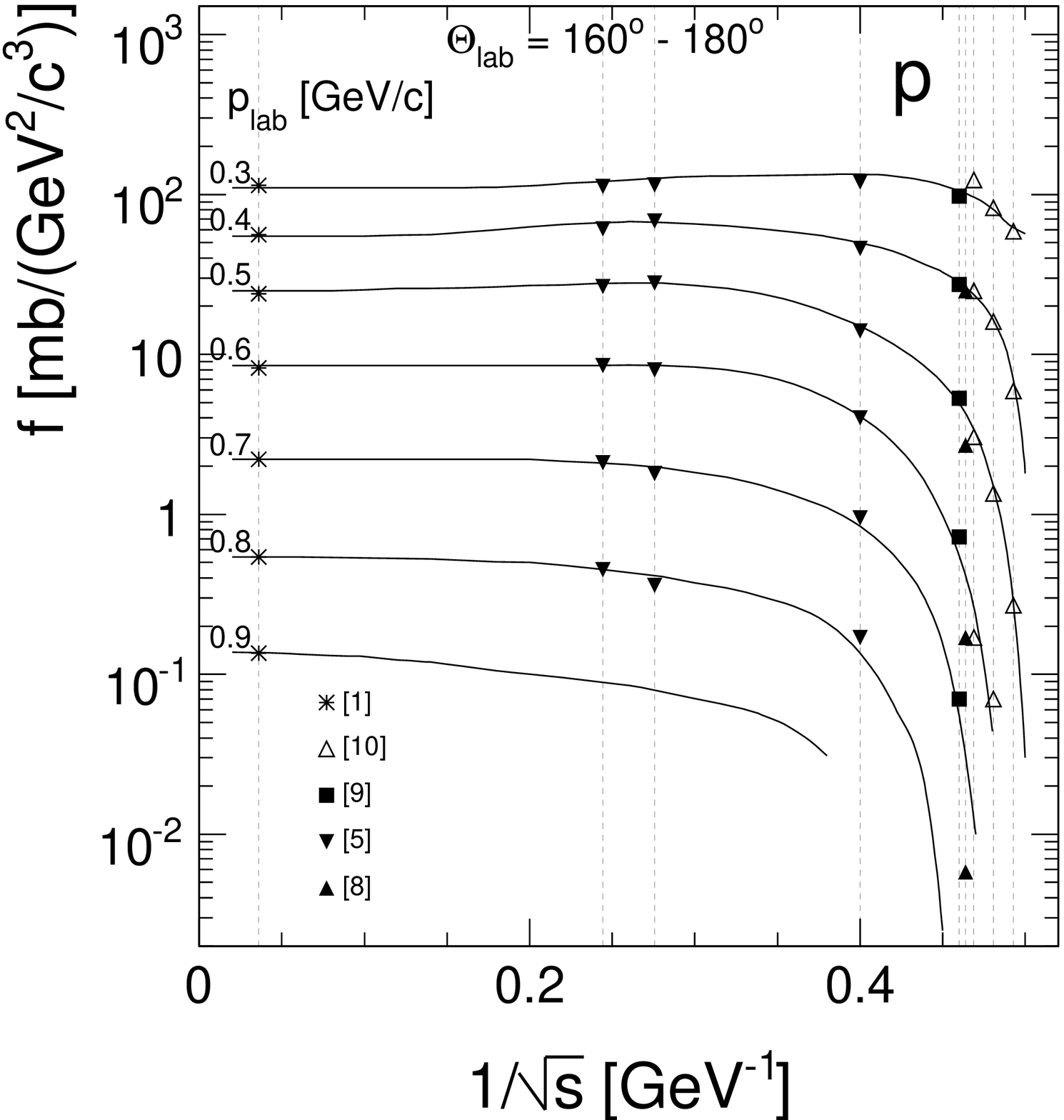}
  \end{center}
  \end{minipage}
  \caption{Invariant cross sections for protons in p+C
      collisions as a function of $1/\sqrt{s}$ at fixed $p_{\textrm{lab}}$ and
      $\Theta_{\textrm{lab}}$. The interpolated data points are indicated
      by symbols corresponding to the respective experiments
      in each panel. The solid lines represent the interpolation
      of the data}
  \label{fig:oneoversq}
\end{figure}

The invariant proton cross sections are shown in Fig.~\ref{fig:oneoversq} as a function of $1/\sqrt{s}$ 
for a grid of ten lab angles between 25 and 180~degrees and constant lab momenta 
between 0.3 and 1.2~GeV/c. The interpolated data points in each panel are identified 
by symbols corresponding to the different experiments.

\begin{figure*}[h]
  \begin{minipage}[h]{0.5\linewidth} 
  \begin{center}
  	\includegraphics[width=7.1cm]{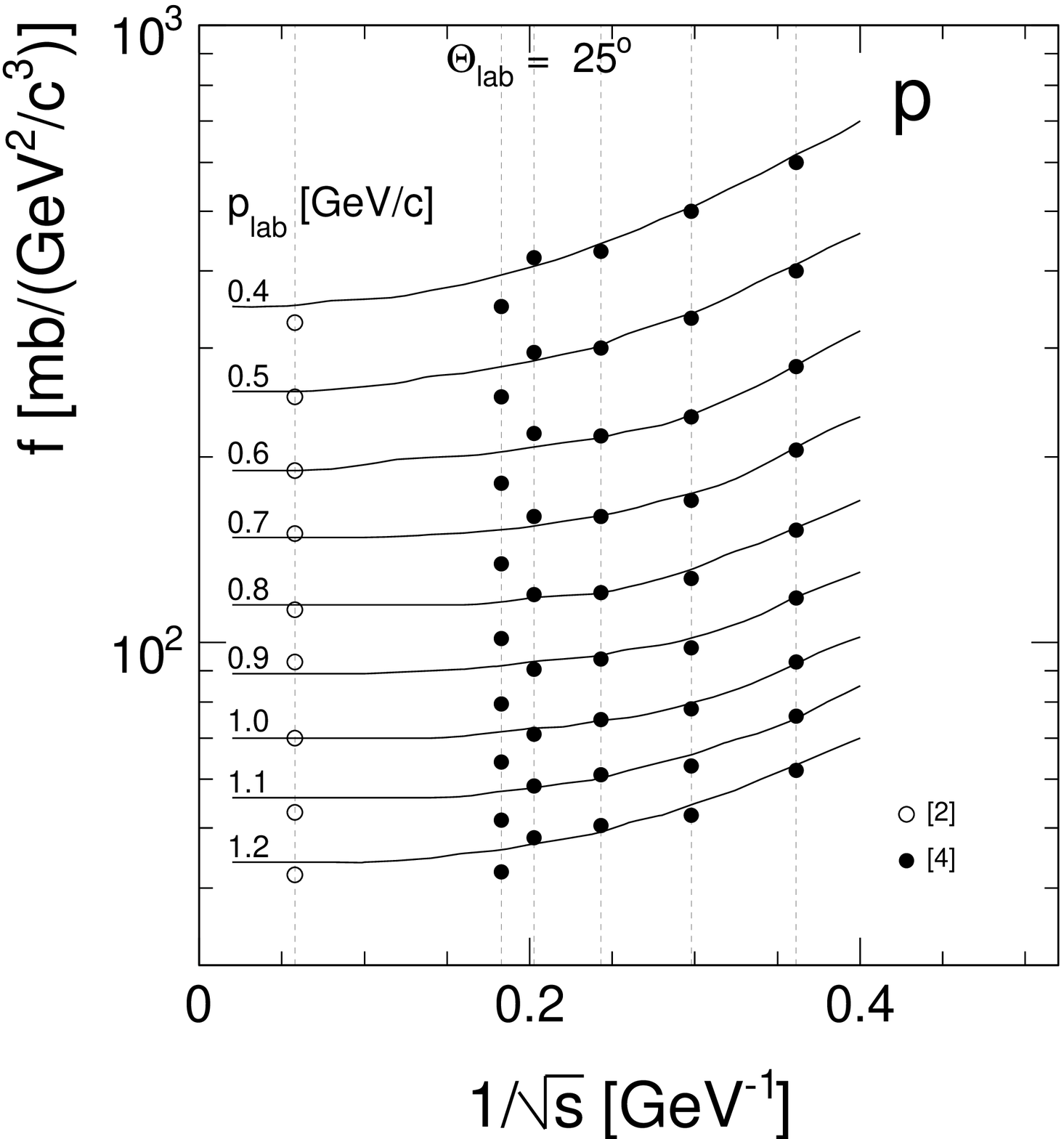}
  \end{center}
  \end{minipage}
  \begin{minipage}[h]{0.5\linewidth} 
  \begin{center}
  	\includegraphics[width=7.1cm]{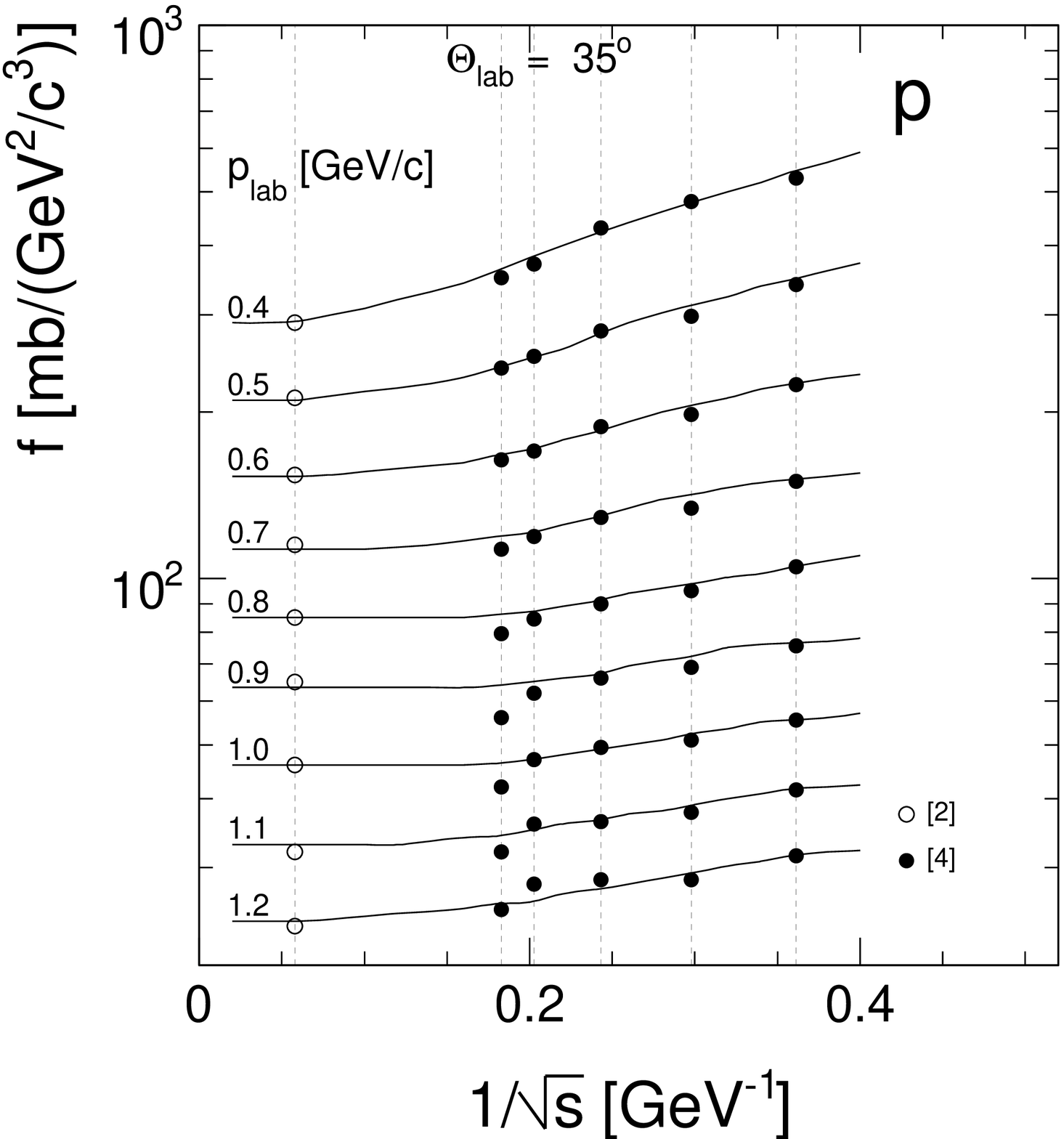}
  \end{center}
  \end{minipage}
\end{figure*}
\begin{figure*}[h]
  \begin{minipage}[h]{0.5\linewidth} 
  \begin{center}
  	\includegraphics[width=7.1cm]{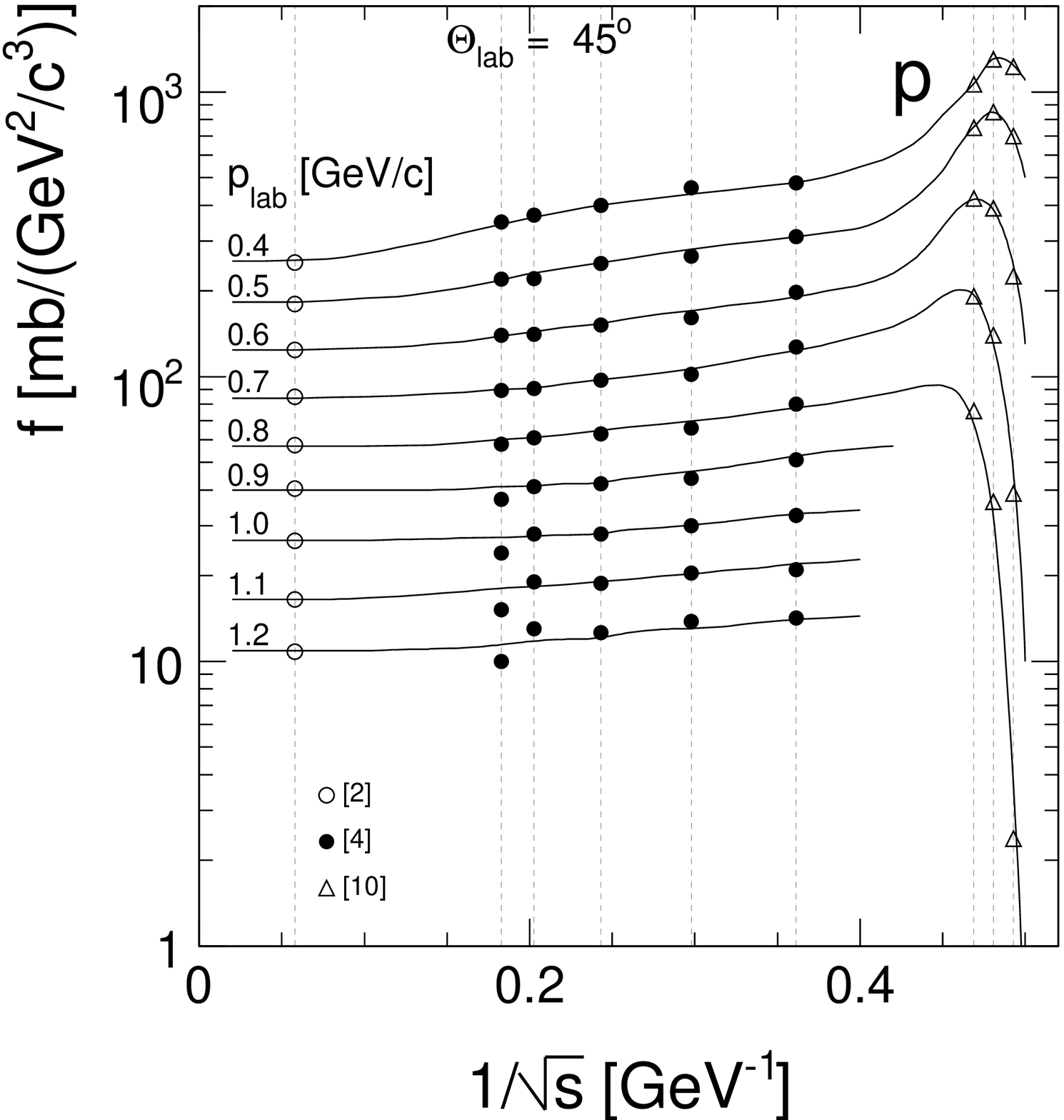}
  \end{center}
  \end{minipage}
  \begin{minipage}[h]{0.5\linewidth} 
  \begin{center}
  	\includegraphics[width=7.1cm]{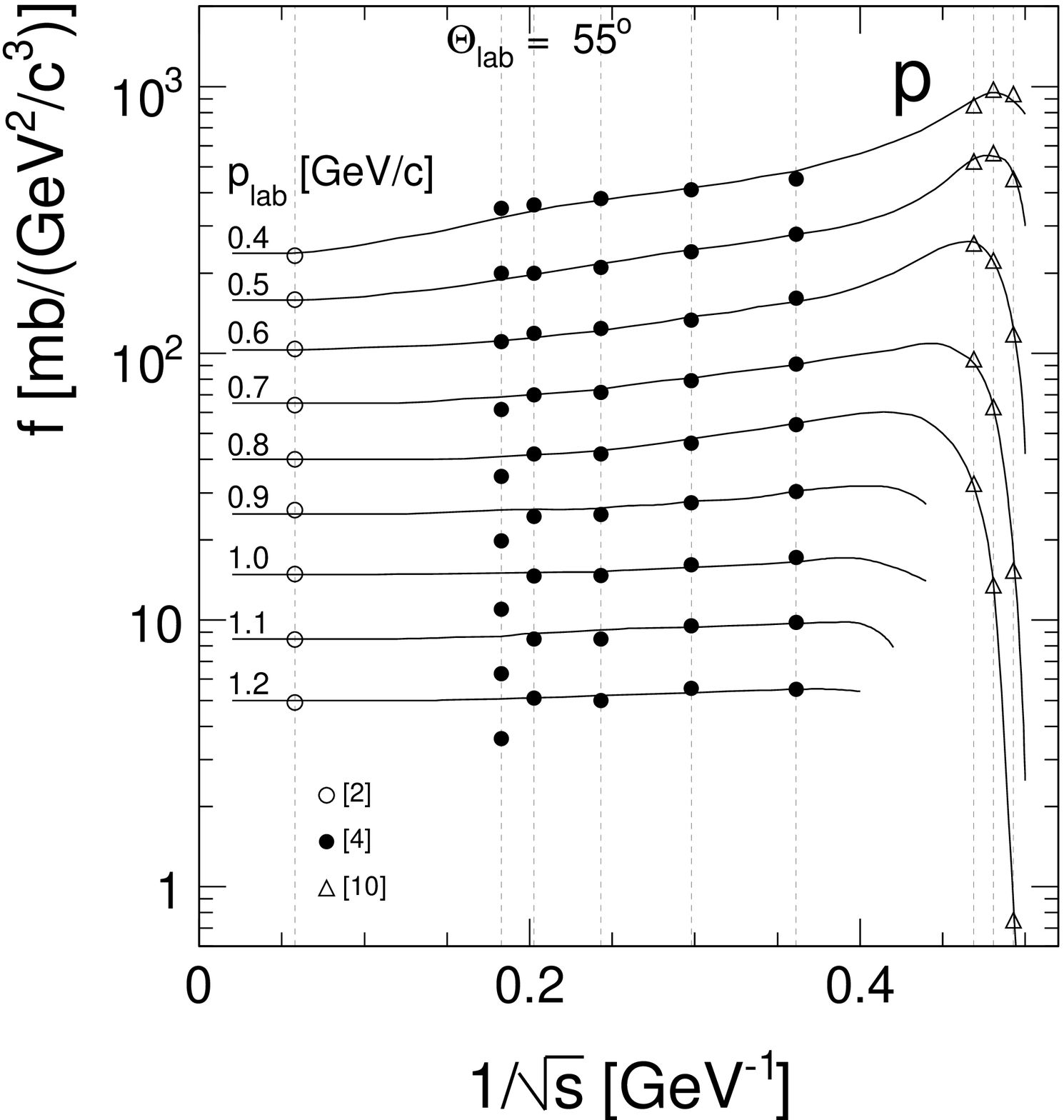}
  \end{center}
  \end{minipage}
\end{figure*}
\begin{figure*}[h]
  \begin{minipage}[h]{0.5\linewidth} 
  \begin{center}
  	\includegraphics[width=7.1cm]{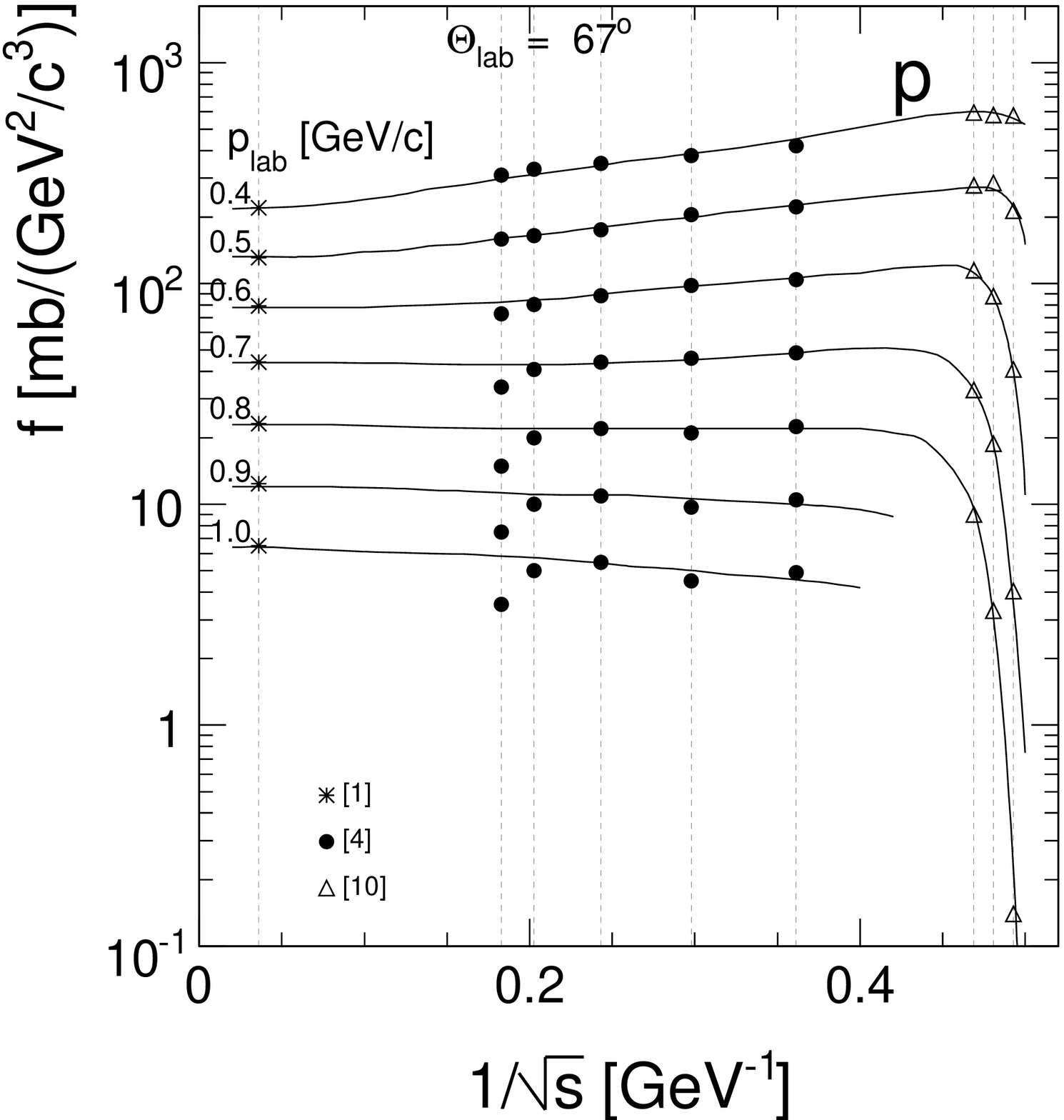}
  \end{center}
  \end{minipage}
  \begin{minipage}[h]{0.5\linewidth} 
  \begin{center}
  	\includegraphics[width=7.1cm]{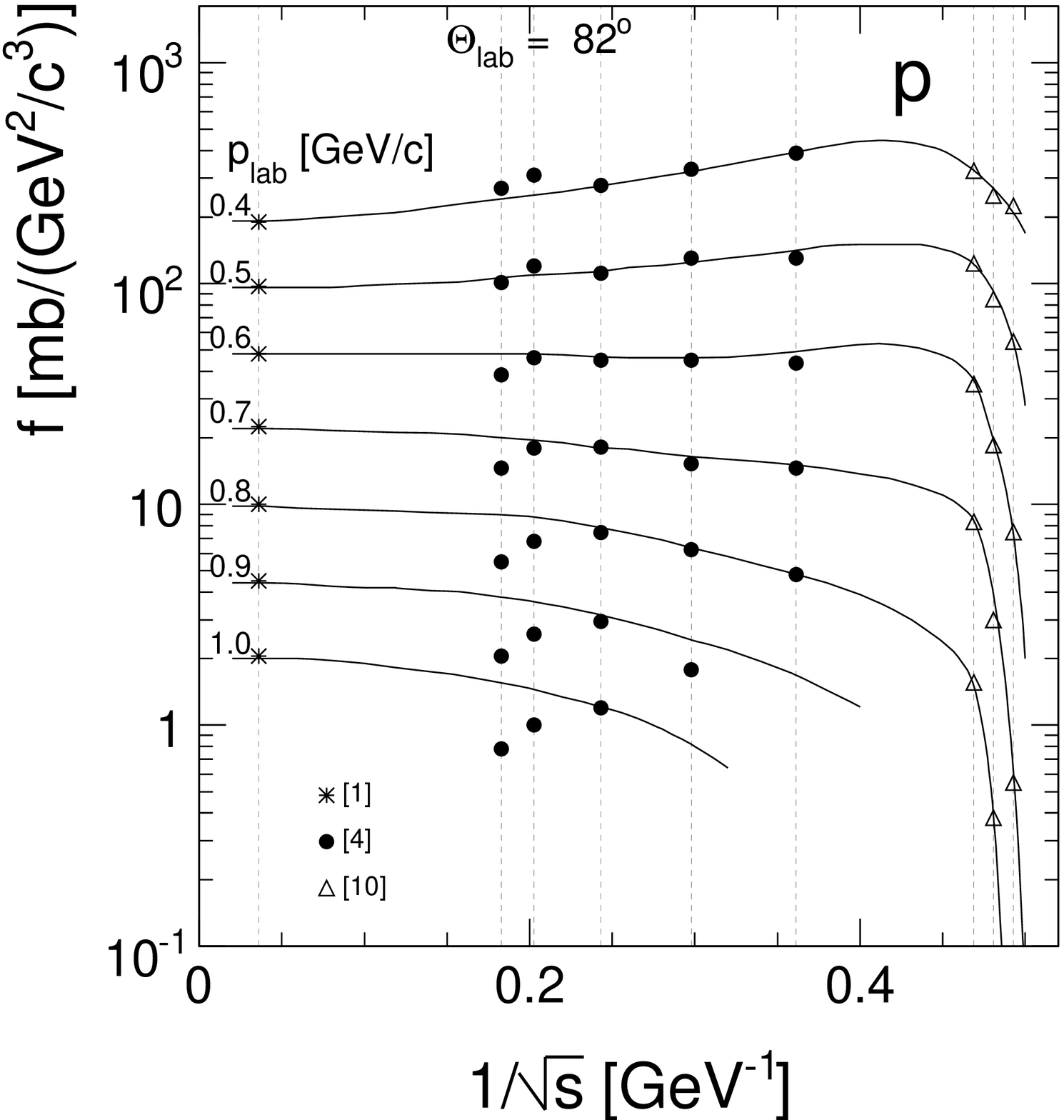}
  \end{center}
  \end{minipage}
\end{figure*}
\begin{figure*}[h]
  \begin{minipage}[t]{0.5\linewidth} 
  \begin{center}
  	\includegraphics[width=7.1cm]{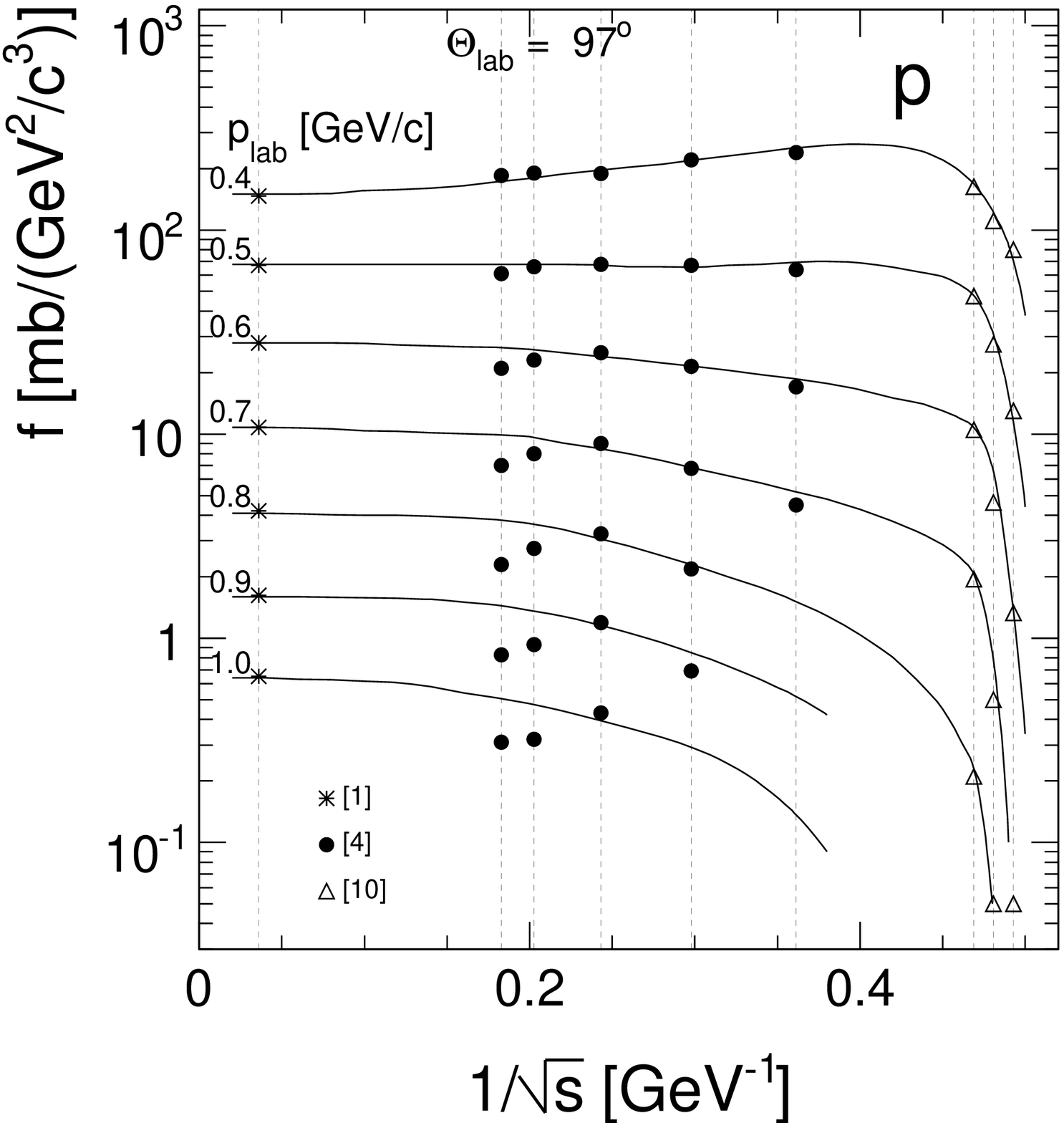}
  \end{center}
  \end{minipage}
  \begin{minipage}[t]{0.5\linewidth} 
  \begin{center}
  	\includegraphics[width=7.1cm]{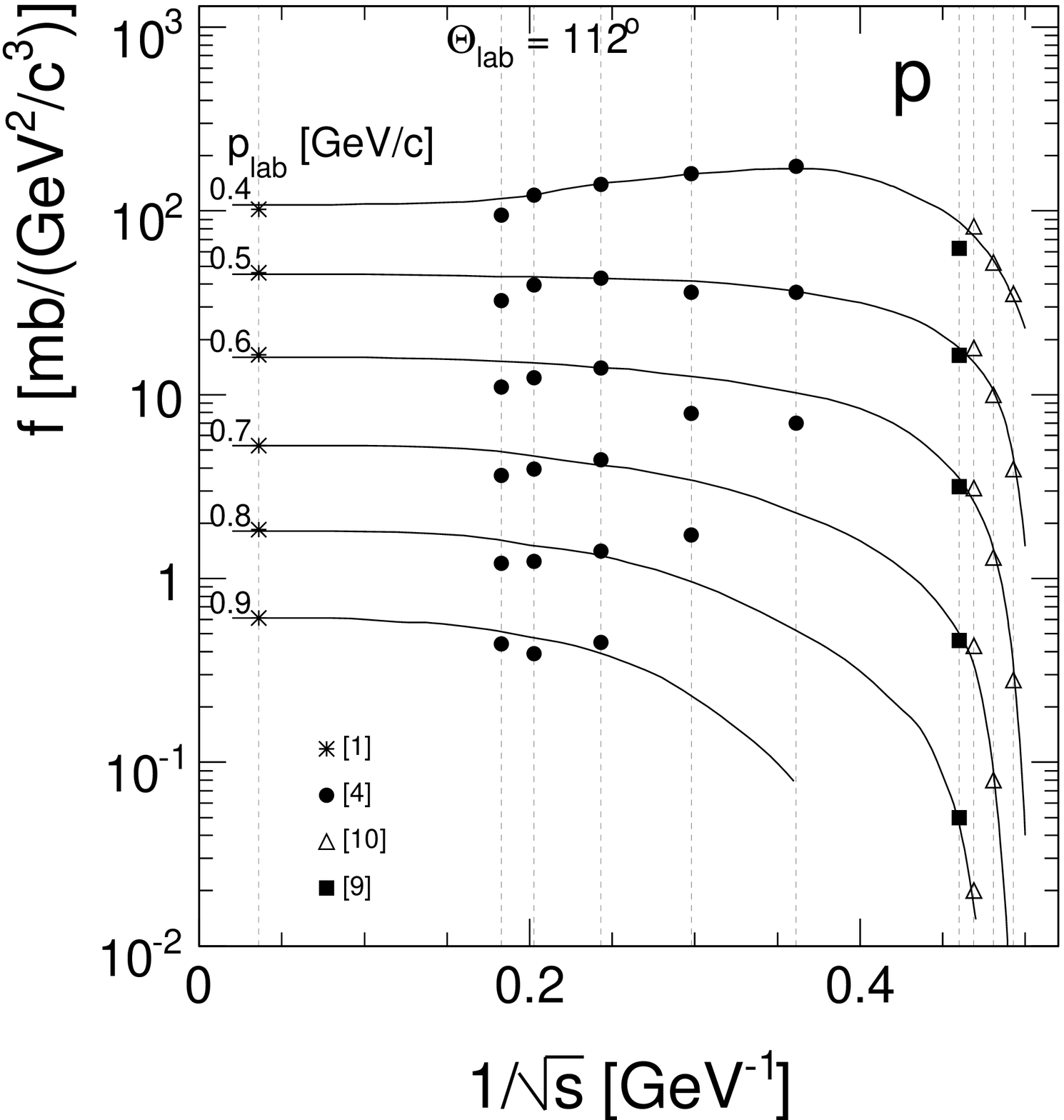}
  \end{center}
  \end{minipage}
\end{figure*}

The solid lines are eyeball interpolations through the data points.
A first remark concerning this Figure concerns the smoothness
and continuity of the $1/\sqrt{s}$ dependences. The achieved overall
consistency of all data is rather impressive even if single
points are deviating in some areas of phase space. The salient features 
of the physics contained in these plots may be summarized
as follows:

\begin{itemize}
 \item A strong yield suppression between 1/$\sqrt{s} \sim$~0.45 and the
       elastic limit at $1/\sqrt{s}$~=~0.53 is evident.
 \item The n+C data \cite{franz} are well consistent with the p+C results
       in the overlap regions; they define a broad maximum of the
       cross sections at 1/$\sqrt{s} \sim$~0.46 at medium angles and low $p_{\textrm{lab}}$.
 \item There is a well-defined asymptotic behaviour of the cross sections
       for $1/\sqrt{s}$ below about 0.2 or beam momenta above about
       12~GeV/c.
 \item For the lower $\Theta_{\textrm{lab}}$ region and/or low $p_{\textrm{lab}}$ the asymptotic
       region is approached from above.
\end{itemize}

\begin{figure}[h]
  \begin{center}
  	\includegraphics[width=9.cm]{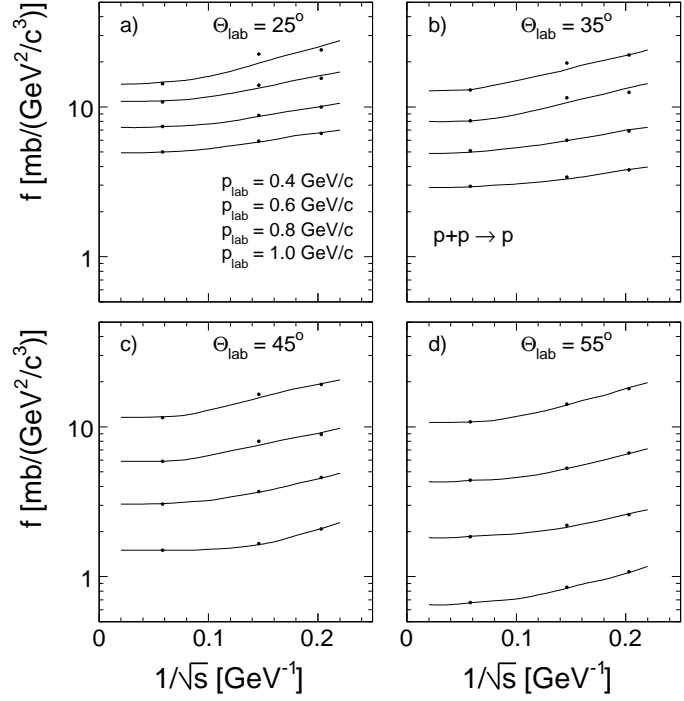}
 	\caption{Invariant proton cross sections as a function
             of $1/\sqrt{s}$ for p+p interactions at the four lab angles  
             a) 25, b) 35, c) 45 and d) 55~degrees for $p_{\textrm{lab}}$ values from 0.4
             to 1~GeV/c. The data are interpolated from Blobel \cite{blobel}
             and NA49 \cite{pp_proton}. The lines are drawn to guide the eye}
  	 \label{fig:oneoversq_pp}
  \end{center}
\end{figure}

The latter point is reminiscent of the behaviour of the proton
yields in p+p interactions, as shown in Fig.~\ref{fig:oneoversq_pp}.

\begin{figure}[h]
  \begin{center}
  	\includegraphics[width=8.cm]{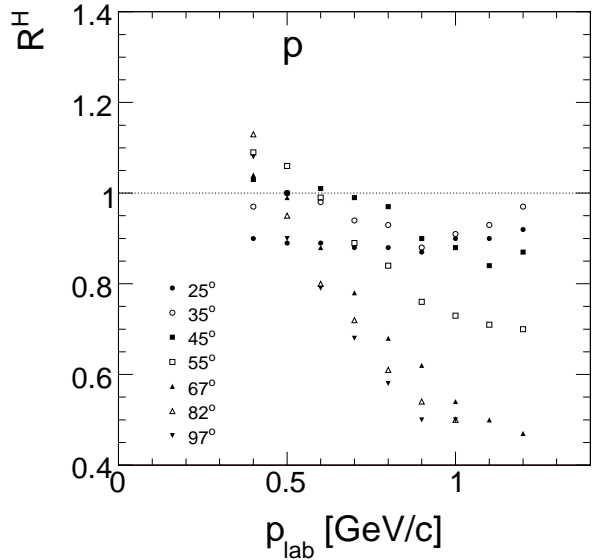}
 	\caption{Ratio $R^H$ between the interpolated invariant
             proton cross sections from HARP-CDP \cite{harp-cdp} and the global
             interpolation as a function of $p_{\textrm{lab}}$ for the angular range
            25~$< \Theta_{\textrm{lab}} <$~97~degrees}
  	 \label{fig:harp15_prot}
  \end{center}
\end{figure}

Another feature of Fig.~\ref{fig:oneoversq} is the systematic droop of the
cross sections from HARP-CDP at their highest beam momentum
of 15~GeV/c or $1/\sqrt{s}$~=~0.18, demonstrating the discriminative
power of the approach. This decrease is quantified in Fig.~\ref{fig:harp15_prot} where
the ratio $R^H$ between the measured invariant cross sections
and the data interpolation is shown as a function of $p_{\textrm{lab}}$ 
for the complete angular range from 25 to 97~degrees. Here
deviations of up to 50\% are visible.

The abruptness of this decrease would necessitate a rather
violent variation of the cross sections with increasing energy
including a minimum between PS and SPS energies. A final
clarification of this situation is given by the proton data
from Serpukhov \cite{belyaev_prot} which, although suffering from a different
and independent problem, at least exclude such variations
in the region between 17 and 67~GeV/c beam momentum, see Sect.~\ref{sec:belyaev} below. 

%
%
\subsection{$\mathbf {cos(\theta_{\textrm{lab}})}$ dependence}
\vspace{3mm}
\label{sec:prot_thdep}

In addition to the description of the energy dependence, the
global interpolation has of course also to result in a smooth 
and continuous verification of the angular dependence presenting
the third dimension of the present study. This constraint has
to be fulfilled at any value of $1/\sqrt{s}$.

\begin{figure}[h]
  \begin{center}
  	\includegraphics[width=12.5cm]{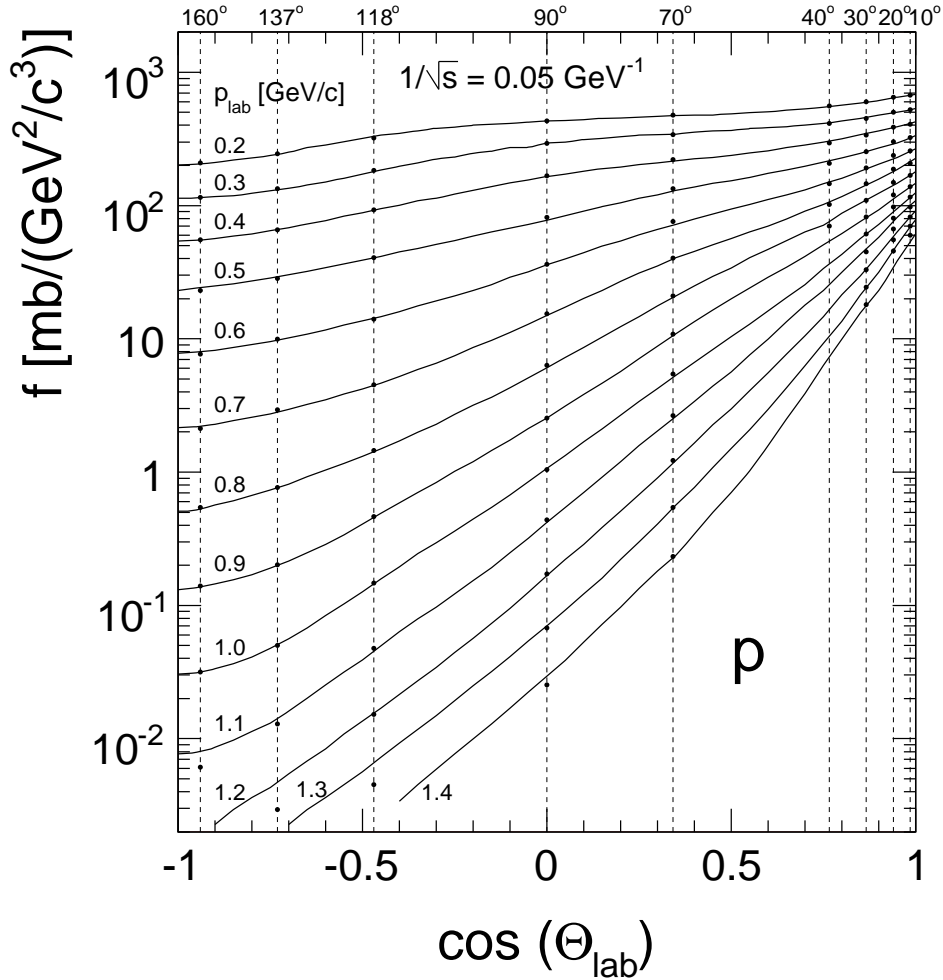}
 	  \caption{Invariant proton cross sections at $1/\sqrt{s}$~=~0.05
               as a function of $\cos(\theta_{\textrm{lab}})$ combining the Fermilab 
               and NA49 data for $p_{\textrm{lab}}$ between 0.2 and 1.4~GeV/c. The global
               interpolation is shown as full lines. The measured cross
               sections in the angular ranges from 70 to 160~degrees
               \cite{bayukov} and from 10 to 40~degrees \cite{pc_proton} are
               given on the vertical broken lines}
  	 \label{fig:pc_prot_theta}
  \end{center}
\end{figure}

In a first example the situation at $1/\sqrt{s}$~=~0.05 is shown in
Fig.~\ref{fig:pc_prot_theta}. This value lies in between 
the Fermilab \cite{bayukov} and NA49 \cite{pc_proton} 
data in the region of negligible $s$-dependence. It therefore 
allows for the direct comparison of the two experiments in 
their respective angular regions which have no overlap.

Several observations are in place here:

\begin{itemize}
 \item The two experimental results connect perfectly through the gap 
       between the NA49 ($\theta_{\textrm{lab}} <$~40~degrees) and the Fermilab 
       ($\theta_{\textrm{lab}} >$~70~degrees) data.
 \item There is at most a few percent variation of the cross sections
       between the angles of 160 and 180~degrees taking into account
       the constraint of continuity through 180~degrees discussed in
       Sect.~\ref{sec:comp} above. This allows the combination of results in this 
       angular region as it is applied in the determination of the 
       $1/\sqrt{s}$ dependence, Fig.~\ref{fig:oneoversq}.
 \item The angular distributions are smooth and close to exponential
       in shape. In particular, no instability in the region around
       90~degrees is visible where an eventual diffractive peak from
       target fragmentation would appear, see also \cite{pc_proton}.
\end{itemize}

\begin{figure}[h]
  \begin{center}
  	\includegraphics[width=14.cm]{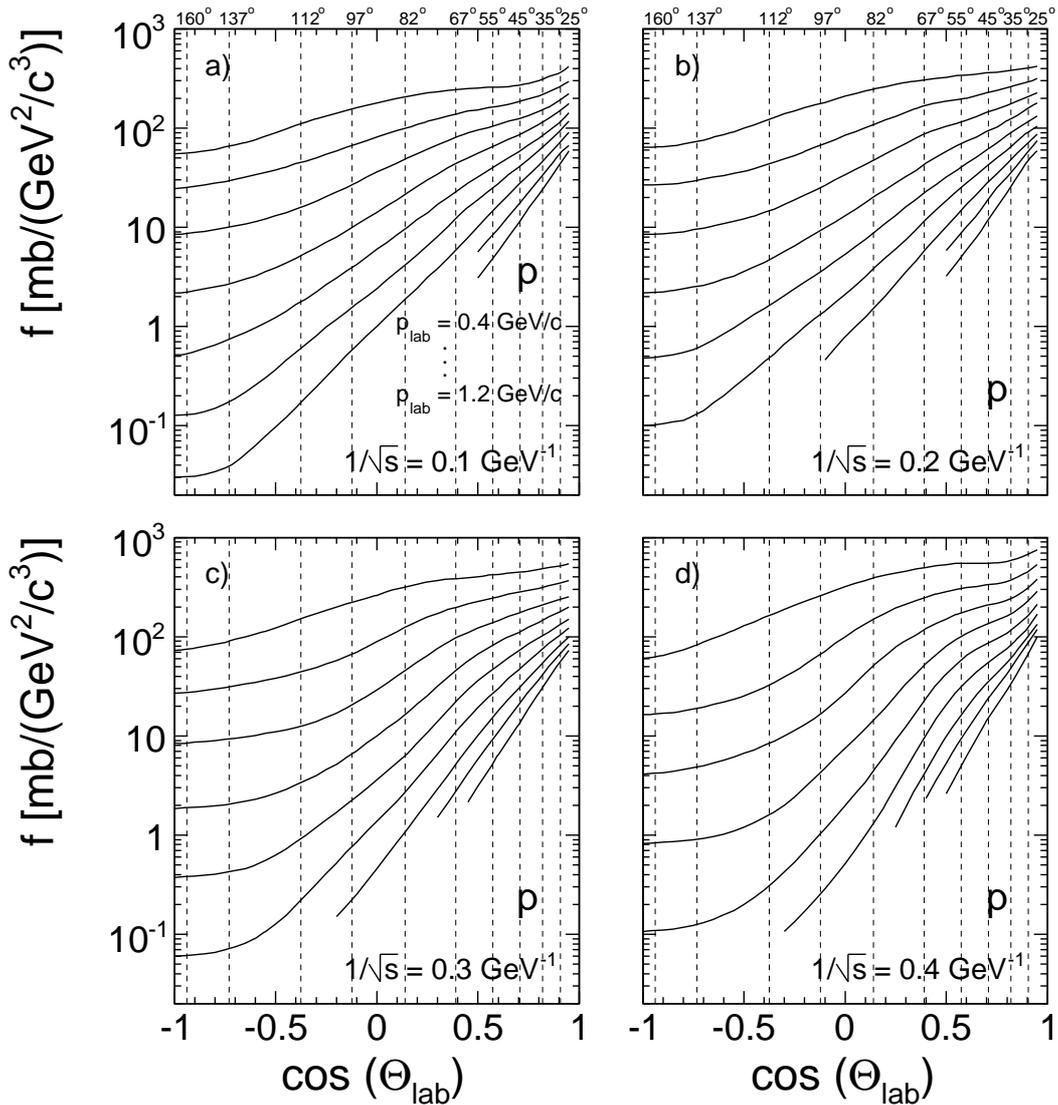}
 	  \caption{Invariant proton cross sections as a function of
               $\cos(\theta_{\textrm{lab}})$ for four values of $1/\sqrt{s}$: a) 0.1, b) 0.2,
               c) 0.3, d) 0.4~GeV$^{-1}$ and for $p_{\textrm{lab}}$ values between 0.4 and
               1.2~GeV/c. The standard grid of 10 angles, Fig.~\ref{fig:oneoversq}, is indicated
               by the vertical broken lines}
  	 \label{fig:costheta}
  \end{center}
\end{figure}

Further angular distributions at four $1/\sqrt{s}$ values between
0.1 and 0.4~GeV$^{-1}$ are given in Fig.~\ref{fig:costheta}. In fact such distributions
at arbitrary values of $1/\sqrt{s}$ may be obtained from the global
interpolation as it is presented in numerical form at the 
NA49 web page \cite{site}.

Evidently the angular distributions maintain their smooth and
continuous shape, specifically through 90~degrees, at all interaction 
energies. With the approach to low beam momenta however, a progressive
rounding of the shape towards higher lab angles manifests itself.
 
%
%
\section{The data for positive pions}
\vspace{3mm}
\label{sec:pip}

The global interpolation of the $\pi^+$ data is presented in this
section in close analogy to the preceding section for protons.

%
%
\subsection{$\mathbf {1/\sqrt{s}}$ dependence}
\vspace{3mm}
\label{sec:pip_sdep}

\begin{figure}[h]
  \begin{minipage}[t]{0.5\linewidth} 
  \begin{center}
  	\includegraphics[width=6.8cm]{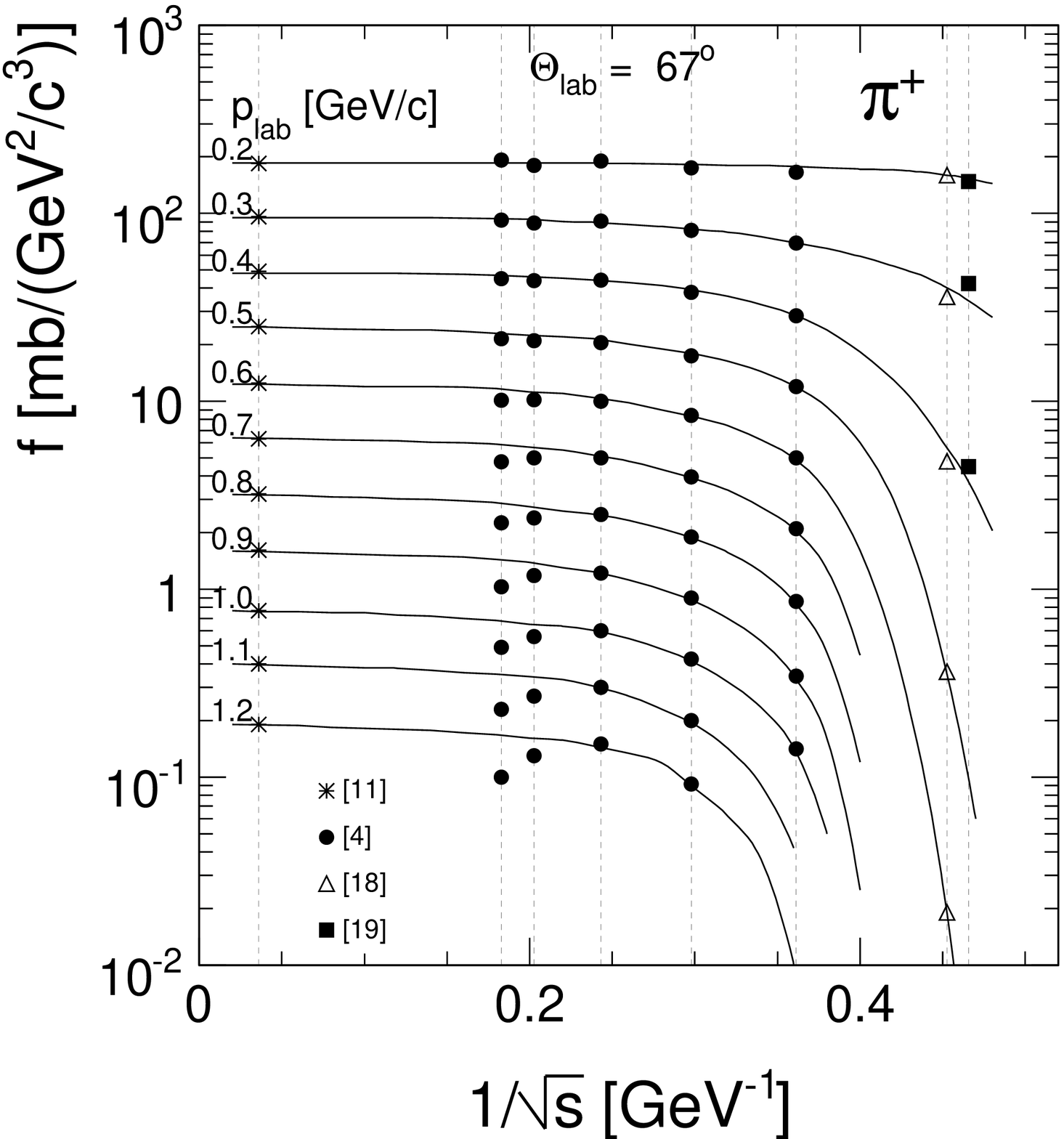}
  \end{center}
  \end{minipage}
  \begin{minipage}[t]{0.5\linewidth} 
  \begin{center}
  	\includegraphics[width=6.8cm]{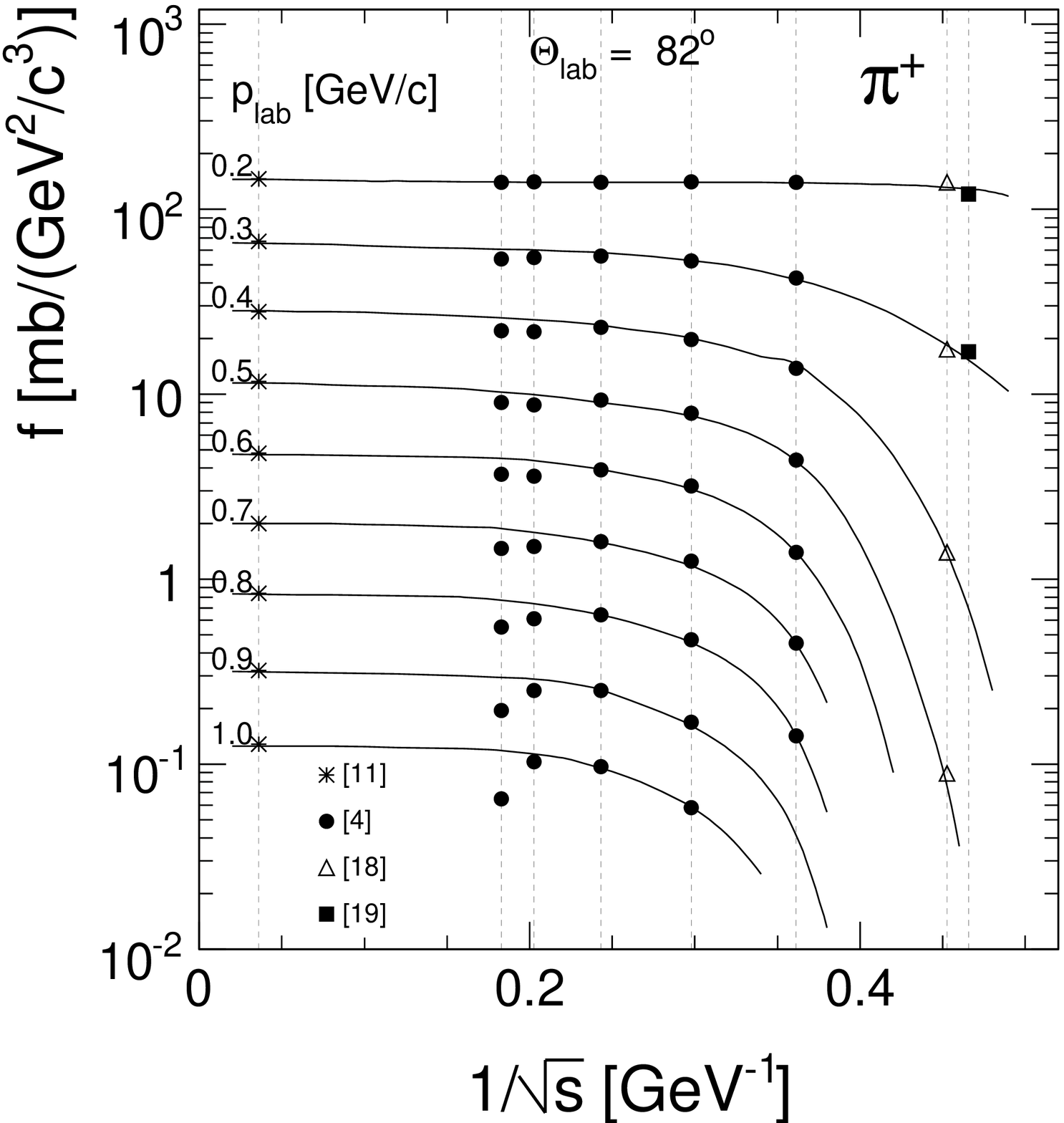}
  \end{center}
  \end{minipage}
  \begin{minipage}[t]{0.5\linewidth} 
  \begin{center}
  	\includegraphics[width=6.8cm]{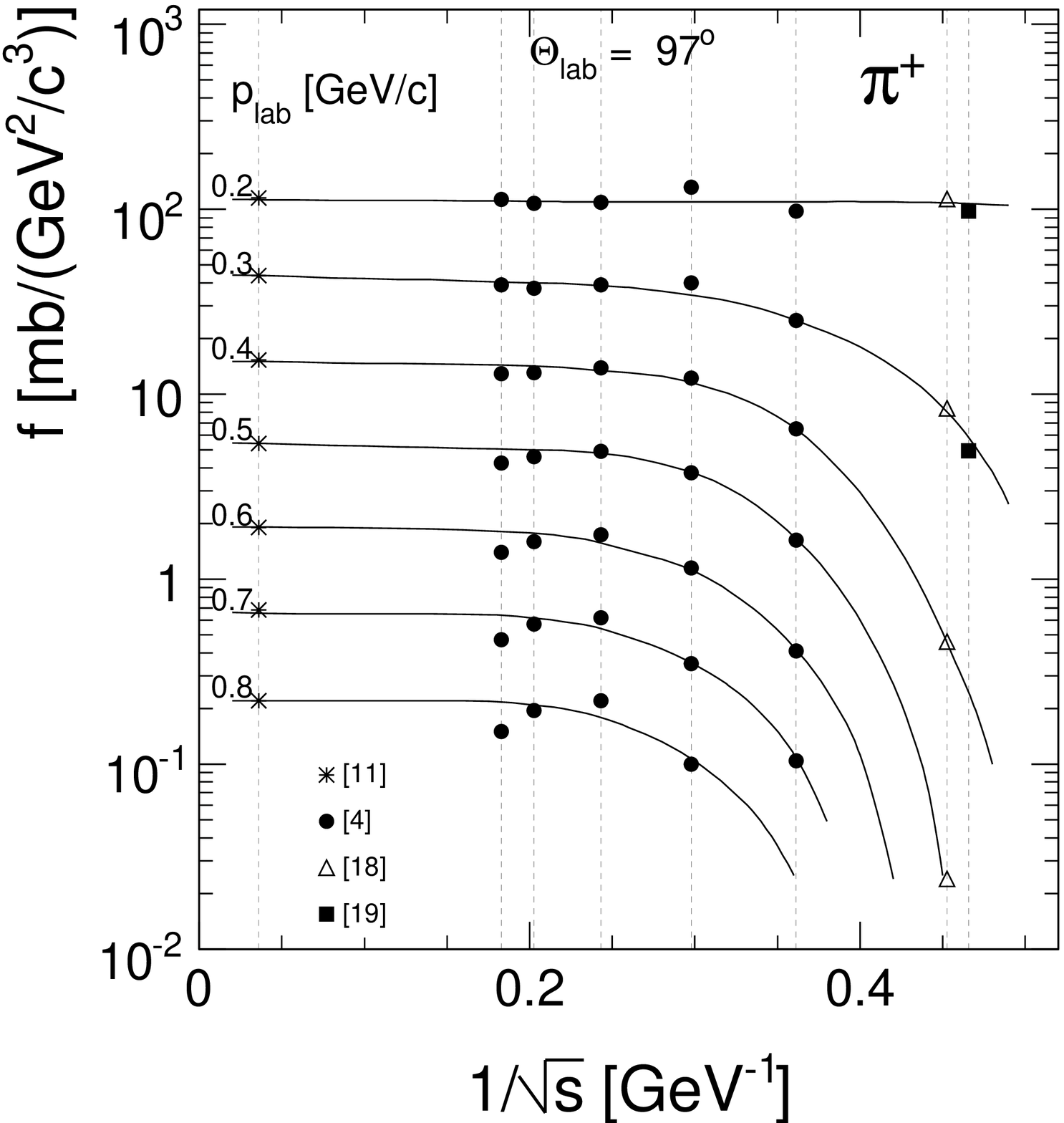}
  \end{center}
  \end{minipage}
  \begin{minipage}[t]{0.5\linewidth} 
  \begin{center}
  	\includegraphics[width=6.8cm]{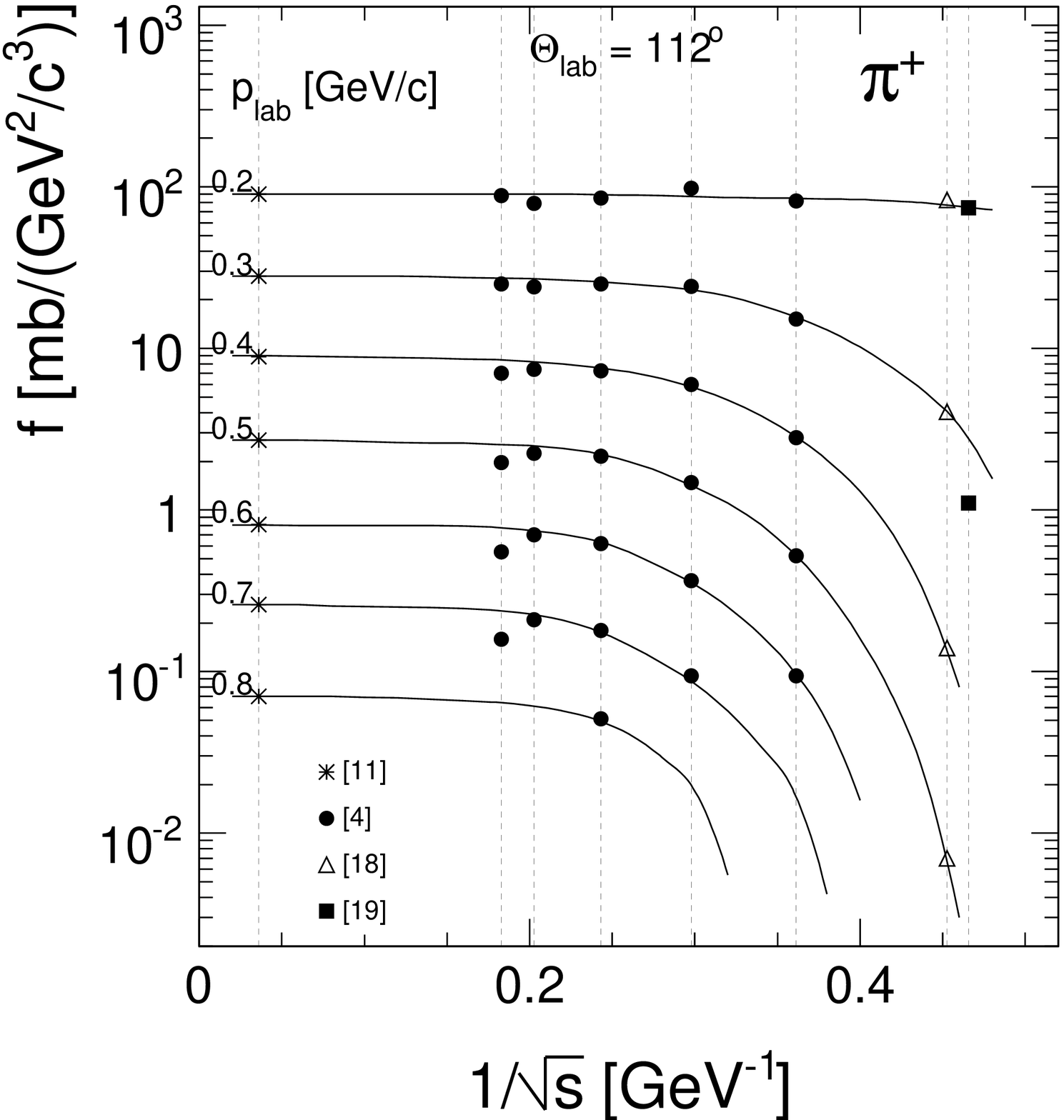}
  \end{center}
  \end{minipage}
  \begin{minipage}[t]{0.5\linewidth} 
  \begin{center}
  	\includegraphics[width=6.8cm]{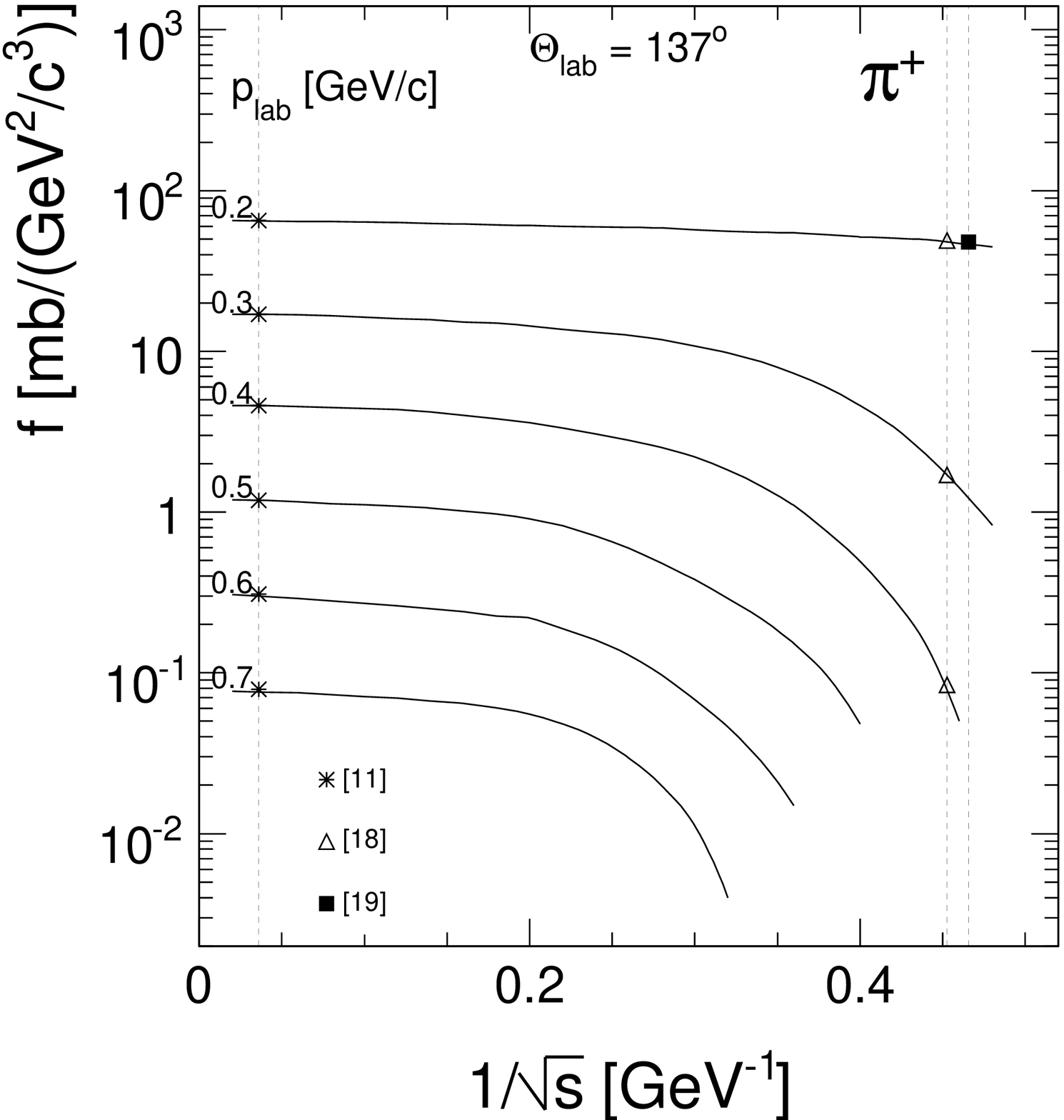}
  \end{center}
  \end{minipage}
  \begin{minipage}[t]{0.5\linewidth} 
  \begin{center}
  	\includegraphics[width=6.8cm]{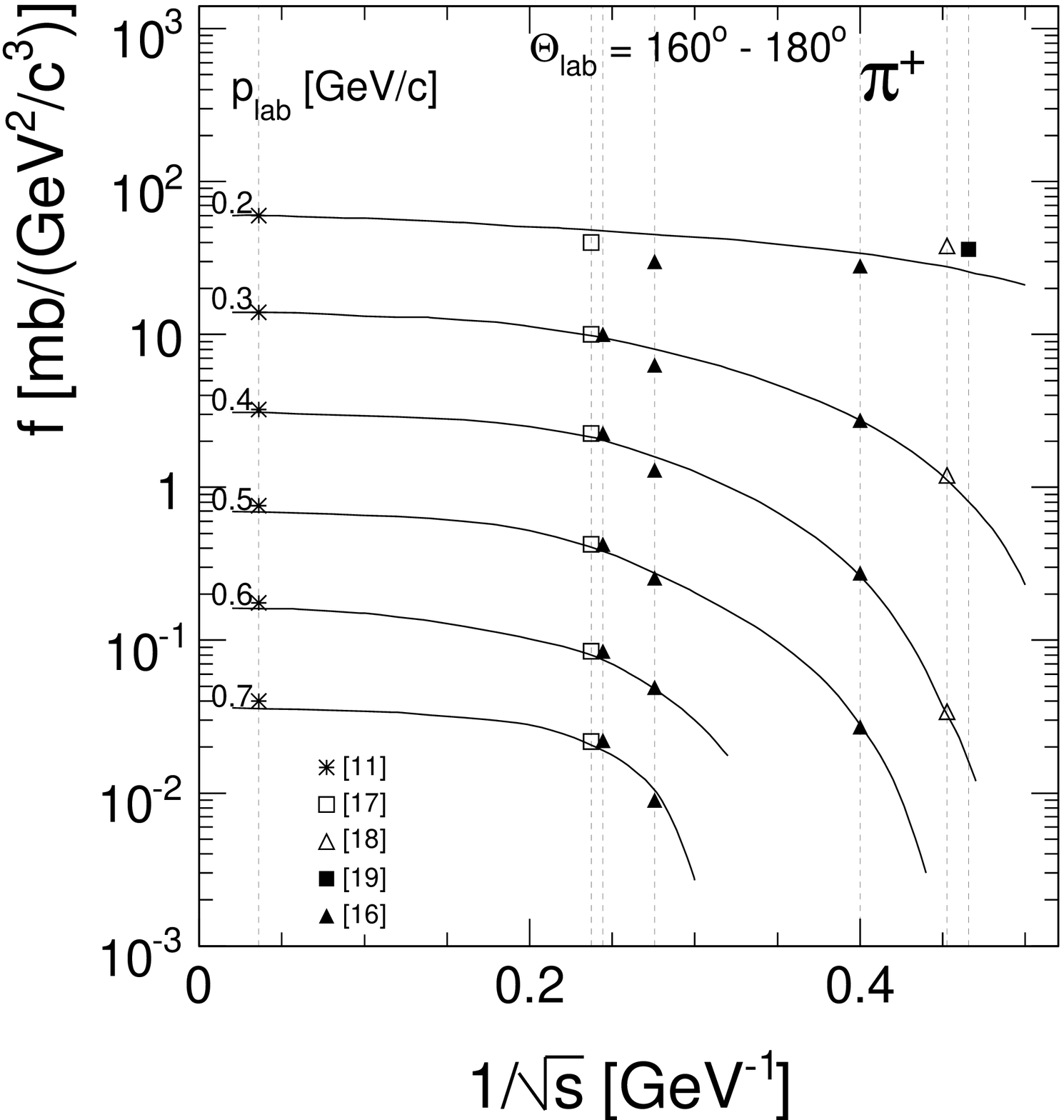}
  \end{center}
  \end{minipage}
  \caption{Invariant cross sections for $\pi^+$ in p+C
           collisions as a function of $1/\sqrt{s}$ at fixed $p_{\textrm{lab}}$ and
           $\theta_{\textrm{lab}}$. The interpolated data points are indicated
           by symbols corresponding to the respective experiments
           in each panel. The solid lines represent the global
           data interpolation}
  \label{fig:sqs_pip}
\end{figure}

The invariant $\pi^+$ cross sections are shown in Fig.~\ref{fig:sqs_pip} as a function 
of $1/\sqrt{s}$ for the standard grid of ten lab angles between 25 and
180~degrees and for constant lab momenta between 0.2 and 1.2~GeV/c. 
The interpolated data points in each panel are identified by symbols 
corresponding to the different experiments.

\begin{figure*}[b]
  \begin{minipage}[t]{0.5\linewidth} 
  \begin{center}
  	\includegraphics[width=6.8cm]{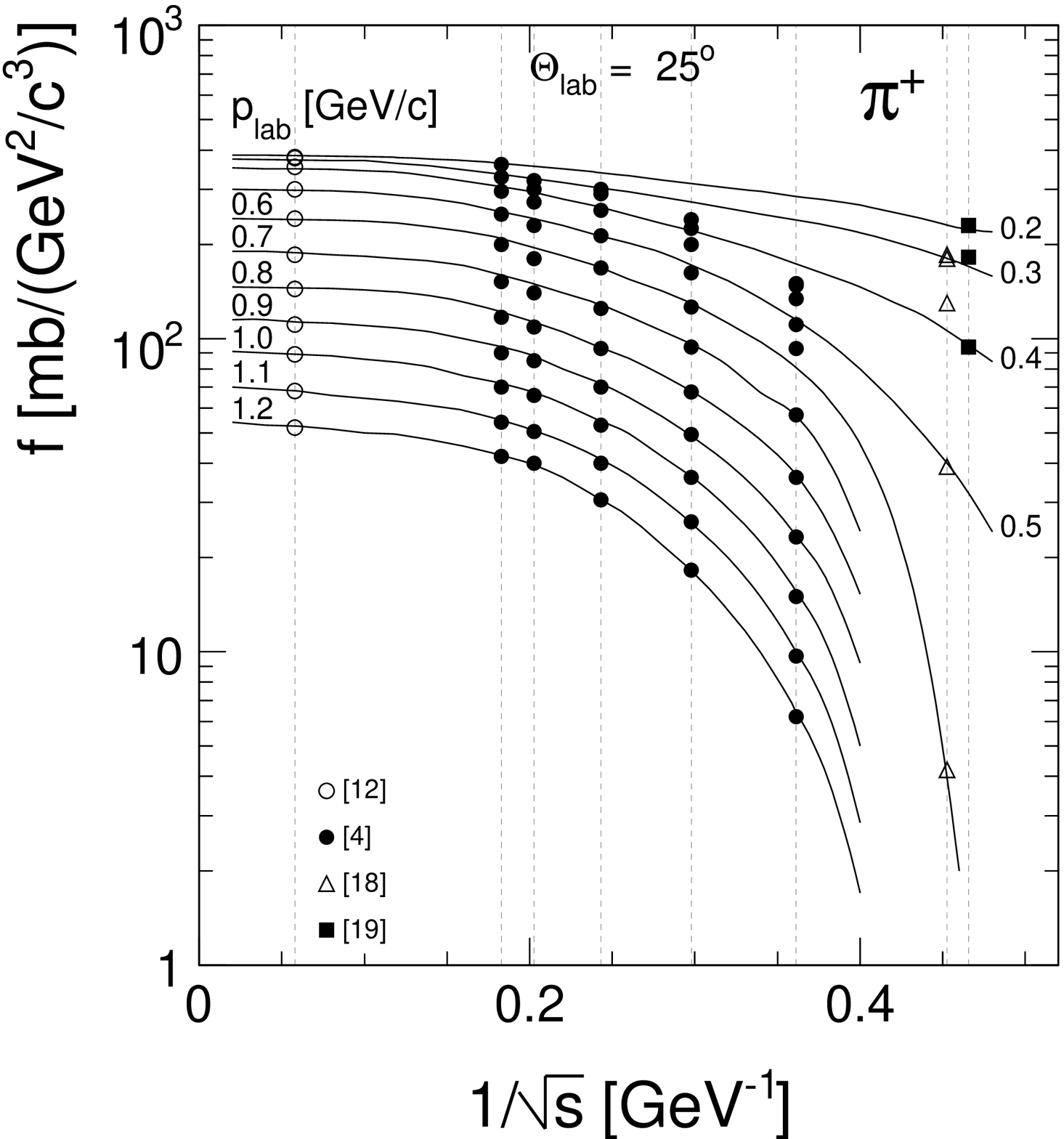}
  \end{center}
  \end{minipage}
  \begin{minipage}[t]{0.5\linewidth} 
  \begin{center}
  	\includegraphics[width=6.8cm]{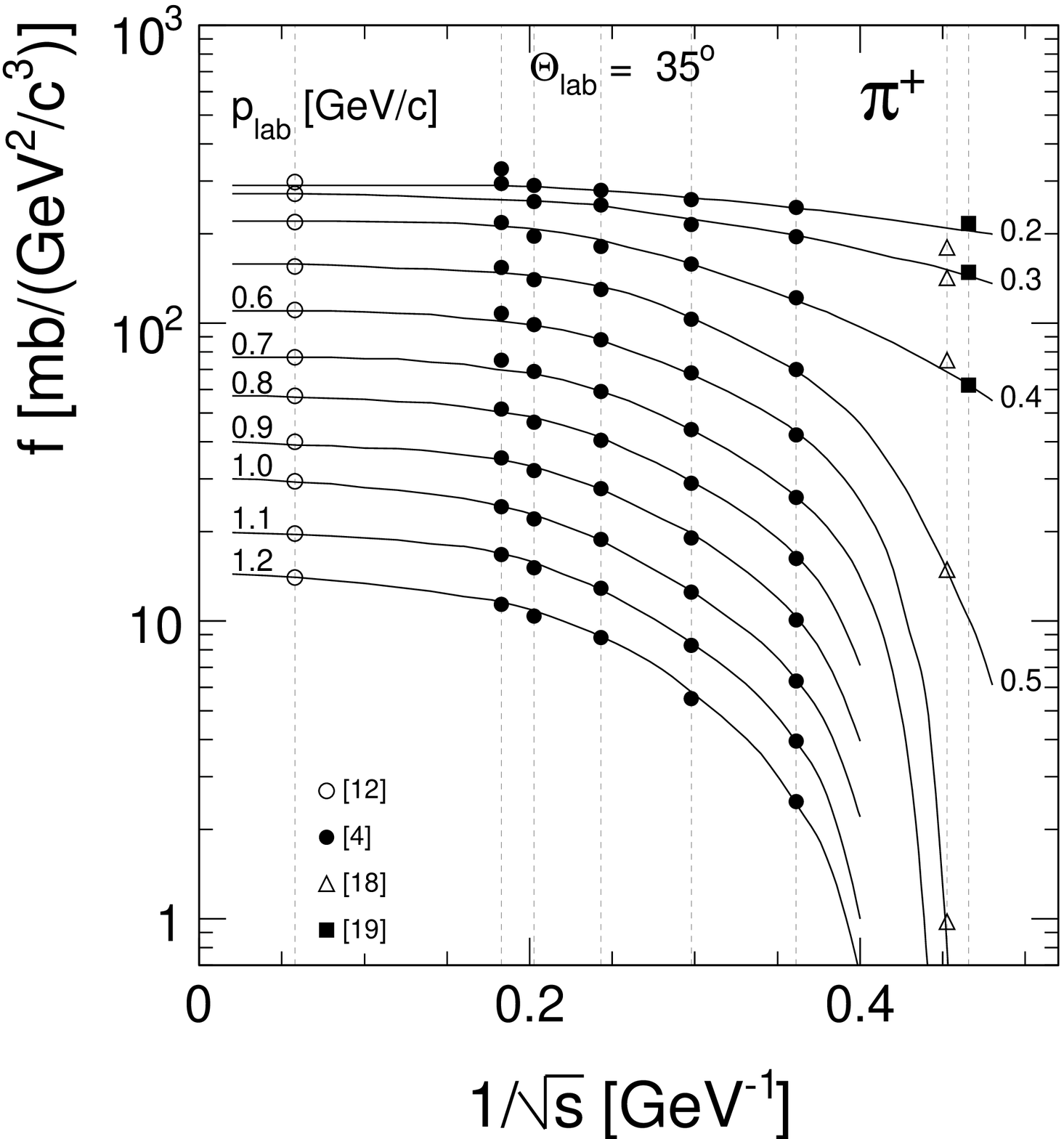}
  \end{center}
  \end{minipage}
  \begin{minipage}[t]{0.5\linewidth} 
  \begin{center}
  	\includegraphics[width=6.8cm]{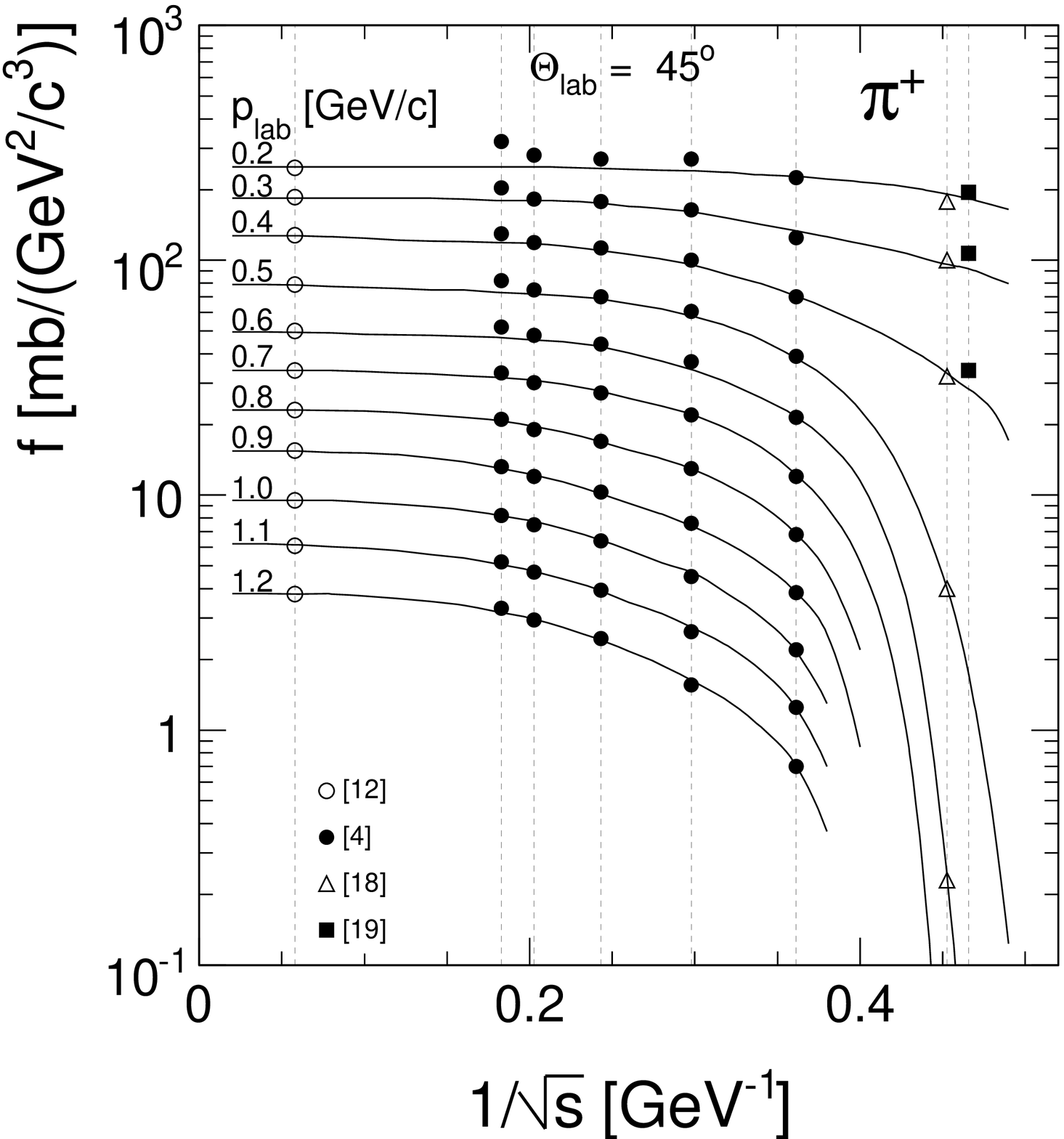}
  \end{center}
  \end{minipage}
  \begin{minipage}[t]{0.5\linewidth} 
  \begin{center}
  	\includegraphics[width=6.8cm]{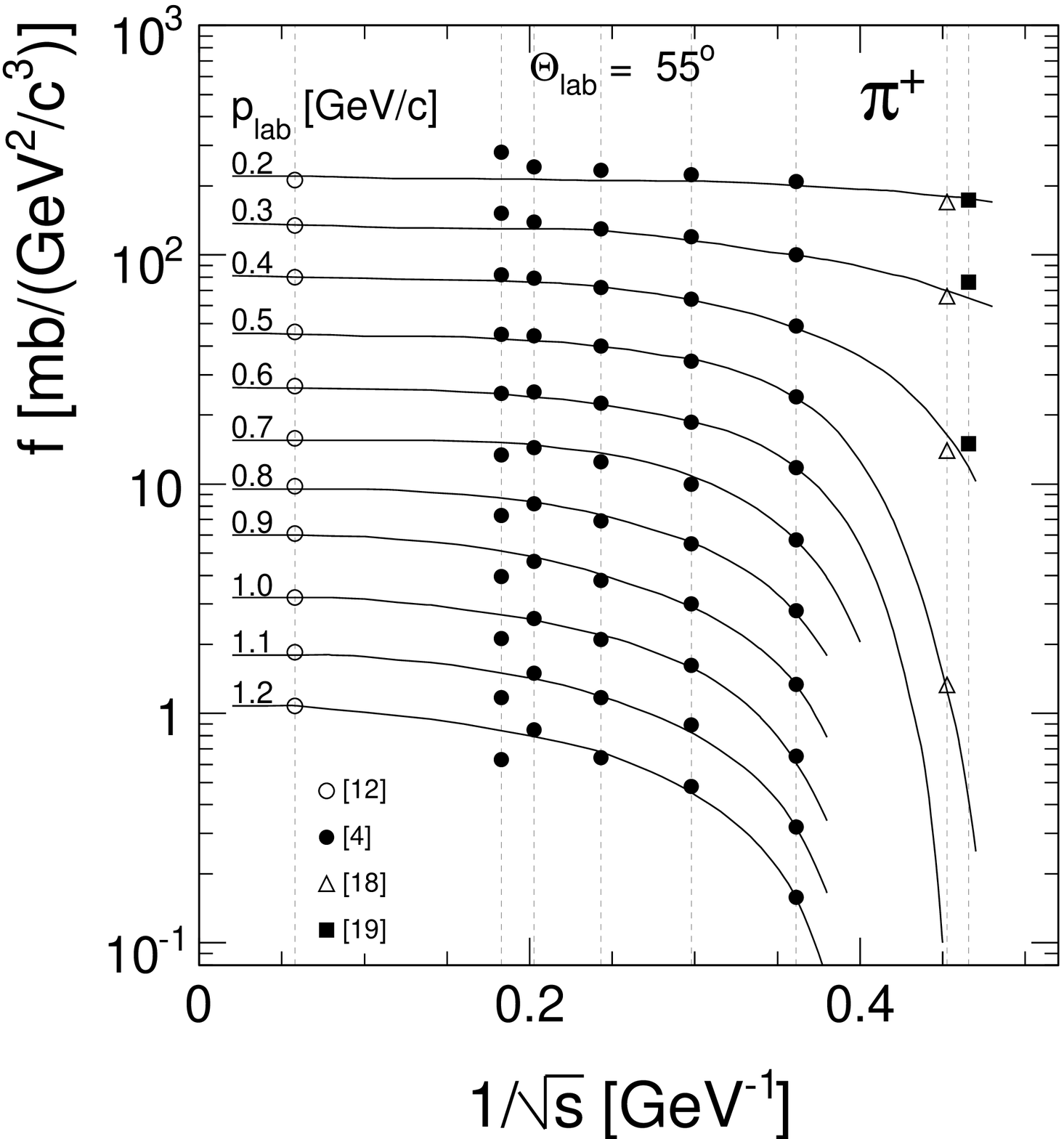}
  \end{center}
  \end{minipage}
\end{figure*}

The solid lines represent the global interpolation by eyeball fits
of both the energy and the angular dependences. Again the $1/\sqrt{s}$
dependence is in general smooth and continuous, with an impressive
overall consistency of all data with only few exceptions discussed
below. There are some general trends to be pointed out:

\begin{itemize}
 \item At the lowest lab momentum, the pion cross sections are practically
       $s$-independent, with variations of only 10--20\% in the range from
       1 to 400~GeV/c beam momentum.
 \item This fact suggests $\pi^+$ production at low momentum transfer in
       the nuclear cascade.
 \item For all lab momenta, the approach to high energies is very flat
       for $1/\sqrt{s} <$~0.2 or beam momenta above 12~GeV/c. 
 \item The high energy cross sections are approached for all angles
       and beam momenta from below.
\end{itemize}

\begin{figure}[h]
  \begin{center}
  	\includegraphics[width=9.cm]{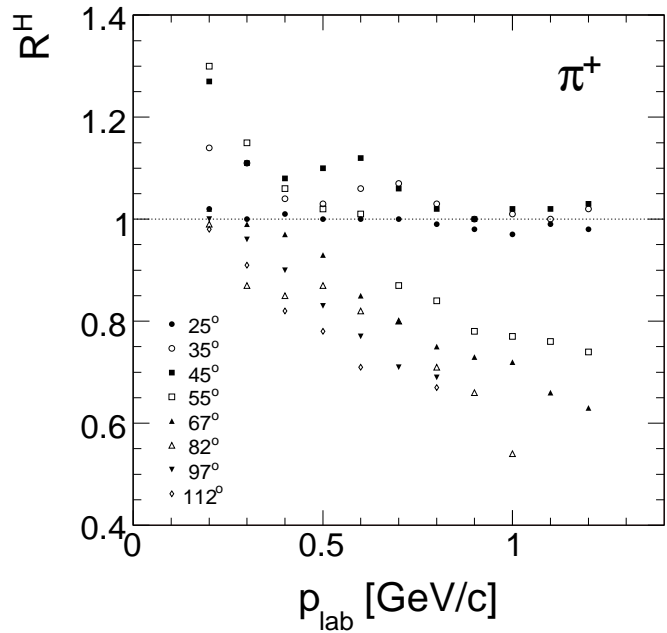}
 	\caption{Ratio $R^H$ between the interpolated invariant
             $\pi^+$ cross sections from HARP-CDP \cite{harp-cdp} and the global
             interpolation as a function of $p_{\textrm{lab}}$ for the angular range
             25~$< \theta_{\textrm{lab}} <$~112~degrees}
  	 \label{fig:harp15_pip}
  \end{center}
\end{figure}

There are two areas of deviation from the global interpolation which 
are both connected to the HARP-CDP data \cite{harp-cdp}. At their lowest angle of
25~degrees, the cross sections are systematically low by up to a
factor of two below $p_{\textrm{lab}} \sim$~0.5~GeV/c and $1/\sqrt{s}$ above 0.2. This
is in contradiction to the available low energy data from other
experiments also shown in Fig.~\ref{fig:sqs_pip}. The second area concerns, as 
for the protons, the data at 15~GeV/c beam momentum where a 
characteristic pattern of deviations is visible: At low angles and 
low $p_{\textrm{lab}}$, the data tend to overshoot the interpolation, whereas at 
angles above 45~degrees a progressive droop with increasing lab 
momentum is evident. This is quantified by the ratio $R^H$ between 
the HARP-CDP data and the global interpolation shown in Fig.~\ref{fig:harp15_pip}.

These deviations are rather consistent with the ones found for
protons. Also in this case a rapid variation of the cross sections
with increasing beam momentum can be excluded by the comparison
with the pion data from the Serpukhov experiment \cite{belyaev_pion} between 17
and 67~GeV/c beam momentum, see Sect.~\ref{sec:belyaev} below.
 
%
%
\subsection{$\mathbf {cos(\theta_{\textrm{lab}})}$ dependence}
\vspace{3mm}
\label{sec:pip_thdep}

\begin{figure}[h]
  \begin{center}
  	\includegraphics[width=12.5cm]{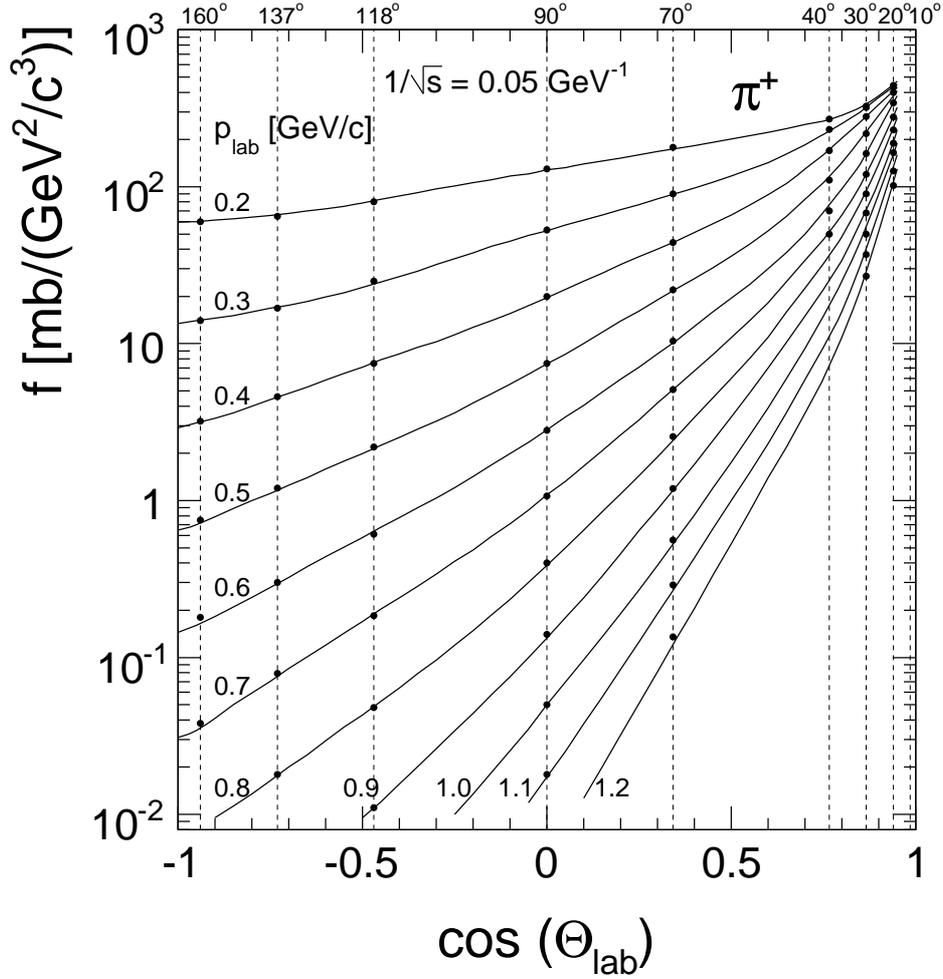}
 	  \caption{Invariant $\pi^+$ cross sections at $1/\sqrt{s}$~=~0.05
              as a function of $\cos(\theta_{\textrm{lab}})$ combining the Fermilab \cite{niki} 
              and NA49 \cite{pc_pion} data for $p_{\textrm{lab}}$ between 0.2 and 1.2~GeV/c. The global
              interpolation is shown as full lines. The measured cross
              sections in the angular ranges from 70 to 160~degrees
              (\cite{niki}) and from 10 to 40~degrees (\cite{pc_pion}) are
              given on the vertical broken lines}
  	 \label{fig:pc_pip_theta}
  \end{center}
\end{figure}

As already shown in Sect.~\ref{sec:prot_thdep} for protons, the angular distributions
at $1/\sqrt{s}$~=~0.05, in between the Fermilab \cite{niki} and NA49 \cite{pc_pion} energies, are
presented in Fig.~\ref{fig:pc_pip_theta}. This allows the comparison of the two data sets
and their connection across the gap in lab angles between 40 and 70~degrees 
which represent the upper and lower limit of the respective experiment.

\begin{figure}[h]
  \begin{center}
  	\includegraphics[width=14.cm]{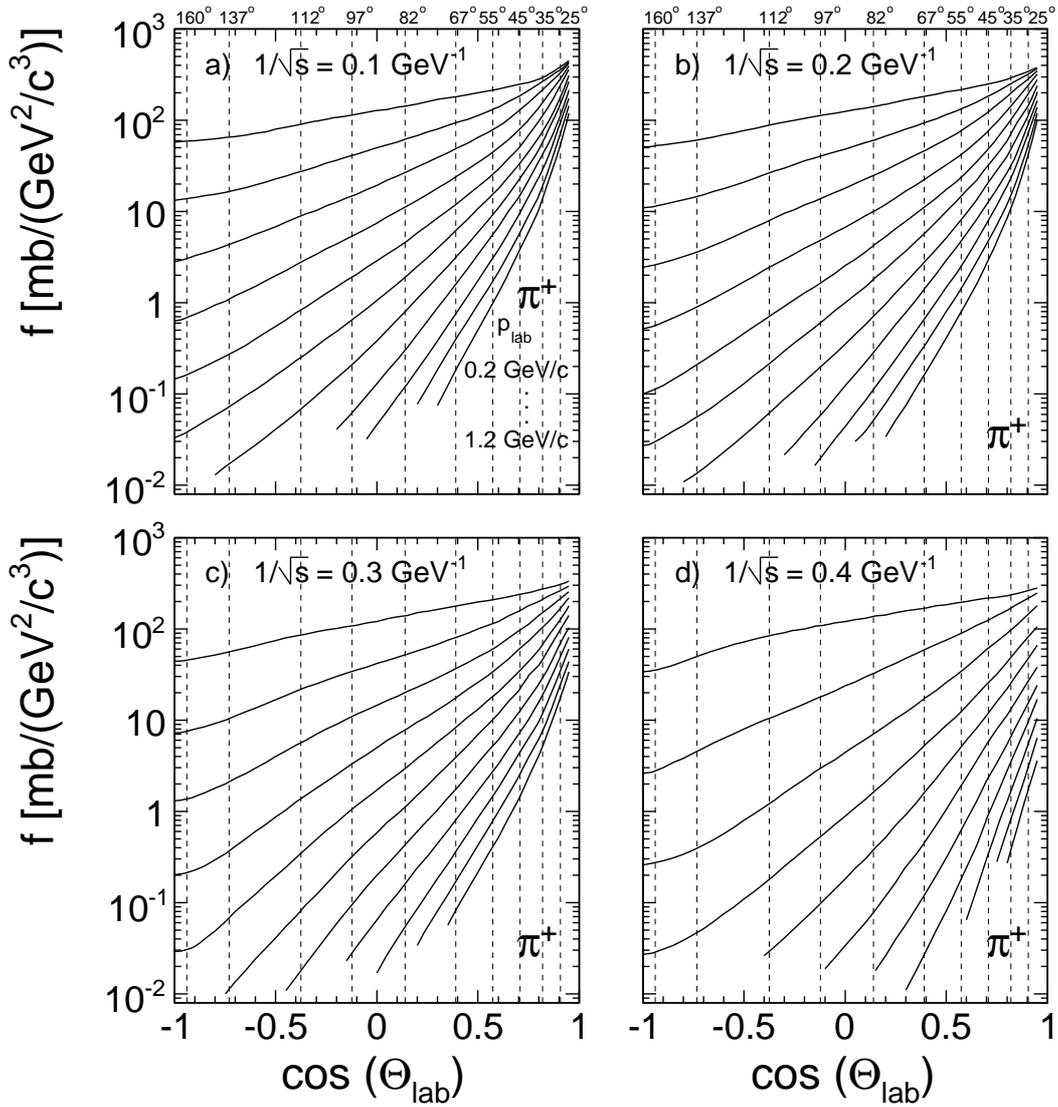}
 	  \caption{Invariant $\pi^+$ cross sections as a function of
               $\cos(\theta_{\textrm{lab}})$ for four values of $1/\sqrt{s}$: a) 0.1, b) 0.2,
               c) 0.3, d) 0.4~GeV$^{-1}$ and for $p_{\textrm{lab}}$ values between 0.2 and
               1.2~GeV/c. The standard grid of 10 angles, Fig.~\ref{fig:oneoversq}, is indicated
               by the vertical broken lines}
  	 \label{fig:costheta_pip}
  \end{center}
\end{figure}

Further angular distributions at four $1/\sqrt{s}$ values between
0.1 and 0.4~GeV$^{-1}$ are given in Fig.~\ref{fig:costheta_pip}.

The angular distributions are characterized by a smooth, close
to exponential shape. At backward angles, the $p_{\textrm{lab}}$ dependence is very
steep with four orders of magnitude already between $p_{\textrm{lab}}$~=~0.2 and
0.8~GeV/c. In forward direction this dependence is much reduced
with less than one order of magnitude between $p_{\textrm{lab}}$~=~0.2 and 1.2~GeV/c.
This is due to the prevailance of target fragmentation in this
region, see Sect.~\ref{sec:separation} for a quantitative study of this phenomenology.

%
%
\section{The data for negative pions}
\vspace{3mm}
\label{sec:pin}

This section follows closely the discussion of the $\pi^+$ cross sections 
in the preceding section.

%
%
\subsection{$\mathbf {1/\sqrt{s}}$ dependence}
\vspace{3mm}
\label{sec:pin_sdep}

\begin{figure}[h]
  \begin{minipage}[t]{0.49\linewidth} 
  \begin{center}
  	\includegraphics[width=7.1cm]{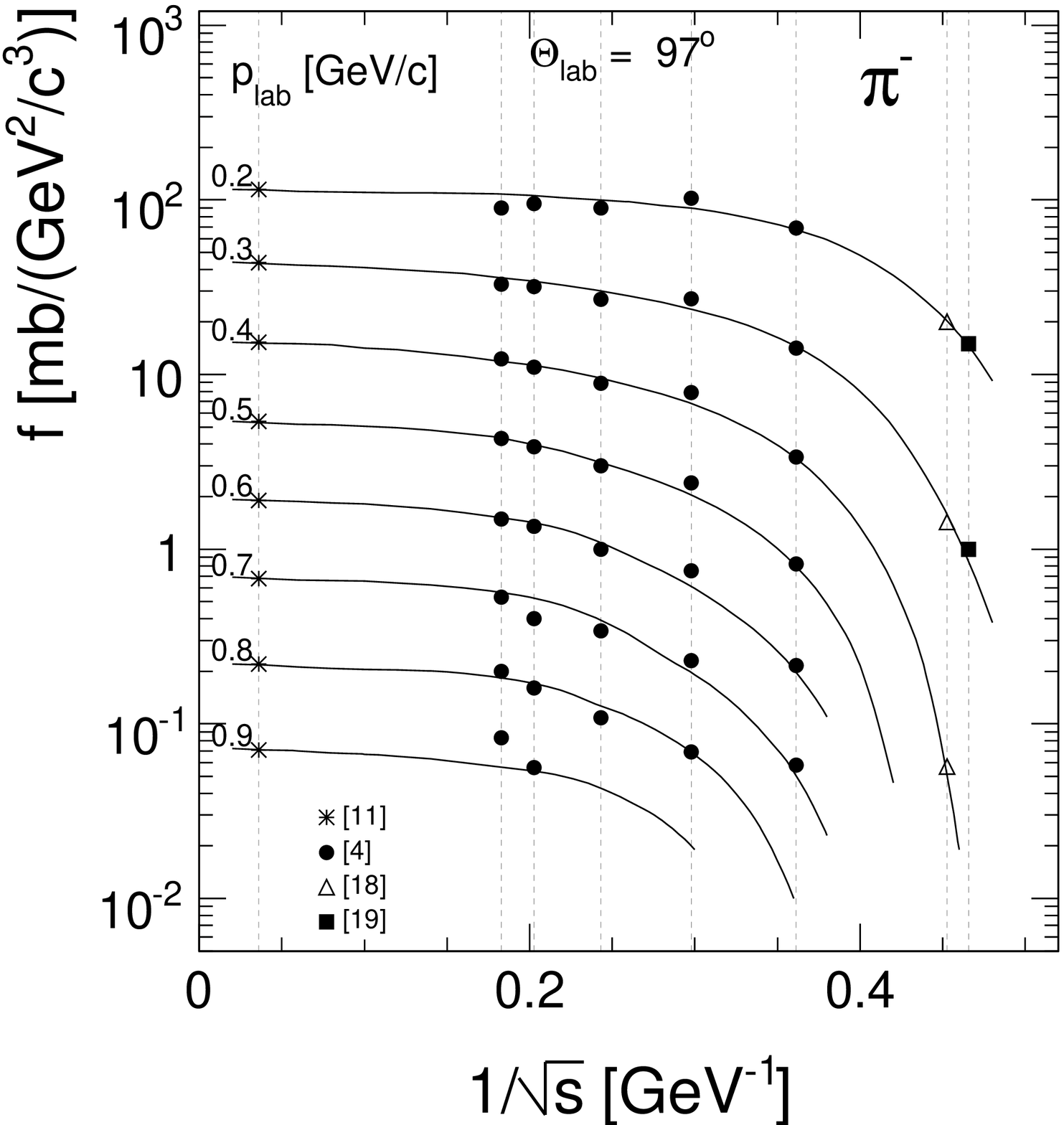}
  \end{center}
  \end{minipage}
  \begin{minipage}[t]{0.49\linewidth} 
  \begin{center}
  	\includegraphics[width=7.1cm]{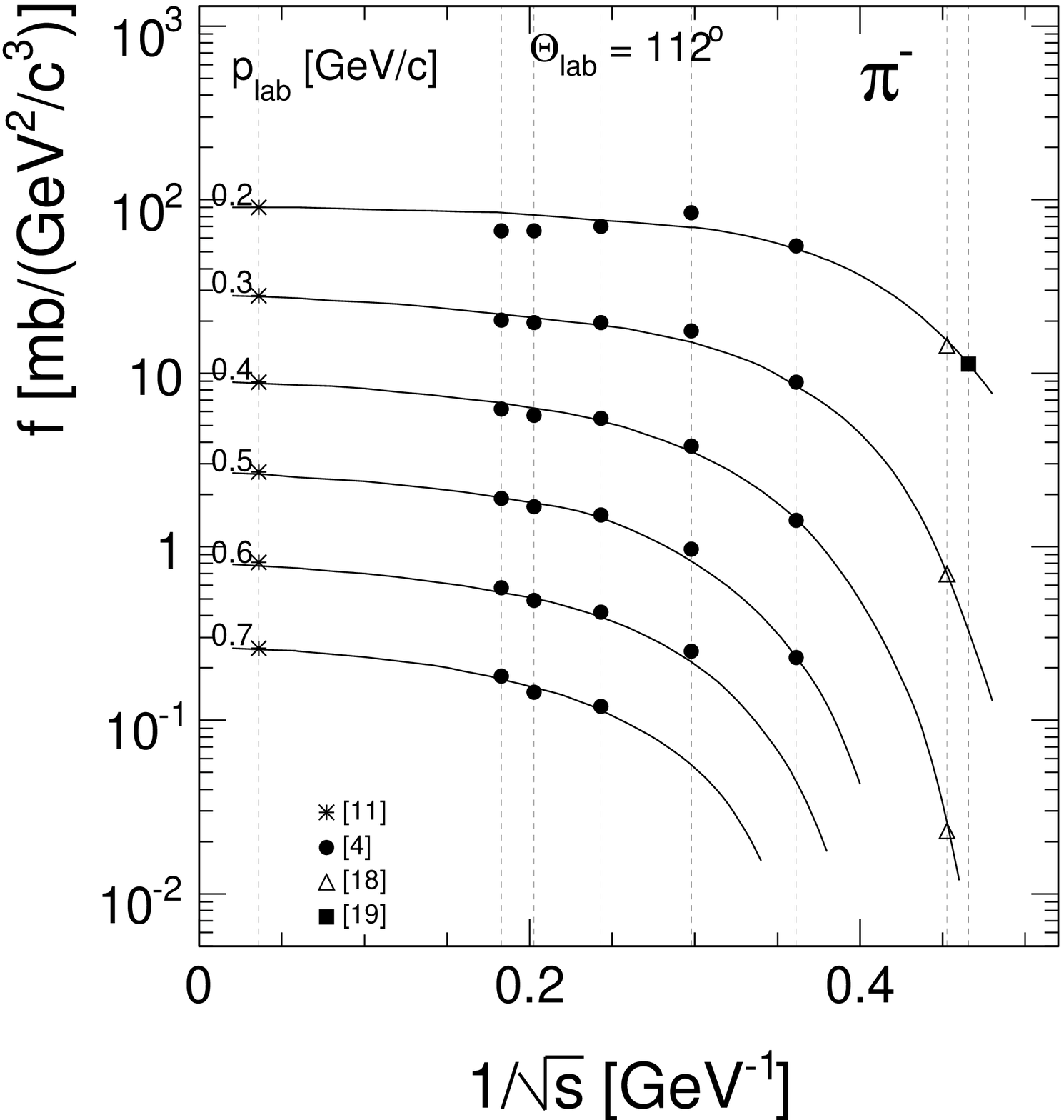}
  \end{center}
  \end{minipage}
  \begin{minipage}[t]{0.49\linewidth} 
  \begin{center}
  	\includegraphics[width=7.1cm]{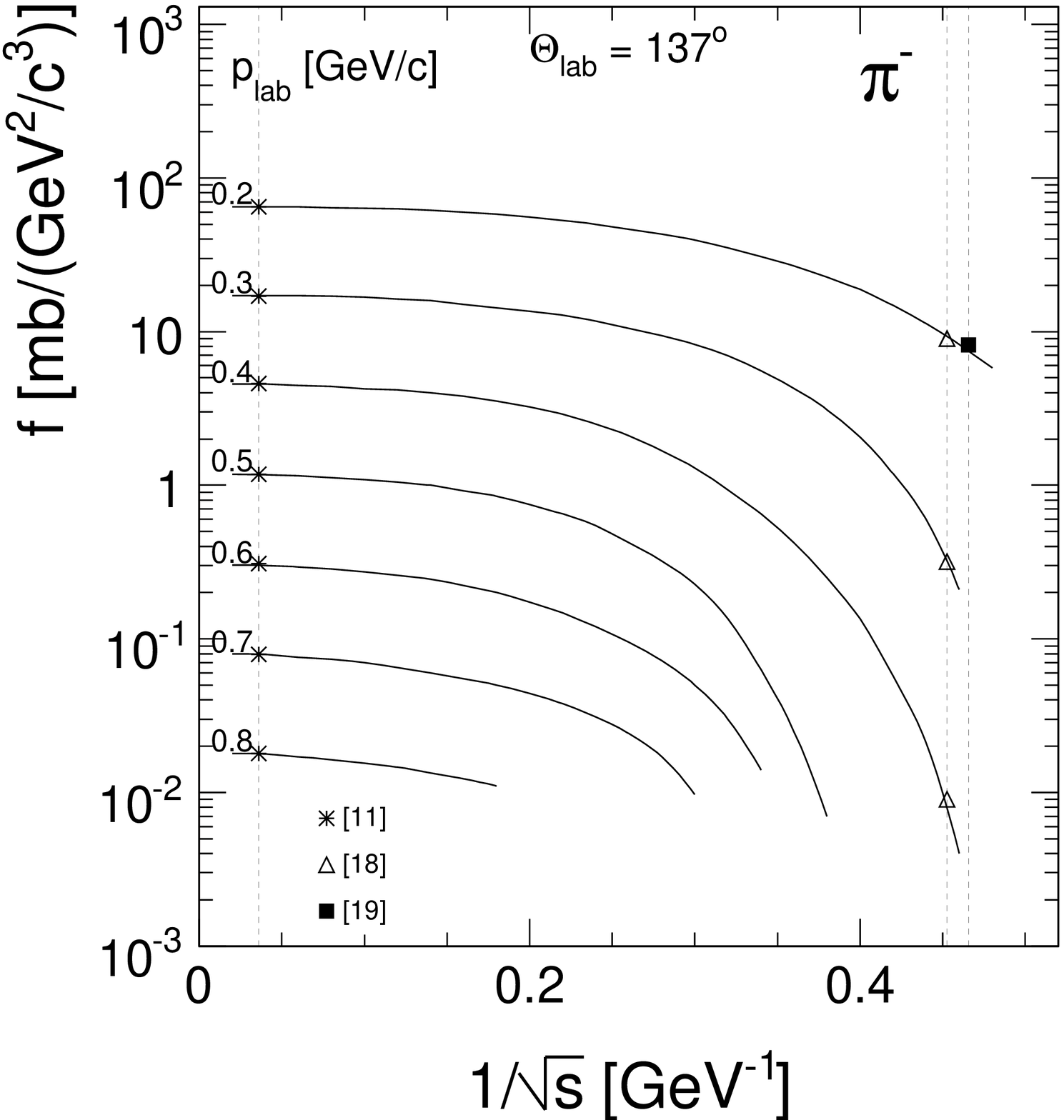}
  \end{center}
  \end{minipage}
  \begin{minipage}[t]{0.49\linewidth} 
  \begin{center}
    \includegraphics[width=7.1cm]{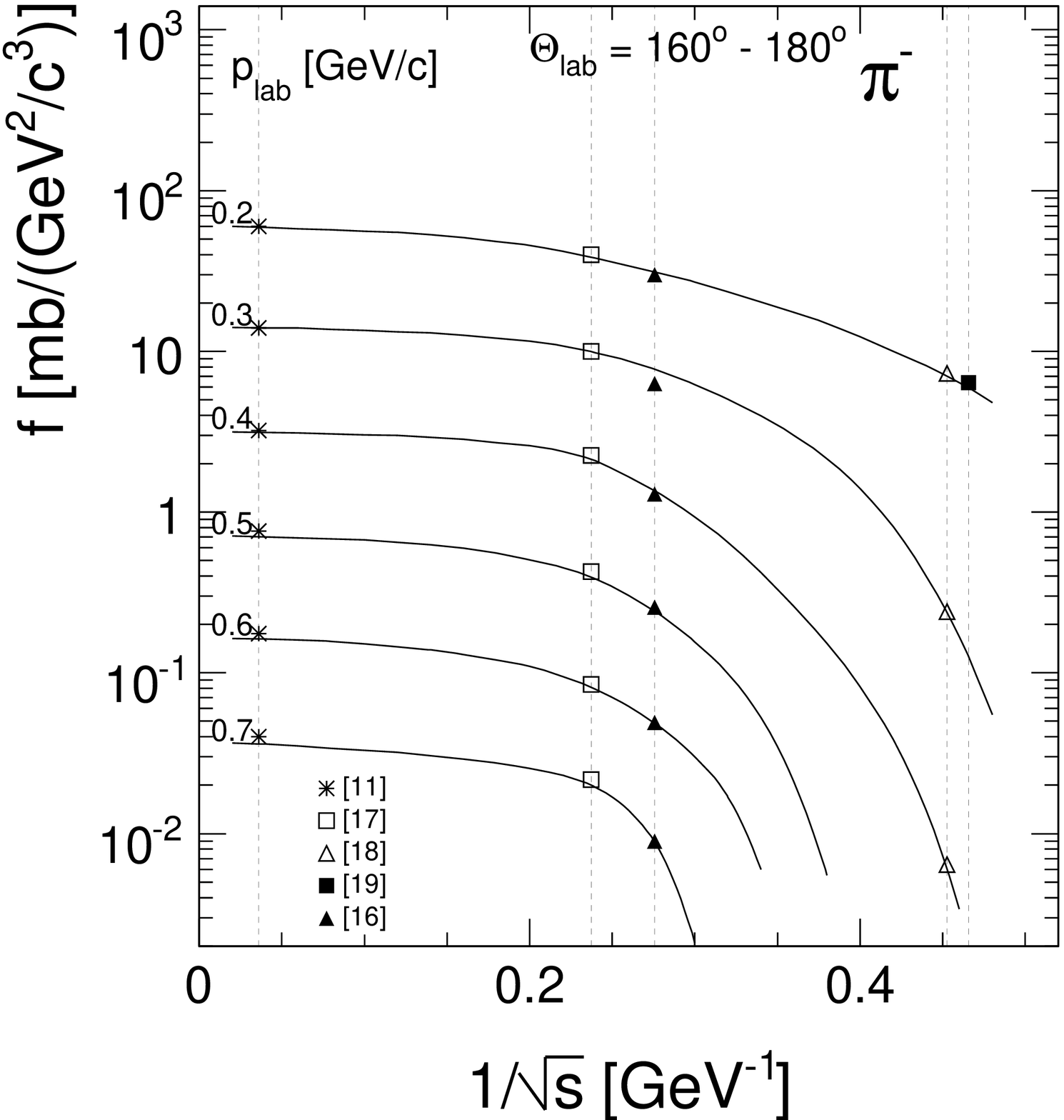}
  \end{center}
  \end{minipage}
  \caption{Invariant cross sections for $\pi^-$ in p+C
           collisions as a function of $1/\sqrt{s}$ at fixed $p_{\textrm{lab}}$ and
           $\theta_{\textrm{lab}}$. The interpolated data points are indicated
           by symbols corresponding to the respective experiments
           in each panel. The solid lines represent the global
           data interpolation}
  \label{fig:sqs_pin}
\end{figure}

The invariant $\pi^-$ cross sections are shown in Fig.~\ref{fig:sqs_pin} as a function 
of $1/\sqrt{s}$ for the standard grid of ten lab angles between 25 and
180~degrees and for constant lab momenta between 0.2 and 1.2~GeV/c. 
The interpolated data points in each panel are identified by symbols 
corresponding to the different experiments.

\begin{figure*}[h]
  \begin{minipage}[t]{0.49\linewidth} 
  \begin{center}
  	\includegraphics[width=7.1cm]{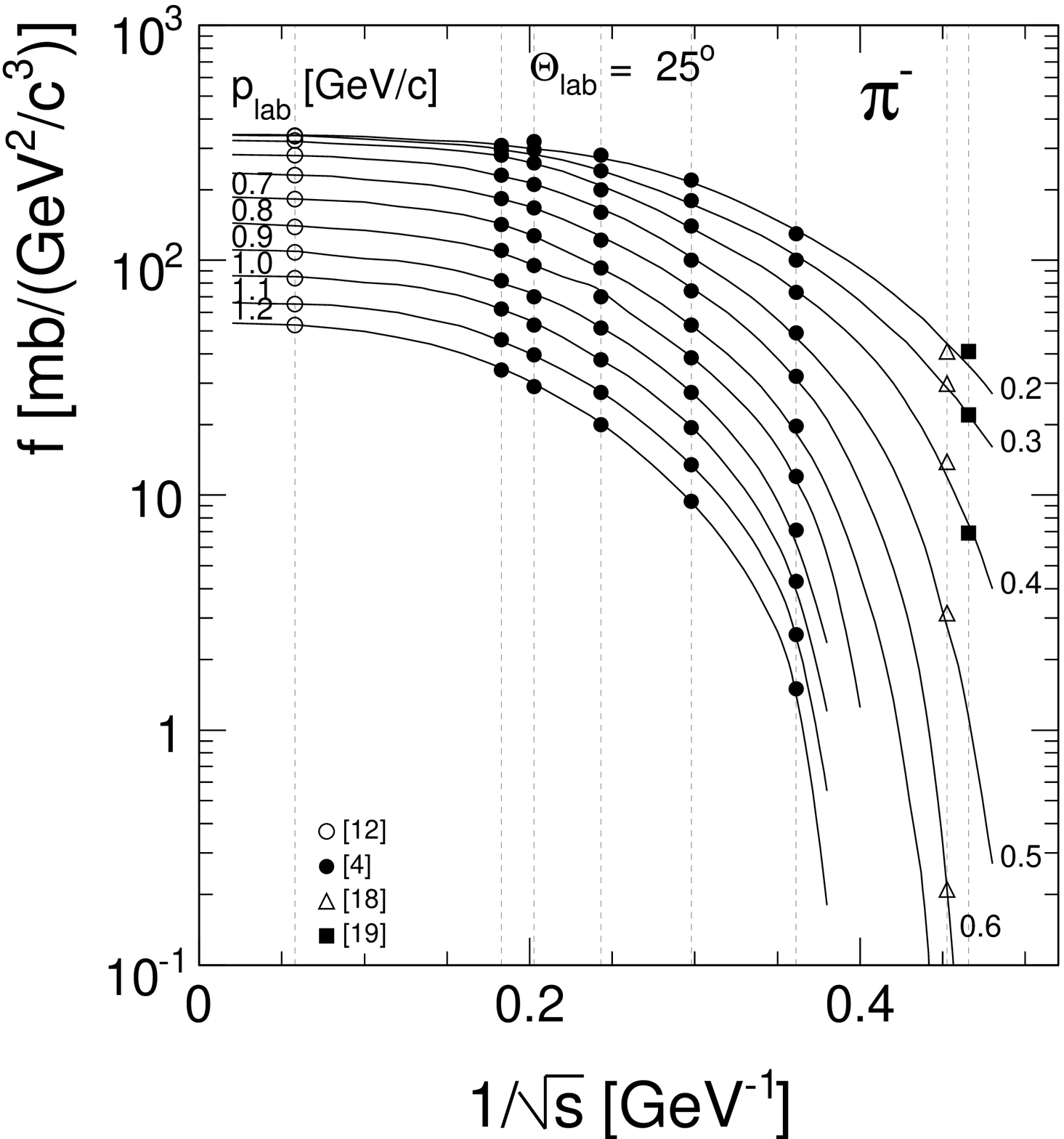}
  \end{center}
  \end{minipage}
  \begin{minipage}[t]{0.49\linewidth} 
  \begin{center}
  	\includegraphics[width=7.1cm]{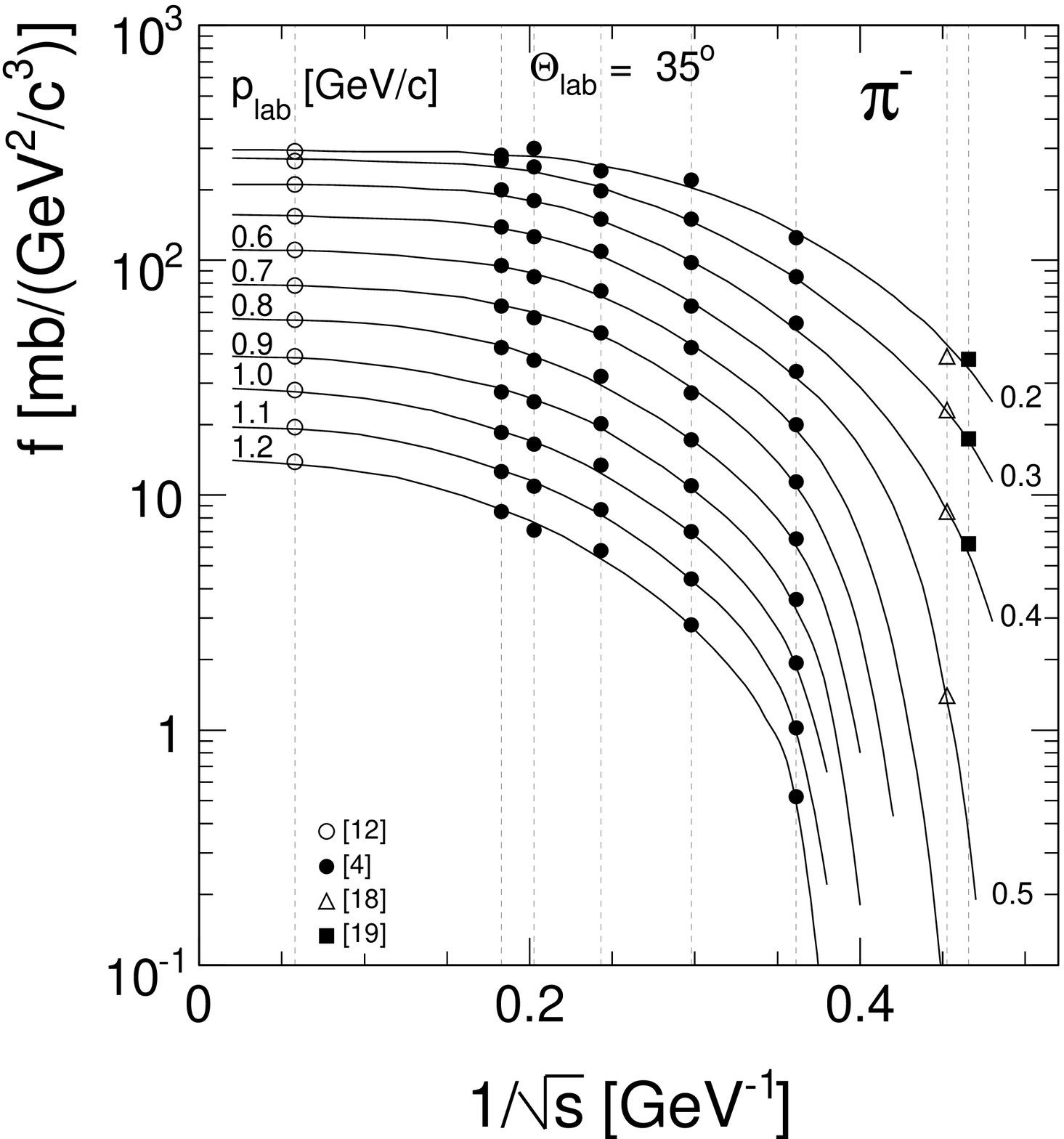}
  \end{center}
  \end{minipage}
\end{figure*}
\begin{figure*}[h]
  \begin{minipage}[t]{0.49\linewidth} 
  \begin{center}
  	\includegraphics[width=7.1cm]{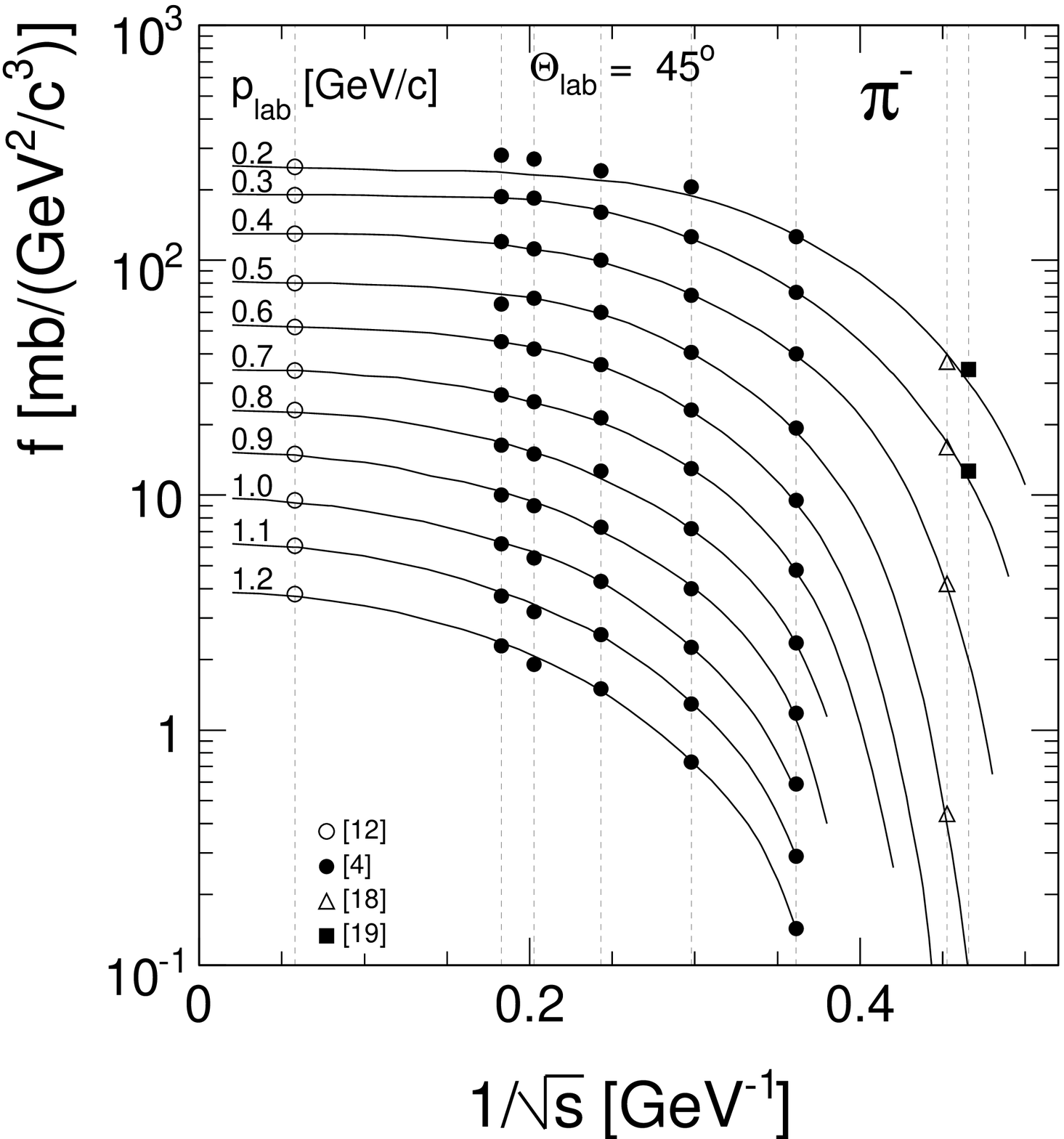}
  \end{center}
  \end{minipage}
  \begin{minipage}[t]{0.49\linewidth} 
  \begin{center}
  	\includegraphics[width=7.1cm]{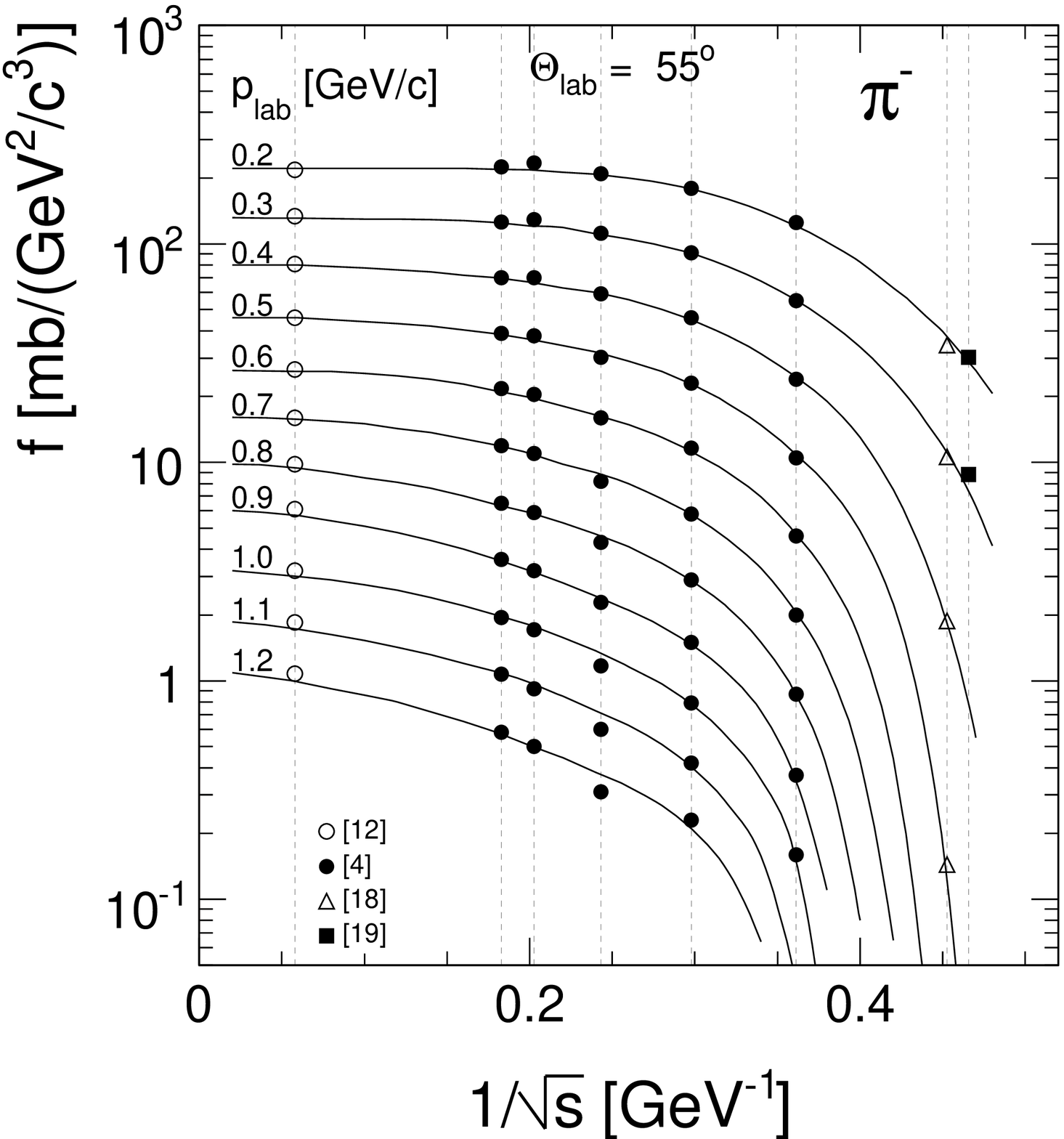}
  \end{center}
  \end{minipage}
\end{figure*}
\begin{figure*}[h]
  \begin{minipage}[t]{0.49\linewidth} 
  \begin{center}
  	\includegraphics[width=7.1cm]{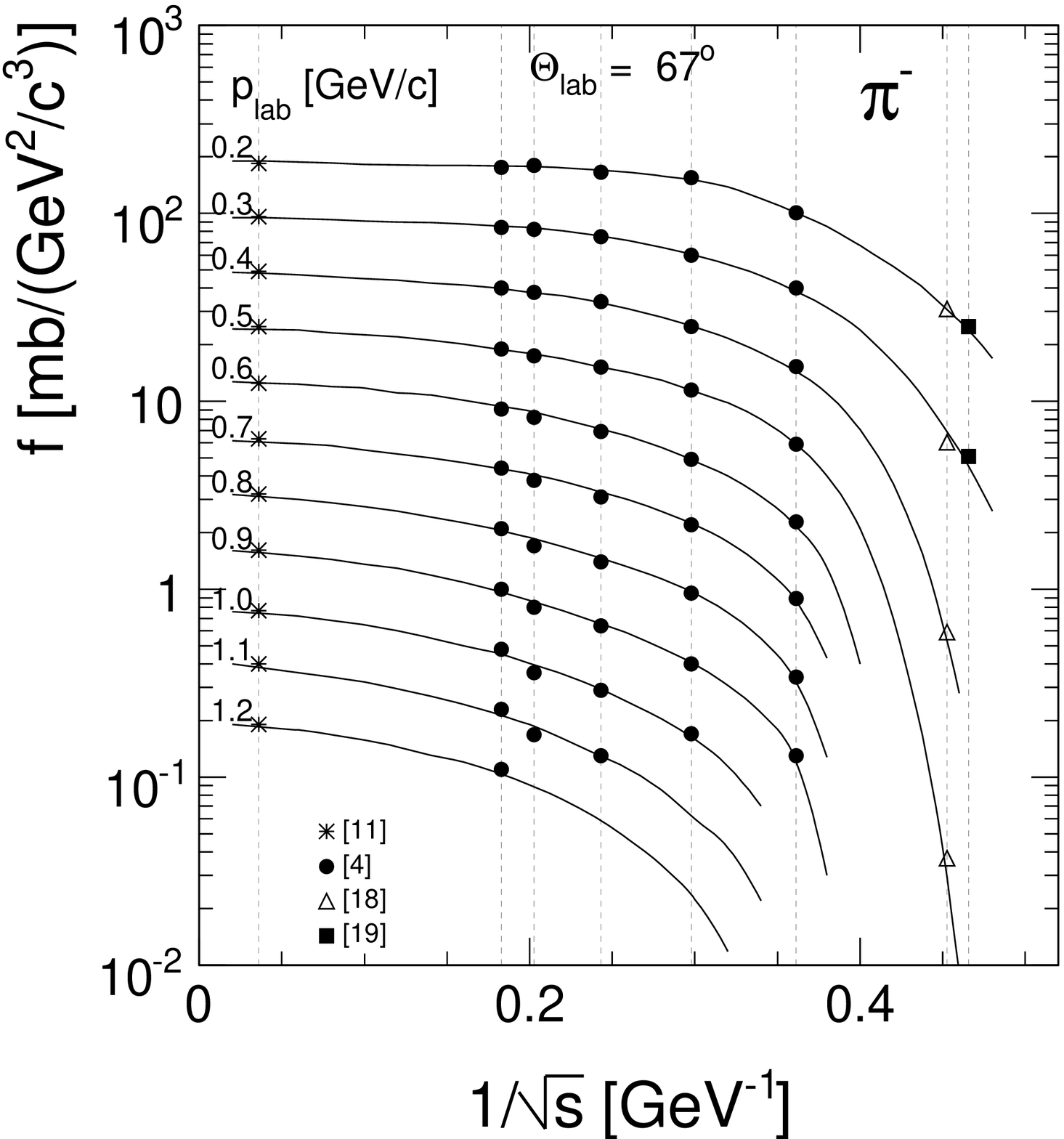}
  \end{center}
  \end{minipage}
  \begin{minipage}[t]{0.49\linewidth} 
  \begin{center}
  	\includegraphics[width=7.1cm]{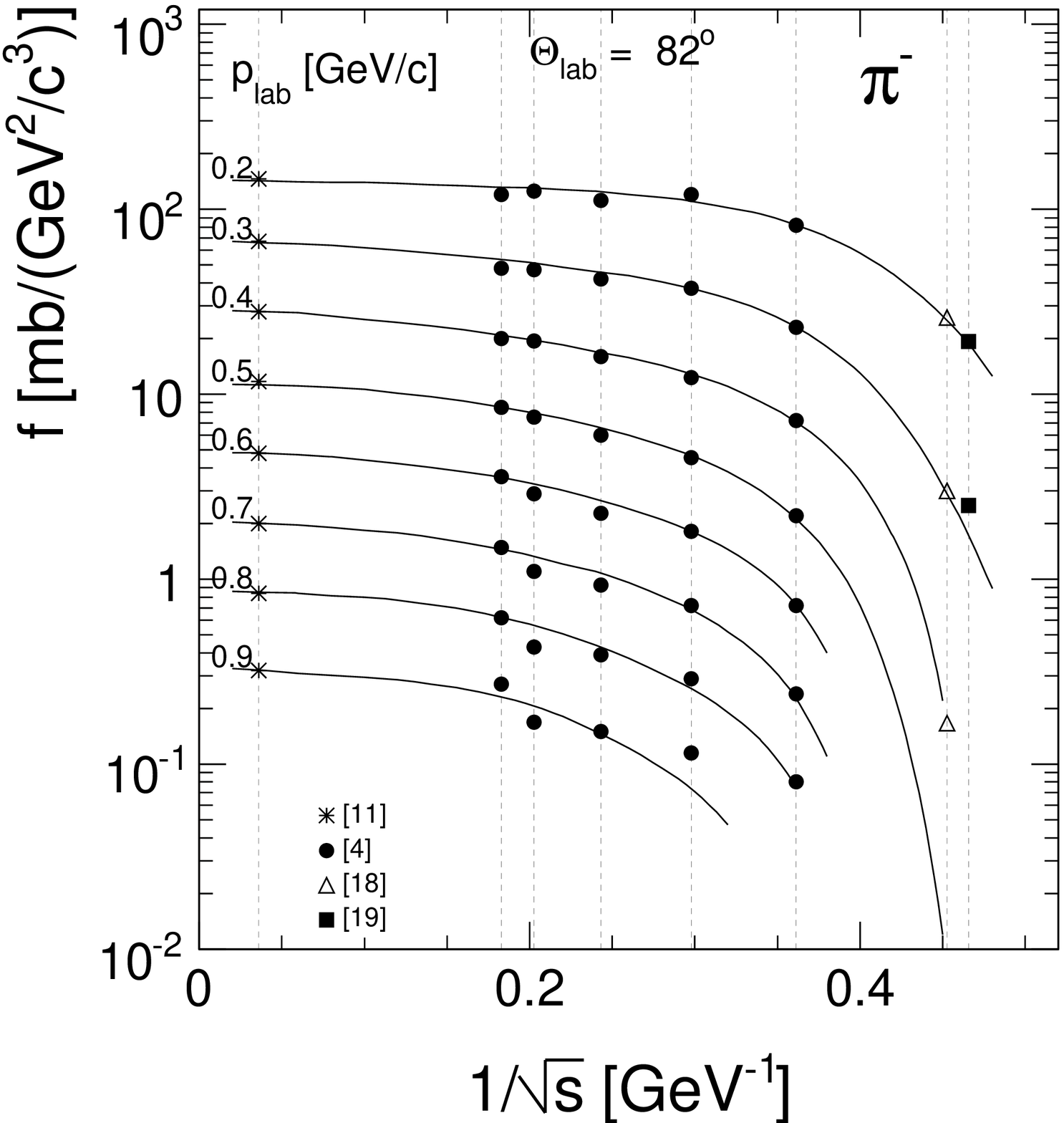}
  \end{center}
  \end{minipage}
\end{figure*}

The solid lines represent the global interpolation by eyeball fits
to the data, with several features which are worth noticing:

\begin{itemize}
 \item All the different data sets form a consistent ensemble without
       the systematic deviations visible in some regions of the
       proton and $\pi^+$ results.
 \item The approach to large beam momenta happens from below for
       all $p_{\textrm{lab}}$.
 \item The $s$-dependence is in general stronger than for $\pi^+$, Fig.~\ref{fig:sqs_pip}.
       If it is again flat up to $1/\sqrt{s} \sim$~0.2 at low $p_{\textrm{lab}}$, it becomes
       more pronounced both towards higher $p_{\textrm{lab}}$ and in the approach
       to the production threshold at large $1/\sqrt{s}$ indicating
       a marked increase of the $\pi^+/\pi^-$ ratio.
 \item This effect has as physics origin the progressive change
       of the production mechanism from pion exchange at low energy
       to gluon or Pomeron exchange at SPS energy. This will be
       discussed in relation to the charge ratios in Sect.~\ref{sec:p2n}.
\end{itemize}

\begin{figure}[h]
  \begin{center}
  	\includegraphics[width=11.cm]{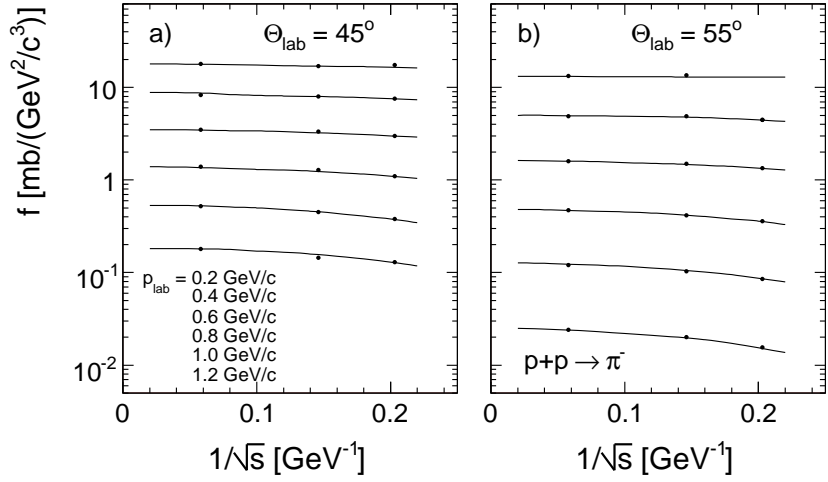}
 	\caption{Invariant $\pi^-$ cross sections as a function
             of $1/\sqrt{s}$ for p+p interactions at the two lab angles
             a) 45, and b) 55~degrees for $p_{\textrm{lab}}$ values from
             0.2 to 1.2~GeV/c. The data are interpolated from Blobel \cite{blobel1} 
             and NA49 \cite{pp_pion}. The lines are drawn to guide the eye}
  	 \label{fig:oneoversq_pp_piminus}
  \end{center}
\end{figure}

It is again interesting to compare the energy dependence to the 
one observed in p+p interactions as presented in Fig.~\ref{fig:oneoversq_pp_piminus}.

Although for both reactions the asymptotic high energy region
is approached from below, this comparison shows a stronger 
$s$-dependence, at the same lab angle, in p+C than in p+p
collisions. This is due to the component of nuclear cascading
which contributes, in the given angular range, with equal strength
than the target fragmentation to the total yield (see Sect.~\ref{sec:separation} 
below). 
 
%
%
\subsection{$\mathbf {cos(\theta_{\textrm{lab}})}$ dependence}
\vspace{3mm}
\label{sec:pin_thdep}

\begin{figure}[h]
  \begin{center}
  	\includegraphics[width=11cm]{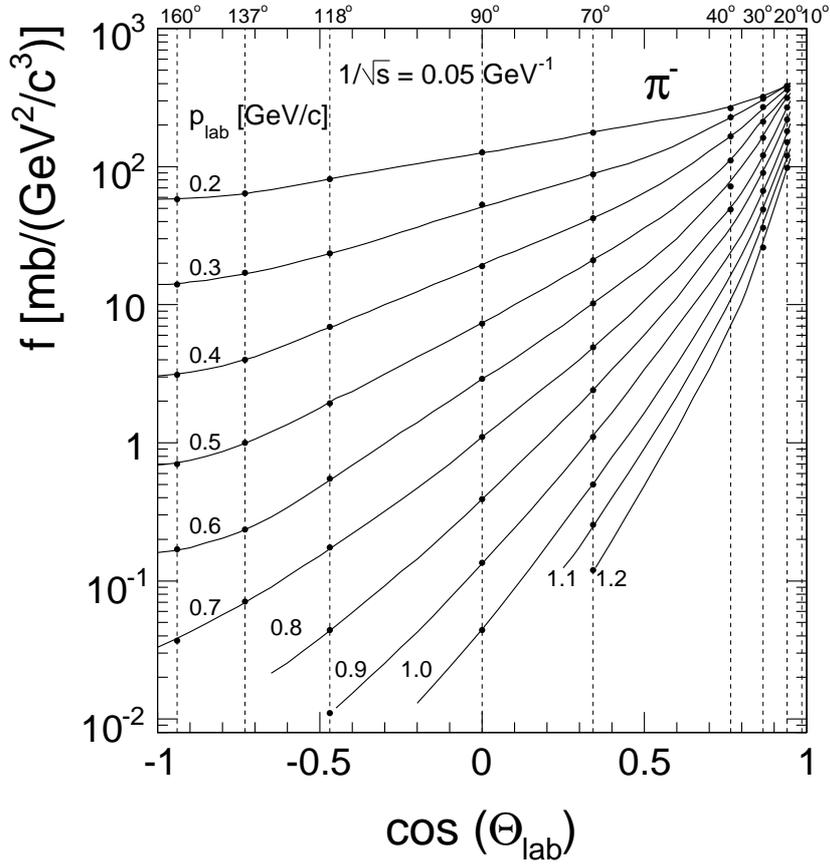}
 	  \caption{Invariant $\pi^-$ cross sections at $1/\sqrt{s}$=~0.05
               as a function of $\cos(\theta_{\textrm{lab}})$ combining the Fermilab \cite{niki} 
               and NA49 \cite{pc_pion} data for $p_{\textrm{lab}}$ between 0.2 and 1.2~GeV/c. The global
               interpolation is shown as full lines. The measured cross
               sections in the angular ranges from 70 to 160~degrees
               (\cite{niki}) and from 10 to 40~degrees (\cite{pc_pion}) are
               given on the vertical broken lines}
  	 \label{fig:pc_pin_theta}
  \end{center}
\end{figure}

As for protons and $\pi^+$ in Figs.~\ref{fig:pc_prot_theta} and \ref{fig:pc_pip_theta}, the $\pi^-$ cross sections
from the Fermilab \cite{niki} and NA49 \cite{pc_pion} experiments are compared and combined
as a function of $\cos(\theta_{\textrm{lab}})$ in Fig.~\ref{fig:pc_pin_theta}.

\begin{figure}[h]
  \begin{center}
  	\includegraphics[width=14cm]{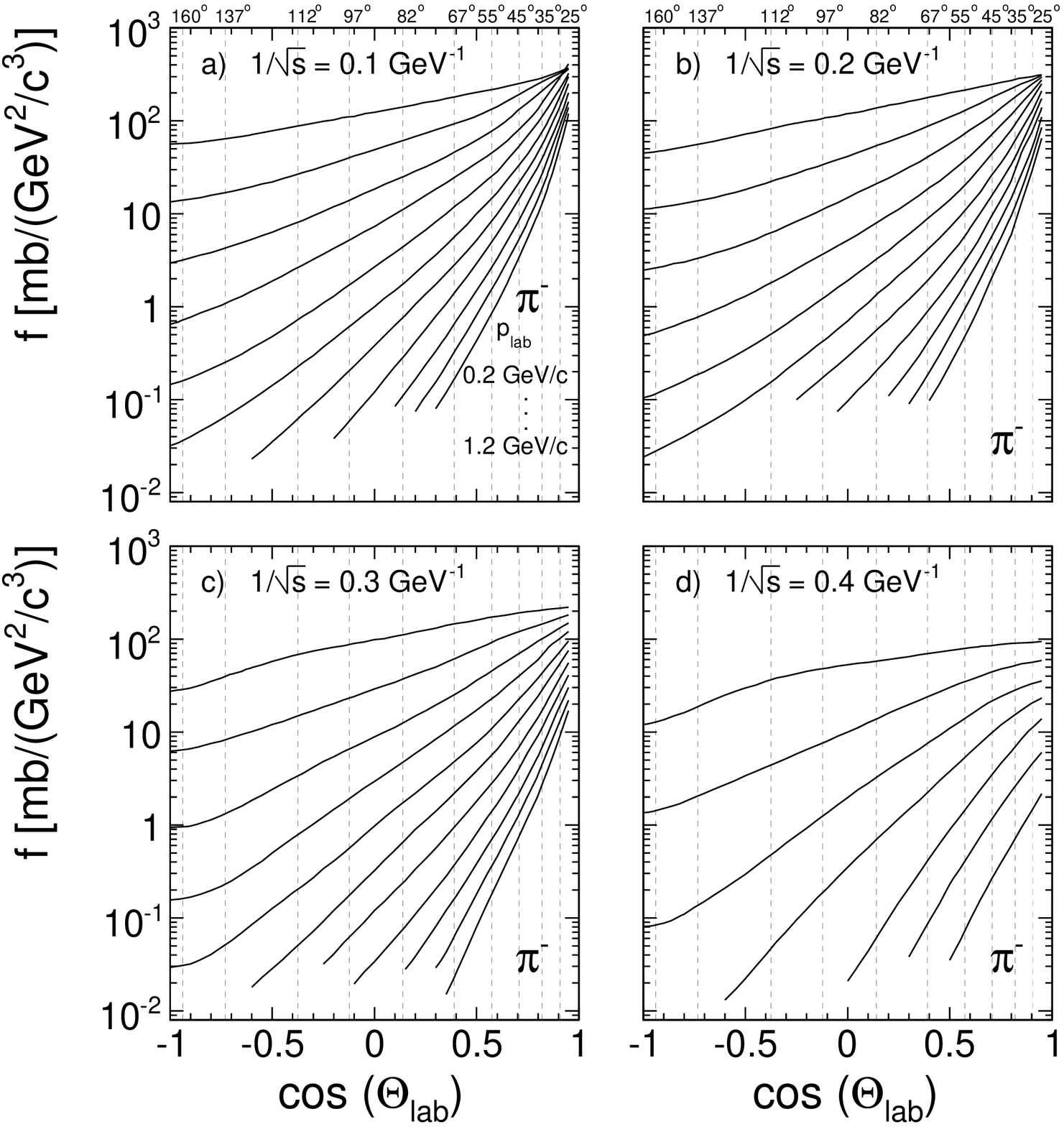}
 	  \caption{Invariant $\pi^-$ cross sections as a function of
               $\cos(\theta_{\textrm{lab}})$ for four values of $1/\sqrt{s}$: a) 0.1, b) 0.2,
               c) 0.3, d) 0.4~GeV$^{-1}$ and for $p_{\textrm{lab}}$ values between 0.2 and
               1.2~GeV/c. The standard grid of 10 angles, Fig.~\ref{fig:oneoversq}, is indicated
               by the vertical broken lines }
  	 \label{fig:costheta_pin}
  \end{center}
\end{figure}

Further angular distributions at four $1/\sqrt{s}$ values between
0.1 and 0.4~GeV$^{-1}$ are given in Fig.~\ref{fig:costheta_pin}.

Concerning smoothness and continuity these distributions are
similar to the $\pi^+$ data, including the large asymmetry between
the forward and backward directions. The reduction of the cross
sections for $\pi^-$ with respect to $\pi^+$ with increasing $1/\sqrt{s}$
is however very apparent. This will be quantified in the
following Section on $\pi^+/\pi^-$ ratios.

%
%
\section{The $\mathbf {\pi^+/\pi^-}$ ratio}
\vspace{3mm}
\label{sec:p2n}

As already evoked in Sect.~\ref{sec:ph3}  above, the study of $\pi^+/\pi^-$
ratios has two main advantages. Firstly, in this ratio a major
fraction of the experimental systematic uncertainties cancels.
Secondly, the ratio is constrained by very fundamental and
model-independent physics arguments like charge conservation
and isospin symmetry. In addition, its $s$-dependence is governed
by the hadronic meson exchange process which leads to a
power-law behaviour that will be shown to be common to a
wide range of interactions. In the following argumentation the
ratio between the global data interpolation for $\pi^+$ and $\pi^-$
as described in the preceding Sects.~\ref{sec:pip} and \ref{sec:pin} will be used:

\begin{equation}
  R_\pm(1/\sqrt{s},p_{\textrm{lab}},\theta_{\textrm{lab}}) = \frac{f(\pi^+)}{f(\pi^-)}
\end{equation}

As a by-product, the fluctuation of this ratio as a function of angle and 
interaction energy will allow for the estimation of the local precision 
of the interpolation procedure.

%
%
\subsection{The high energy limit}
\vspace{3mm}
\label{sec:p2n_high}

It has been established by numerous experimental results that
at collision energies in the SPS/Fermilab range and above
the hadronic interactions are characterized by the absence of charge and 
flavour exchange. It has also been shown that the feed-over of pions from 
the projectile hemisphere into the backward region of $x_F$ is sharply limited 
to the range of $x_F \gtrsim$~-0.05, see \cite{pc_discus} for a detailed discussion. This range
is outside the coverage in $\Theta_{\textrm{lab}}$ and $p_{\textrm{lab}}$ considered in
this publication.

It is therefore to be expected that the backward production of pions off 
an isoscalar nucleus should be charge-symmetric at high energy. This is indeed 
verified by the results on pion production shown in the preceding sections. 
It is quantified in Fig.~\ref{fig:costheta_pin} which shows the $\pi^+/\pi^-$ ratio at $1/\sqrt{s}$~=~0.04 or 
330~GeV/c beam momentum for all lab angles and lab momenta treated in this publication.
This number distribution has a mean value of 1.0125 with an
rms deviation of 3.2\%. This rms value may be seen as a first estimate 
of the local precision of the three-dimensional interpolation scheme at this energy 
which has been established independently for both pion charges.

\begin{figure}[h]
  \begin{center}
  	\includegraphics[width=8.cm]{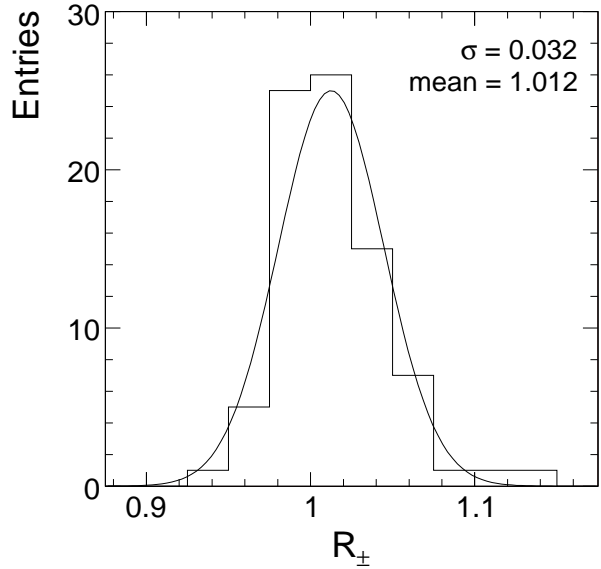}
 	\caption{$\pi^+$/$\pi^-$ ratio $R^{\pm}$ at $1/\sqrt{s}$~=~0.04~GeV$^{-1}$ for 
             25~$< \Theta_{\textrm{lab}}< $~162~degrees and 0.2~$< p_{\textrm{lab}} <$~1.2~GeV/c}
  	 \label{fig:rpm_hist}
  \end{center}
\end{figure}

%
%
\subsection{Energy, momentum and angle dependence of $R_\pm$}
\vspace{3mm}
\label{sec:p2n_endep}

\begin{figure}[h]
  \begin{center}
  	\includegraphics[width=12.cm]{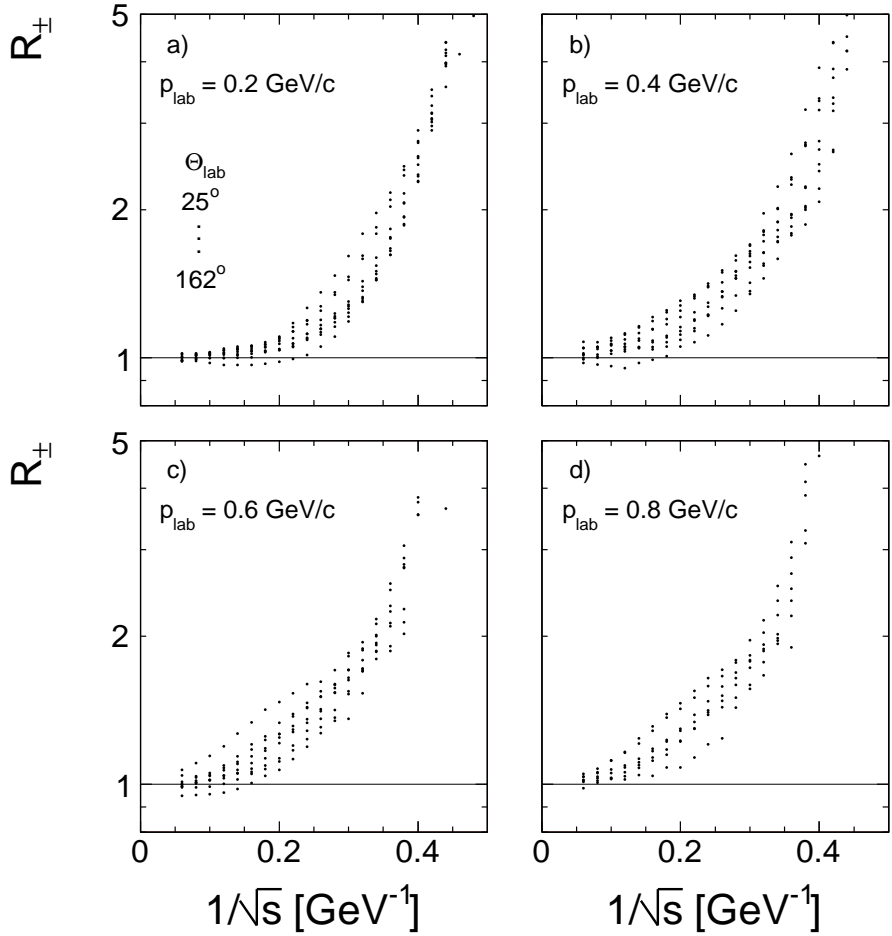}
 	\caption{$R_\pm$ as a function of $1/\sqrt{s}$ for four values of $p_{\textrm{lab}}$. 
 	         The dots represent the ratio of the global data interpolation for $\pi^+$ and $\pi^-$ in steps 
             of 0.02 in $1/\sqrt{s}$ and for the chosen grid of 10 angles.
             a) $p_{\textrm{lab}}$~=~0.2, b) $p_{\textrm{lab}}$~=~0.4, c) $p_{\textrm{lab}}$~=~0.6, 
             d) $p_{\textrm{lab}}$~=~0.8~GeV/c}
  	 \label{fig:rpm_plab}
  \end{center}
\end{figure}

With decreasing interaction energy or increasing $1/\sqrt{s}$ the
$\pi^+/\pi^-$ ratio develops a strong increase at all lab momenta and
lab angles. This is shown in Fig.~\ref{fig:rpm_plab} which gives $R_\pm$ as a function of $1/\sqrt{s}$ 
for four lab momenta. The ratio of the global data interpolation is given in steps 
of 0.02 in $1/\sqrt{s}$. At each value of $1/\sqrt{s}$ the number of points corresponds 
to the standard grid of angles available at this energy.

Several features of Fig.~\ref{fig:rpm_plab} are noteworthy:

\begin{itemize}
 \item Considering the wide range of lab angles, $R_\pm$ is at each value of
       $1/\sqrt{s}$ confined to a narrow band indicating an approximative
       angle independence.
 \item Large $R_\pm$ values in excess of 5 are reached at the upper limit
       of the available scale in $1/\sqrt{s}$.
 \item There is a systematic increase of $R_\pm$ with $p_{\textrm{lab}}$.
\end{itemize}

%
%
\subsubsection{Mean $\mathbf {\pi^+/\pi^-}$ ratios and estimation of the local systematic
               fluctuations of the interpolation process}
\vspace{3mm}
\label{sec:p2n_high1}

\begin{figure}[h]
  \begin{center}
  	\includegraphics[width=7.5cm]{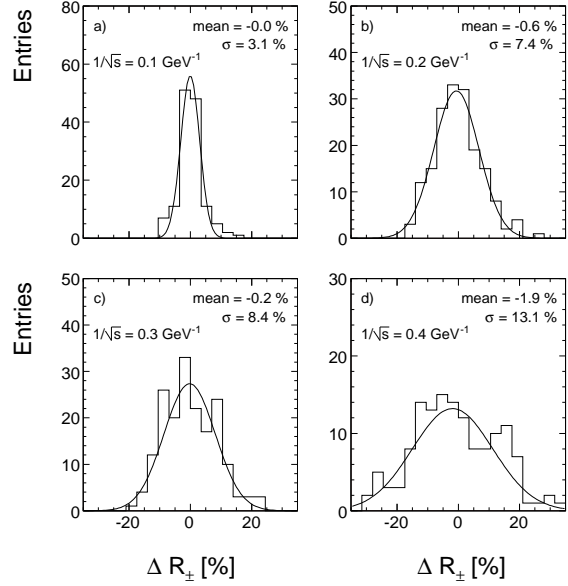}
 	\caption{Number distribution of the normalized percent 
             deviations $\Delta R_\pm$ from the mean over the angular range,
             a) $1/\sqrt{s}$~=~0.1, b) $1/\sqrt{s}$~=~0.2, c) $1/\sqrt{s}$~=~0.3, 
             d)$1/\sqrt{s}$~=~0.4. The four $p_{\textrm{lab}}$ values shown in Fig.~\ref{fig:rpm_plab} are summed up}
  	 \label{fig:p2n_angdiff}
  \end{center}
\end{figure}

The features pointed out above may be quantified and at the same
time the local systematic fluctuations of the interpolation may be
estimated by establishing the mean values $\langle R_\pm \rangle$ averaged over
the angular range at each $1/\sqrt{s}$. These mean values are well defined as shown 
in Fig.~\ref{fig:p2n_angdiff} which presents the normalized distribution 
of the point-by-point deviations from the mean in percent,

\begin{equation}
  \Delta R_\pm = 100 \frac{R_\pm - \langle R_\pm \rangle }{\langle R_\pm \rangle} 
\end{equation}

for four values of $1/\sqrt{s}$, summing the four $p_{\textrm{lab}}$ values used in Fig.~\ref{fig:rpm_plab}.

\begin{figure}[h]
  \begin{center}
  	\includegraphics[width=6.cm]{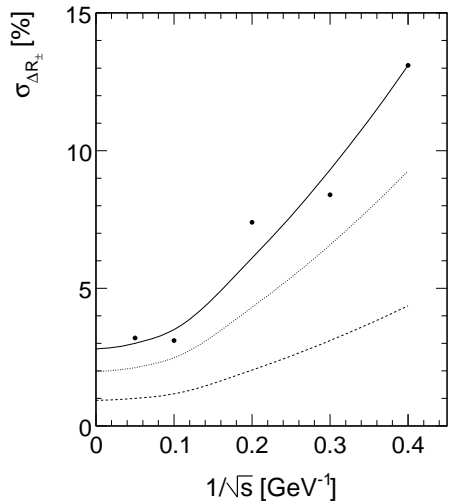}
 	\caption{Rms values of the number distributions of the
      		 normalized point-by-point deviations from the mean charge
      		 ratio $\langle R_\pm \rangle$ as a function of $1/\sqrt{s}$. The full line
             represents a hand interpolation. Broken line: corresponding
             error margin of the mean values $\langle R_\pm \rangle$. Dotted lines: corresponding
             errors for $\pi^+$ and $\pi^-$ separately}
  	 \label{fig:sigsqs}
  \end{center}
\end{figure}

These distributions are of Gaussian shape with an rms which increases
with $1/\sqrt{s}$ as indicated in Fig.~\ref{fig:sigsqs}.

The observed energy dependence of the rms deviations is due to the fact that 
the invariant pion cross sections decrease, after a relatively flat behaviour 
up to $1/\sqrt{s} \sim$~0.15, progressively steeper towards the
production threshold, see Figs.~\ref{fig:sqs_pip} and \ref{fig:sqs_pin}. This leads inevitably
to larger variations in the corresponding energy interpolation.

From the rms values given in Fig.~\ref{fig:sigsqs} the error of $\langle R_\pm \rangle$ may be derived which varies 
between 1\% and 5\% for the highest and lowest interaction energy, respectively (broken line). 
Also the corresponding error margins for the mean pion yields may be extracted as indicated 
by the dotted lines in Fig.~\ref{fig:sigsqs}. From these plots it appears that the global interpolation
induces fluctuations which increase from a few percent in the high $s$
region to about 10\% in the approach to the pion threshold.

%
%
\subsubsection{Dependence of $R_\pm$ on $\Theta_{\textrm{lab}}$}
\vspace{3mm}
\label{sec:p2n_high2}

\begin{figure}[h]
  \begin{center}
  	\includegraphics[width=13.5cm]{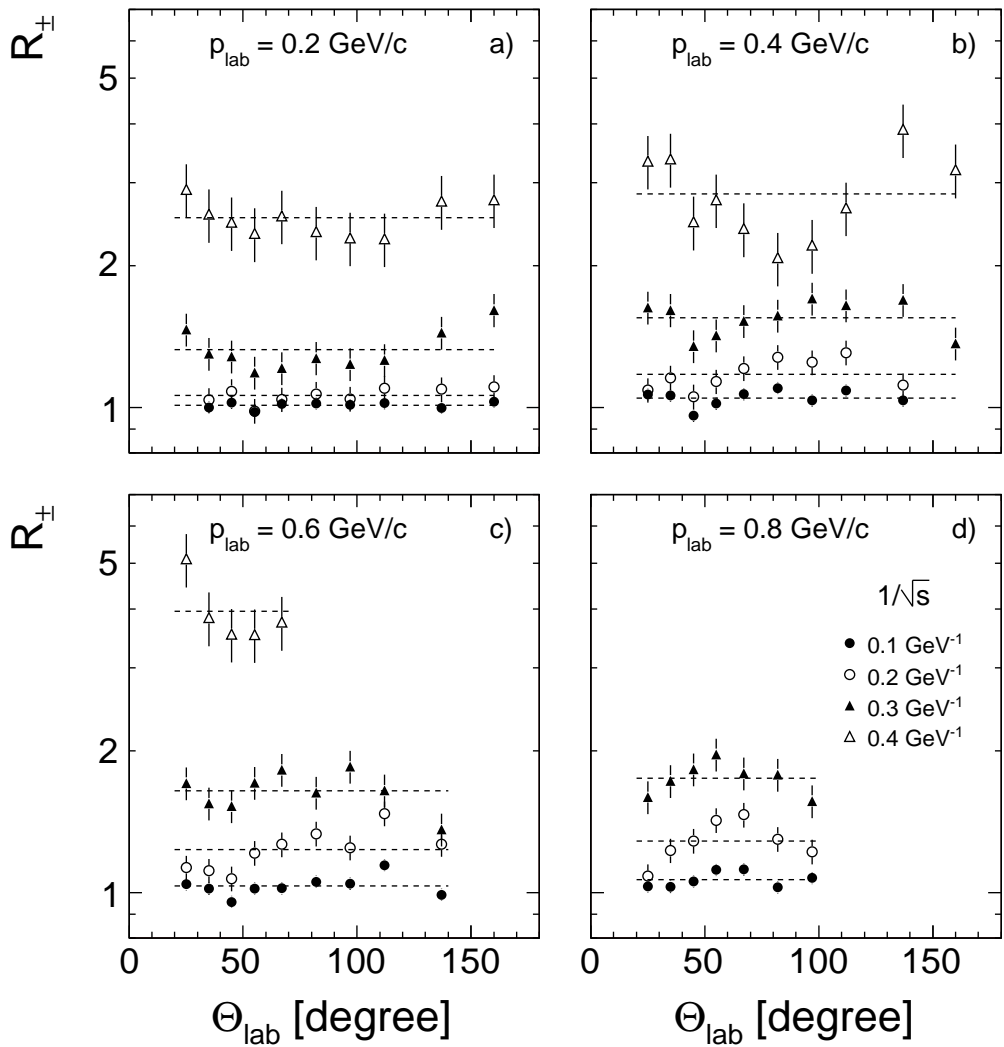}
 	\caption{$R_\pm$ as a function of $\Theta_{\textrm{lab}}$ for four values of
             $1/\sqrt{s}$, a) $p_{\textrm{lab}}$~=~0.2, b) $p_{\textrm{lab}}$~=~0.4, c) $p_{\textrm{lab}}$~=~0.6, 
             and d) $p_{\textrm{lab}}$~=~0.8~GeV/c. The mean values $\langle R_\pm \rangle$ are
             indicated as the horizontal broken lines in each panel}
  	 \label{fig:angdep}
  \end{center}
\end{figure}

The dependence of $R_\pm$ on $\Theta_{\textrm{lab}}$ is shown in Fig.~\ref{fig:angdep} for four values
of $1/\sqrt{s}$ and four values of $p_{\textrm{lab}}$.

Evidently no systematic $\Theta_{\textrm{lab}}$ dependence is visible over the
complete angular range within the quoted errors.

%
%
\subsubsection{Dependence of $\langle R_\pm \rangle$ on $1/\sqrt{s}$ and $p_{\textrm{lab}}$}
\vspace{3mm}
\label{sec:p2n_high3}

\begin{figure}[h]
  \begin{center}
  	\includegraphics[width=12.cm]{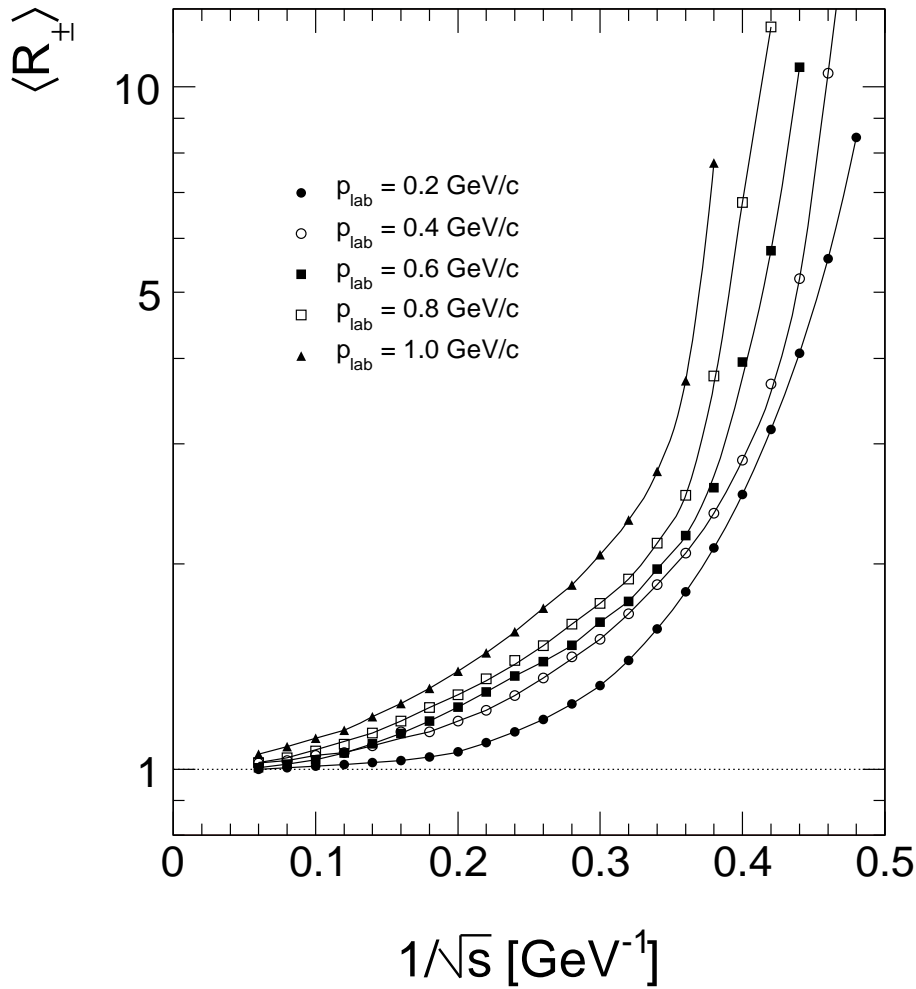}
 	\caption{$\langle R_\pm \rangle$ as a function of $1/\sqrt{s}$ for five values
            of $p_{\textrm{lab}}$ between 0.2 and 1.0~GeV/c. The full lines are
            hand interpolations through the data points}
  	 \label{fig:p2n_meanang}
  \end{center}
\end{figure}

In the absence of angular dependence of $R_\pm$ as shown above, the
mean values $\langle R_\pm \rangle$ may now be used in order to establish a precise
view of the $1/\sqrt{s}$ dependence for different $p_{\textrm{lab}}$ values. This
dependence is presented in Fig.~\ref{fig:p2n_meanang}.

Within the errors of $\langle R_\pm \rangle$ extracted above, a clear $p_{\textrm{lab}}$ dependence
is evident superposing itself to the strong common increase of
$\langle R_\pm \rangle$ with $1/\sqrt{s}$. This increase may be parametrized up to
$1/\sqrt{s} \sim$~0.3 by the functional from $1 + c/s^{\beta(p_{\textrm{lab}})}$ which is,
as discussed below, typical of meson exchange processes. Indeed
the exponent beta varies from 2 to 1.2 for $p_{\textrm{lab}}$ increasing from 0.2 to 0.8~GeV/c.

%
%
\subsection{Interpretation of the observed energy and momentum dependences}
\vspace{3mm}
\label{sec:p2n_epdep}

The strong increase of $R_\pm$ with $1/\sqrt{s}$ merits a detailed study
as it is directly connected to the basic hadronic production mechanisms in p+A interactions. 
The fact that the pion yields in the complete
backward fragmentation region of an isoscalar nucleus remember
the isospin of the projectile is clearly incompatible with charge and 
flavour independent exchange processes. Instead a
meson exchange mechanism may be invoked which has indeed been
used successfully in a wide range of work at low projectile momenta,
see for instance \cite{cochran} and references therein. Close to the pion
production threshold in the nuclear hemisphere, single excitation processes 
via pion exchange of the type

\begin{alignat}{3}
   \textrm{p} + (\textrm{p}) & \quad\rightarrow\quad \Delta^{++} && + (\textrm{n})  && \quad\rightarrow\quad  \pi^+          \\
   \textrm{p} + (\textrm{p}) & \quad\rightarrow\quad \Delta^+    && + (\textrm{p})  && \quad\rightarrow\quad  \pi^+, \pi^0   \\
   \textrm{p} + (\textrm{n}) & \quad\rightarrow\quad \Delta^+    && + (\textrm{n})  && \quad\rightarrow\quad  \pi^0, \pi^+ 
\end{alignat}

only allow $\pi^+$ and $\pi^0$ production, whereas $\pi^-$ production needs
double excitation like

\begin{alignat}{3}
   \textrm{p} + (\textrm{p}) & \quad\rightarrow\quad \Delta^{++} && + (\Delta^0)  && \quad\rightarrow\quad \pi^+, \pi^0, \pi^-  \\
   \textrm{p} + (\textrm{n}) & \quad\rightarrow\quad \Delta^+    && + (\Delta^0)  && \quad\rightarrow\quad \pi^+, \pi^0, \pi^-  \\
   \textrm{p} + (\textrm{n}) & \quad\rightarrow\quad \Delta^{++} && + (\Delta^-)  && \quad\rightarrow\quad \pi^+, \pi^-  
\end{alignat}
with in general an additional penalty for $\pi^-$ due to the isospin Clebsch--Gordan coefficients. 
All meson exchange mechanisms are characterized by a strong decrease with projectile energy.
This energy dependence and its interplay with processes governing
the high energy sector is studied here for the first time in
p+A collisions using the $\pi^+/\pi^-$ ratio.

In this context it seems mandatory to first refer to the study of
exclusive charge exchange reactions in elementary nucleon-nucleon
collisions as the complete energy range discussed here has been
covered there by a number of experiments \cite{miller,kreisler,bohmer,babaev,barton,dekerret,goggi}.

%
%
\subsubsection{The charge exchange mechanism in elementary nucleon-nucleon collisions}
\vspace{3mm}
\label{sec:p2n_epdep1}

Charge exchange processes may be cleanly isolated experimentally in nucleon-nucleon 
interactions by studying the following exclusive
channels:

\begin{itemize}
 \item Charge exchange scattering of the elastic type
  \begin{flalign}
     &\textrm{n} + \textrm{p} \quad\rightarrow\quad \textrm{p} + \textrm{n}&
  \end{flalign}
 \item Single dissociation with pion production
  \begin{flalign}
     \label{eq:single}
     &\textrm{p} + \textrm{p} \quad\rightarrow\quad \textrm{n} + \Delta^{++} \quad\rightarrow\quad \textrm{n} + (\textrm{p} + \pi^+)&
  \end{flalign}
 \item Double dissociation with pion production
  \begin{flalign}
     \label{eq:double}
     &\textrm{p} + \textrm{p} \quad\rightarrow\quad (\textrm{p} + \pi^-) + (\textrm{p} + \pi^+)&
  \end{flalign}
\end{itemize}

These channels are characterized by a very steep energy dependence.

This is to be confronted with non-charge-exchange exclusive channels like:

\begin{itemize}
 \item Elastic scattering
  \begin{flalign}
     &\textrm{p} + \textrm{p} \quad\rightarrow\quad \textrm{p} + \textrm{p}&
  \end{flalign}
 \item Single dissociation
  \begin{flalign}
     \label{eq:single0}
     &\textrm{p} + \textrm{p} \quad\rightarrow\quad \textrm{p} + (\textrm{p} + \pi^+ + \pi^-)&
  \end{flalign}
 \item Double dissociation
  \begin{flalign}
     \label{eq:double0}
     &\textrm{p} + \textrm{p} \quad\rightarrow\quad (\textrm{p} + \pi^+ + \pi^-) + (\textrm{p} + \pi^+ + \pi^-)&
  \end{flalign}
\end{itemize}
which show a constant or logarithmically increasing $s$-dependence.

Charge exchange scattering has been measured by five experiments
in the range of neutron beam momenta from 3 to 300~GeV/c. \cite{miller,kreisler,bohmer,babaev,barton}.
This is exactly covering the energy range discussed in this paper.
The single and double dissociation has been studied at the
CERN ISR by two experiments \cite{dekerret,goggi} extending the energy scale to $s$~=~3700~GeV$^2$. 
The two ISR experiments may be directly compared
to the charge exchange measurements after appropriate re-normalization
of the cross sections in the overlap region at the lowest ISR energy.

\begin{figure}[h]
  \begin{center}
  	\includegraphics[width=13cm]{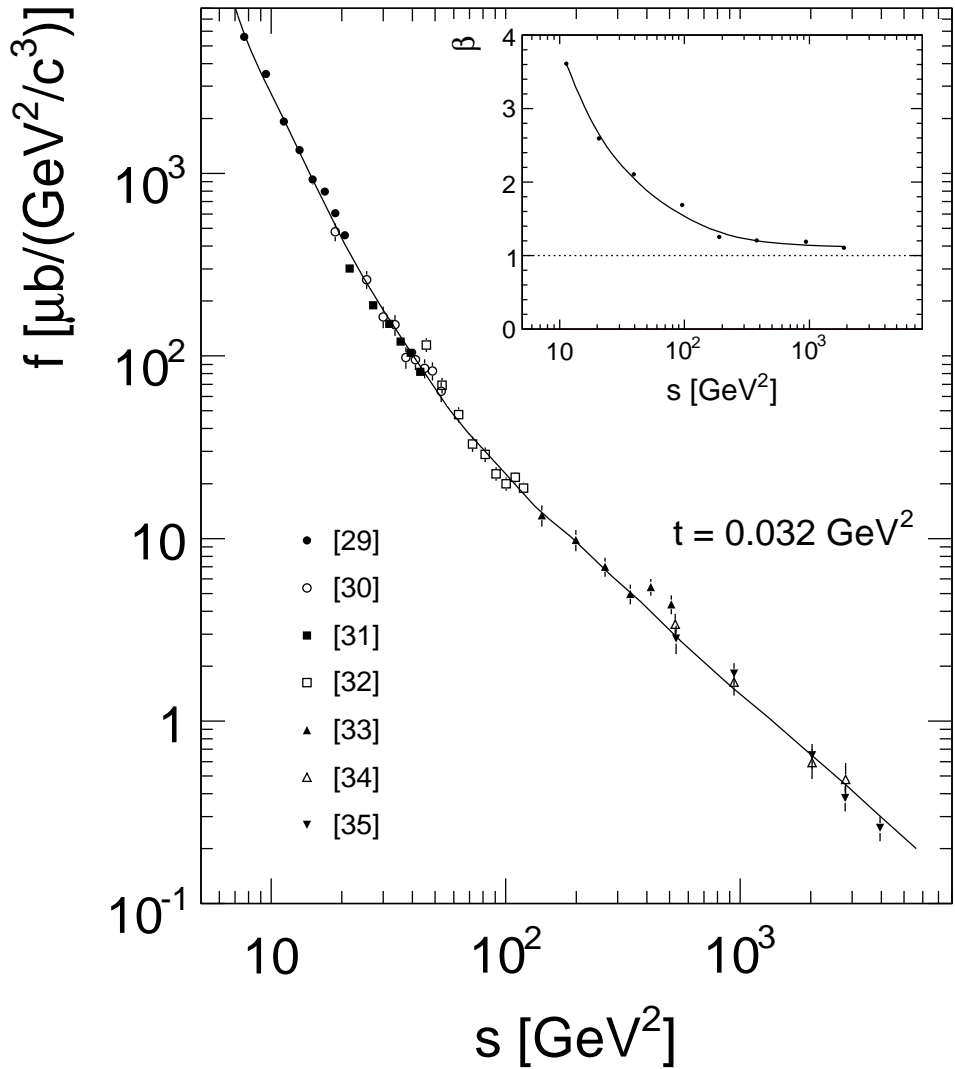}
 	\caption{Invariant cross sections of charge exchange and
             single and double dissociation in nucleon-nucleon interactions
             as a function of $s$ at a momentum transfer $t$~=~0.032~GeV$^2$. The
             full line represents an interpolation of the data points.
             The insert gives the local slope $\beta$ in the parametrization
             $f \sim s^{-\beta}$ as a function of $s$}
  	 \label{fig:chargeexch}
  \end{center}
\end{figure}

The resulting $s$-dependence at a momentum transfer $t$~=~0.032~GeV$^2$
is presented in Fig.~\ref{fig:chargeexch}.

Several features of Fig.~\ref{fig:chargeexch} are of interest:

\begin{itemize}
 \item There is a decrease of about 4 orders of magnitude in cross section
       between the lowest and highest $s$ value. This decrease is to be
       compared to the constant or logarithmically increasing elastic
       and nucleon diffraction cross sections. The charge exchange
       contribution is therefore negligible compared to the inclusive
       baryon yields already at SPS energy.
 \item There is a steady decrease of the local slope $df/ds$ with energy,
       from about 3.6 at 3~GeV/c to about 1.1 above 80~GeV/c beam
       momentum.
 \item A characteristic change of slope manifests itself at around
       30~GeV/c beam momentum.
\end{itemize}

These features have been interpreted in the 1970's when the relevant
experiments were performed, in the framework of Regge theory which
predicts an $s$-dependence of the form

\begin{equation}
   \label{eq:regge}
   f \sim s^{2\alpha -2} = s^{-\beta},
\end{equation}
where $\alpha$ is the intercept of the leading trajectory. This should
in the case of one-meson exchange at low energy be given by the pion trajectory 
with zero intercept. The actual beta values above 3 at low $s$ seem to contradict 
however this expectation. Here threshold effects may play a role which 
are not included in the parametrization (\ref{eq:regge}).

With increasing energy the slopes move through the region of pion
exchange with $\beta \sim$~2 down to values of about 1.1 at high energy which could 
be connected to $\rho$ and a$_2$ exchange with correspondingly
higher intercepts $\alpha$ in the region of 0.5. At ISR energy the
ratio of $\rho/\pi$ contributions has indeed been estimated to be
about 2 \cite{goggi}. Anyway the simple parametrization given by (\ref{eq:regge}) should not be expected 
to hold over the full energy scale. What is interesting here is rather the strong decline 
of the charge exchange cross sections with energy and 
the experimentally rather precisely determined slope variation.

%
%
\subsubsection{A remark concerning baryon resonance production in hadronic interactions}
\vspace{3mm}
\label{sec:p2n_epdep2}

The single (\ref{eq:single}) and double (\ref{eq:double}) dissociation processes defined above
are determined by the formation of $\Delta$ resonances in the final states. 
They therefore constitute a source of direct $\Delta$ production in
nucleon-nucleon interactions. These channel cross sections decrease
rapidly to the $\mu$barn level at SPS energies. In contrast, the
non-charge exchange channels like (\ref{eq:single0}) and (\ref{eq:double0}) have no $s$-dependence
and stay on the mb level of cross sections. Their final states
have been shown to be governed by N$^*$ resonances \cite{goggi1} which may be
excited by Pomeron exchange. Moreover, the p+$\pi^+$ combination
of the p+$\pi^+$+$\pi^-$ final states has been shown to be dominated by
$\Delta^{++}$ \cite{conta}. This is an indirect source of $\Delta$ resonances as a
decay product of N$^*$ states which have large decay branching
fractions into $\Delta$+$\pi$ and $\Delta$+$\rho$. It is therefore questionable
if, at SPS energies and above, any direct $\Delta$ production is
persisting. This is an interesting question for the majority of
microscopic models which produce final states by string fragmentation.
In the baryonic sector, diquark fragmentation is generally invoked
with a prevailing direct production of $\Delta$ resonances which by
isospin counting will dominate over N$^*$. Indeed in practically all
such models there is no or only negligible N$^*$ production. As shown
below, the decrease of charge exchange processes can be traced well
into the non-diffractive, inelastic region of particle production.
The multi-step, cascading decay of primordial N$^*$ resonances into
$\Delta$ resonances and final state baryons should therefore be seriously
considered, in particular also concerning the consequences for
the evolution of final state energy densities with time.

%
%
\subsubsection{The charge exchange mechanism in p+C interactions as a function
      of interaction energy}
\vspace{3mm}
\label{sec:p2n_epdep3}

\begin{figure}[h]
  \begin{center}
  	\includegraphics[width=11.cm]{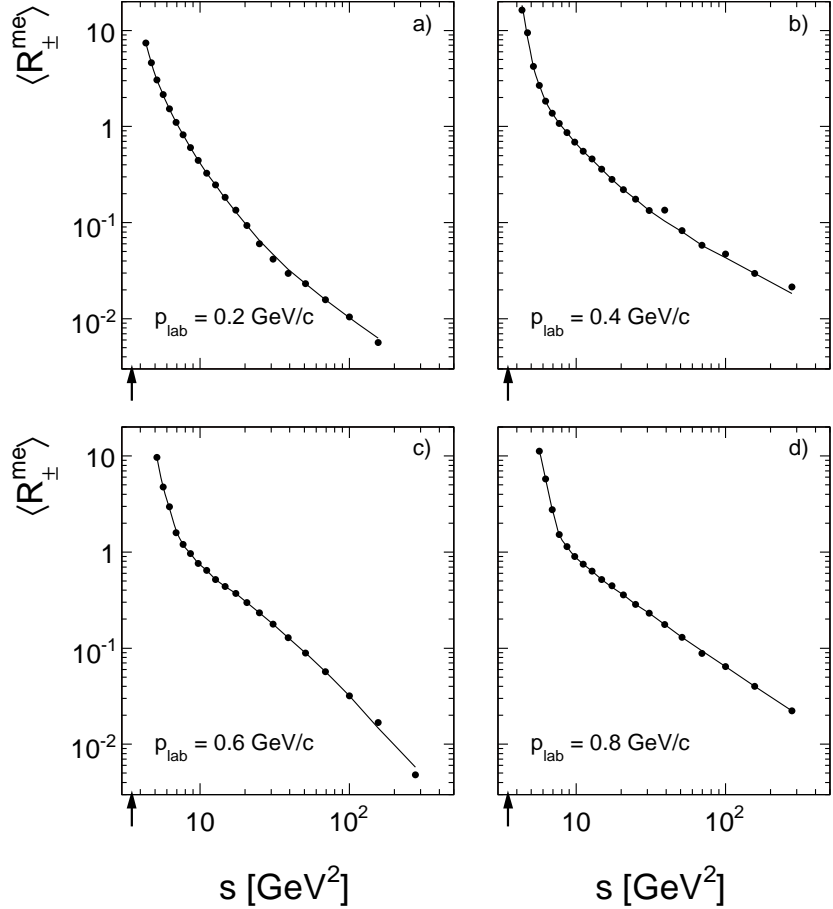}
 	\caption{$\langle R_\pm^{me} \rangle$ as a function of $s$ for different
             values of $p_{\textrm{lab}}$, a) $p_{\textrm{lab}}$~=~0.2, b) $p_{\textrm{lab}}$~=~0.4, 
             c) $p_{\textrm{lab}}$~=~0.6, and d) $p_{\textrm{lab}}$~=~0.8~GeV/c. The elastic limit is
             indicated by the arrows}
  	 \label{fig:p2n_meanang1}
  \end{center}
\end{figure}

The very characteristic decrease of $\langle R_\pm \rangle$ with increasing $s$ derived from 
the global data interpolation, Fig.~\ref{fig:p2n_meanang}, offers a tempting possibility of comparison 
to the phenomenology discussed above for the elementary nucleon-nucleon sector. 
Indeed, two components should contribute to the observed $\pi^+/\pi^-$ ratios: at high energy 
this ratio should approach unity due to the absence of charge and flavour exchange in this region. 
At low energy on the contrary it should be governed by meson exchange with its strong $s$-dependence. 
These two components may be tentatively separated by using instead of $\langle R_\pm \rangle$ the quantity

\begin{equation}
  \langle R_\pm^{me} \rangle = \langle R_\pm \rangle - 1
\end{equation}
in order to extract the meson exchange contribution. This quantity is plotted in Fig.~\ref{fig:p2n_meanang1} 
as a function of $s$ for four $p_{\textrm{lab}}$ values from 0.2 to 0.8~GeV/c.

A very characteristic pattern emerges which resembles the $s$-dependence for the charge exchange 
in elementary interactions described above, see Fig.~\ref{fig:chargeexch}. In general $\langle R_\pm^{me} \rangle$ follows 
a power law dependence on $s$

\begin{equation}
    \langle R_\pm^{me} \rangle \sim cs^{-\beta^{me}}
\end{equation}
with local slopes $\beta^{me}$ which are in turn a function of $s$. Three different regions with distinct local slopes can be
identified in Fig.~\ref{fig:p2n_meanang1}:

\begin{itemize}
 \item A first region with large slopes is located at $s$ below about
       6~GeV$^2$. This region is strongly influenced by threshold effects
       as the threshold for inelastic production is placed at the elastic
       limit $s = 4m_p^2 = 3.5 GeV^2$ indicated in Fig.~\ref{fig:p2n_meanang1}. In the approach
       to pion threshold the $\pi^+/\pi^-$ ratio has to diverge as $\pi^-$ is
       progressively suppressed, see above. With increasing $p_{\textrm{lab}}$ this
       suppression will of course be more pronounced. 
 \item An intermediate region between about 8 and 40~GeV$^2$ with an
       $s$ dependence decreasing with increasing $p_{\textrm{lab}}$.
 \item A third region with flattening $s$-dependence above about 40~GeV$^2$.
\end{itemize}

\begin{figure}[h]
  \begin{center}
  	\includegraphics[width=14.cm]{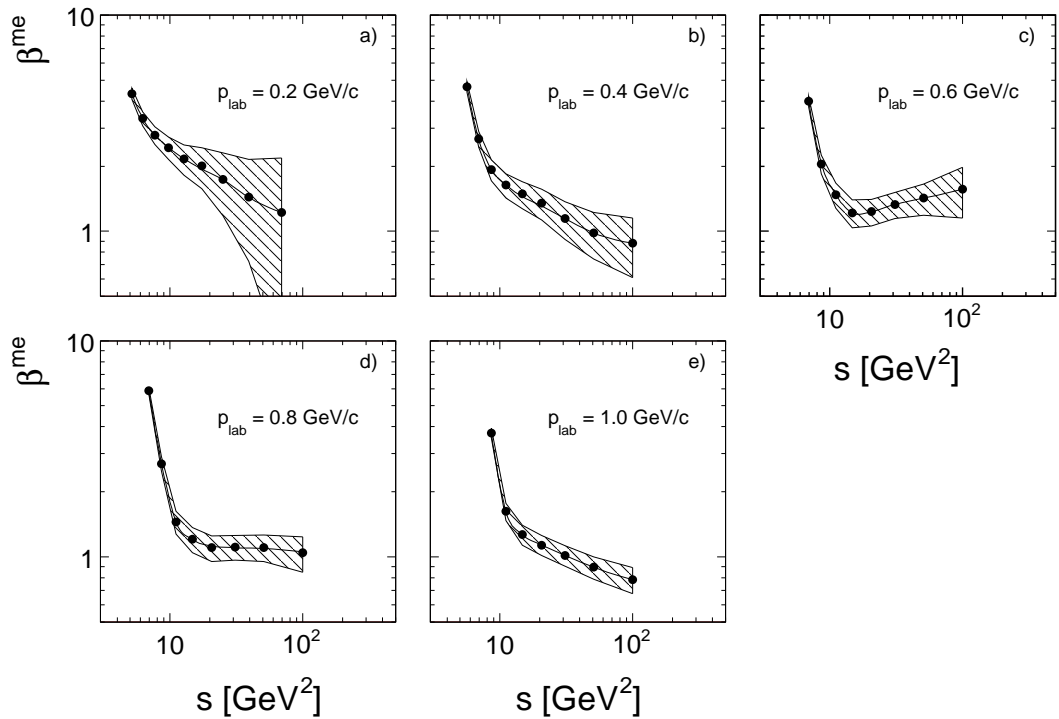}
 	\caption{Slopes $\beta^{me}$ of the $s$-dependence of $\langle R_\pm^{me} \rangle$
             as a function of s: a) $p_{\textrm{lab}}$~=~0.2, b) $p_{\textrm{lab}}$~=~0.4, c) $p_{\textrm{lab}}$~=~0.6,
             d) $p_{\textrm{lab}}$~=~0.8, and e) $p_{\textrm{lab}}$~=~1.0~GeV/c. The shaded regions mark the error margins}
  	 \label{fig:beta}
  \end{center}
\end{figure}

At the lowest $p_{\textrm{lab}}$ value of 0.2~GeV/c corresponding to the lowest momentum transfer, 
the similarity to the charge exchange process
in nucleon-nucleon interactions, Fig.~\ref{fig:chargeexch}, is absolutely striking.
This concerns both the detailed shape and the overall suppression factors. With increasing $p_{\textrm{lab}}$, 
the $s$ dependence is modified in
a systematic way by a general reduction of slopes, with the exception
of the threshold enhancement. This is quantified in Fig.~\ref{fig:beta} which
shows the local slopes as a function of $s$ for $p_{\textrm{lab}}$ values between
0.2 and 1~GeV/c.

\begin{figure}[h]
  \begin{center}
  	\includegraphics[width=9.cm]{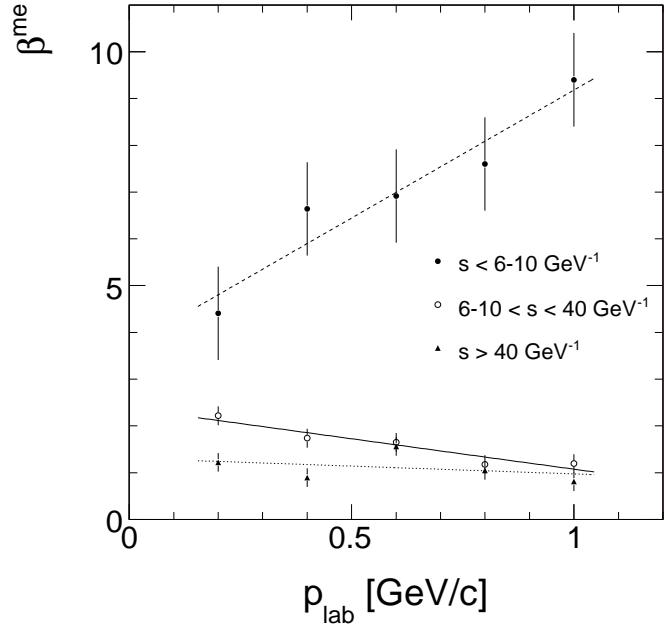}
 	\caption{Local slopes in the regions $s <$~6~GeV$^2$ (broken line),
             6~$< s <$~40~GeV$^2$ (full line) and $s >$~40~GeV$^2$ (dotted line) as
             a function of $p_{\textrm{lab}}$ between 0.2 and 1~GeV/c}
  	 \label{fig:beta_pldep}
  \end{center}
\end{figure}

With the exception of the threshold region, the slopes are confined
to the region between 2 and 1 typical of meson exchange processes.
The dependence on $p_{\textrm{lab}}$ is given in Fig.~\ref{fig:beta_pldep} where the slopes in
the three regions of $s$ specified above are presented.

This Figure shows clearly the different nature of the low $s$ enhancement
where the slopes increase strongly with $p_{\textrm{lab}}$. The two other regions,
full and dotted lines, are compatible with a Regge parametrization
with trajectory intercepts which increase with $p_{\textrm{lab}}$. This is insofar
interesting as the region of measurements regarded here covers the
complete backward angular range and the corresponding interactions
are by no means confined to diffractive or low momentum transfer
collisions. It is shown in Sect.~\ref{sec:separation} of this paper that in the
backward hemisphere the pion yields from nuclear cascading and
target fragmentation are comparable. If the nuclear component is
characterized by low momentum transfer reactions \cite{pc_proton} the target fragmentation 
is manifestly inelastic and non-diffractive.
It governs the total yield at all angles below about 70~degrees.

In conclusion of this study of $\pi^+/\pi^-$ ratios in p+C interactions
the following points should be stressed:

\begin{itemize}
 \item The global data interpolation leads to a precise and consistent
       description of the behaviour of the $\pi^+/\pi^-$ ratios in the full
       backward hemisphere, thus offering an additional tool for the
       discrimination of experimental deviations.
 \item The inspection of the detailed $s$-dependence of the ratios
       opens a new window on the underlying exchange processes.
 \item In particular the comparison to the elementary nucleon-nucleon
       collisions establishes a close relation between apparently
       disjoint sectors of the different hadronic interactions.
\end{itemize}

%
%
\section{Data sets not used in the global interpolation}
\vspace{3mm}
\label{sec:dev_exp}

As mentioned in Sect.~\ref{sec:ph} four of the 19 investigated data sets 
are incompatible with the attempt at generating an overall
consistent description of the experimental situation. These
data will be shortly discussed below.

%
%
\subsection{The proton data of ref. \cite{geaga}}
\vspace{3mm}
\label{sec:geaga}

\begin{figure}[h]
  \begin{center}
  	\includegraphics[width=10cm]{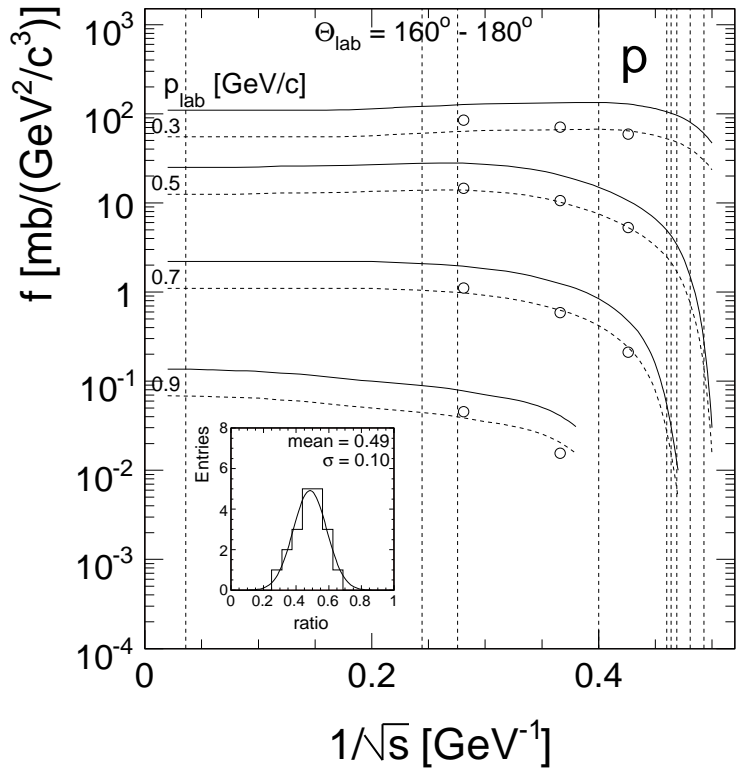}
 	\caption{The proton cross sections from \cite{geaga}
             in comparison to the global data interpolation (full lines)
             in the lab angle bin from 160 to 180~degrees. The broken
             lines correspond to a reduction of the interpolation by a
             factor of two. The inserted histogram gives the number
             distribution of the ratio between data and interpolation}
  	 \label{fig:bevalac}
  \end{center}
\end{figure}

These data have been obtained at the Bevalac using beam momenta of 
1.75, 2.89 and 5.89~GeV/c, spanning a lab momentum range from
0.3 to 0.9~GeV/c at a lab angle of 180~degrees. The resulting cross
sections trace the shape of the $1/\sqrt{s}$ dependence rather precisely
but are consistently about a factor of two below the global
interpolation as shown in Fig.~\ref{fig:bevalac}. Here the full lines correspond 
to the global interpolation and the broken lines give the 
interpolation divided by a factor of two.

\begin{figure}[b]
  \begin{center}
    \includegraphics[width=15.5cm]{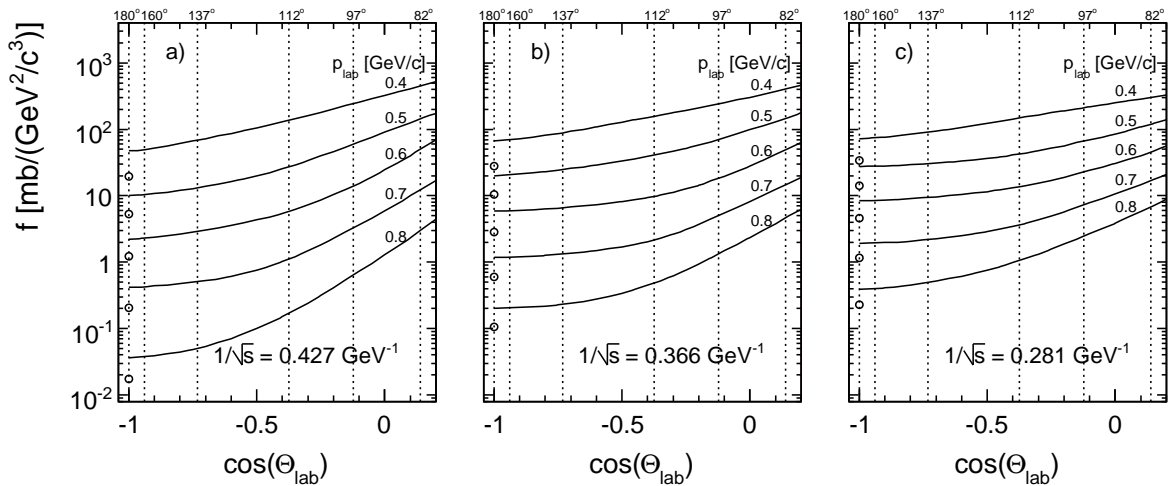}
  \caption{Invariant proton cross sections as a function of
             $\cos(\Theta_{\textrm{lab}})$ for three values of $1/\sqrt{s}$; a) 0.427, b)
             0.366 and c) 0.281 GeV$^{-1}$. The full lines give the global data
             interpolation, the open circles the data from \cite{geaga}}
     \label{fig:geaga}
  \end{center}
\end{figure}

As the angular bin from 160 to 180~degrees is mostly covered by
data around 160--162~degrees, a steep angular dependence in this
region cannot a priori be excluded. The smooth and gentle angular
dependence of the interpolated data shown in Fig.~\ref{fig:geaga} for the
angular range from 82 to 180~degrees and for the three $1/\sqrt{s}$
values of ref. \cite{geaga}, together with the constraint of the approach
to 180~degrees with tangent zero, excludes however a drop of the 
cross sections by a factor of two between 160 and 180~degrees. 

%
%
\subsection{The proton and pion data of refs. \cite{belyaev_prot,belyaev_pion}}
\vspace{3mm}
\label{sec:belyaev}

\begin{figure}[h]
  \begin{center}
    \includegraphics[width=16cm]{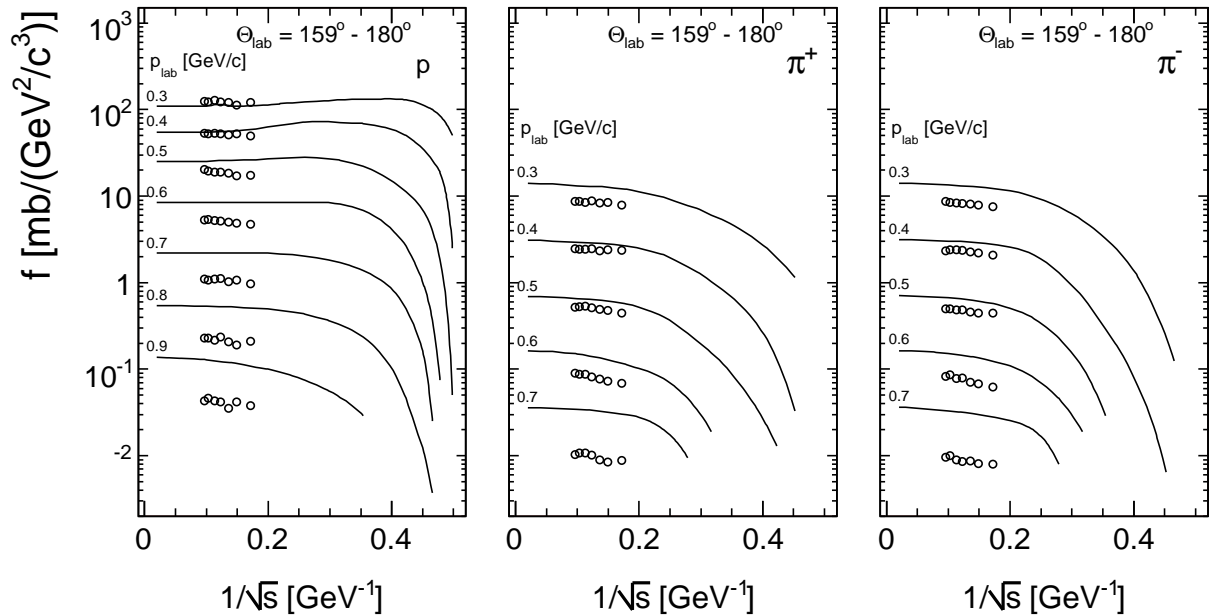}
  \caption{The data of \cite{belyaev_prot,belyaev_pion}  as a function of
             $1/\sqrt{s}$ at $\Theta_{\textrm{lab}}$~=~159~degrees (open circles) in comparison
             to the global data interpolation at 160 degrees (full lines)}
     \label{fig:serp_sdep}
  \end{center}
\end{figure}

A sizeable set of data on proton \cite{belyaev_prot} and pion \cite{belyaev_pion} production
has been obtained at the Serpukhov accelerator spanning the range
of beam momenta between 17 and 57~GeV/c. This fills the gap
between the PS and SPS energies where no other data are available.
The data cover the $p_{\textrm{lab}}$ range from 0.25 to 1.2~GeV/c at 
$\Theta_{\textrm{lab}}$~=~159~degrees. They are presented in Fig.~\ref{fig:serp_sdep} in comparison to the
global data interpolation. 

Several features of this comparison are noteworthy:

\begin{itemize}
 \item The shape of the $1/\sqrt{s}$ dependences complies precisely with
       the global interpolation. This is compatible with the absence 
       of rapid variations of the cross sections with energy in the 
       region between PS and SPS.
 \item There is a pronounced suppression of these data with respect
       to the interpolation with increasing $p_{\textrm{lab}}$ reaching factors of
       three at the upper ranges for protons and pions.
 \item The $\pi^+$ and $\pi^-$ data show an identical behaviour.
 \item The proton data are tracing the interpolation up to 
       $p_{\textrm{lab}}$~=~0.4~GeV/c whereas the pion data are already suppressed in this
       $p_{\textrm{lab}}$ range.
 \item The suppression factors are generally bigger for the pions at equal $p_{\textrm{lab}}$.
 \item These features might be compatible with a momentum scale error.
\end{itemize}

\begin{figure}[h]
  \begin{center}
  	\includegraphics[width=9cm]{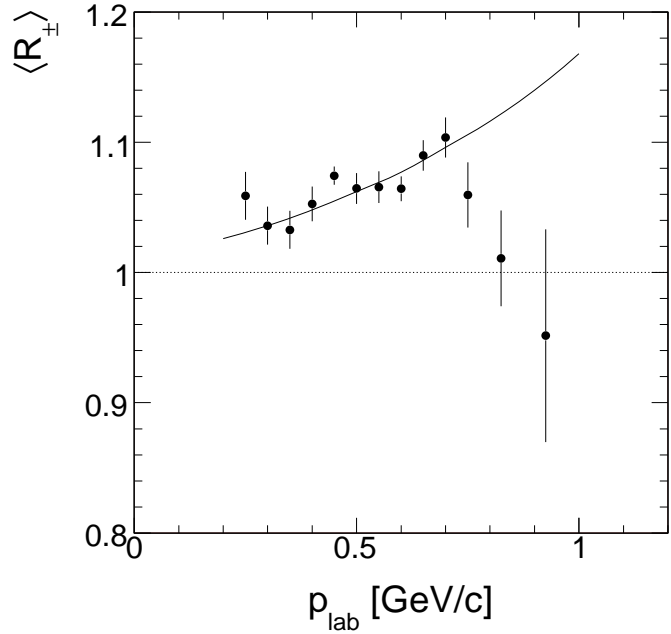}
 	\caption{Pion charge ratio $R_\pm$ from \cite{belyaev_pion} as a function of $p_{\textrm{lab}}$
             averaged over beam momentum between 17 and 57~GeV/c. The full
             line gives the result of the global interpolation averaged
              over the same beam momentum scale, Sect.~\ref{sec:p2n}}
  	 \label{fig:serp_p2n}
  \end{center}
\end{figure}

In addition to the reproduction of the shape of the $1/\sqrt{s}$
dependence, also the $\pi^+/\pi^-$ ratio complies exactly with the
one extracted from the global interpolation, Fig.~\ref{fig:serp_p2n}, up to
$p_{\textrm{lab}}$~=~0.7~GeV/c. Above this value there is a sharp drop of
$R_\pm$ reaching unphysical values at the upper limit of $p_{\textrm{lab}}$.
This drop of about 20\% has however to be compared to a drop 
of 300\% of the invariant cross sections at this limit.

%
%
\subsection{The pion data of ref. \cite{harp}}
\vspace{3mm}
\label{sec:harp}

\begin{figure}[h]
  \begin{center}
    \includegraphics[width=13.cm]{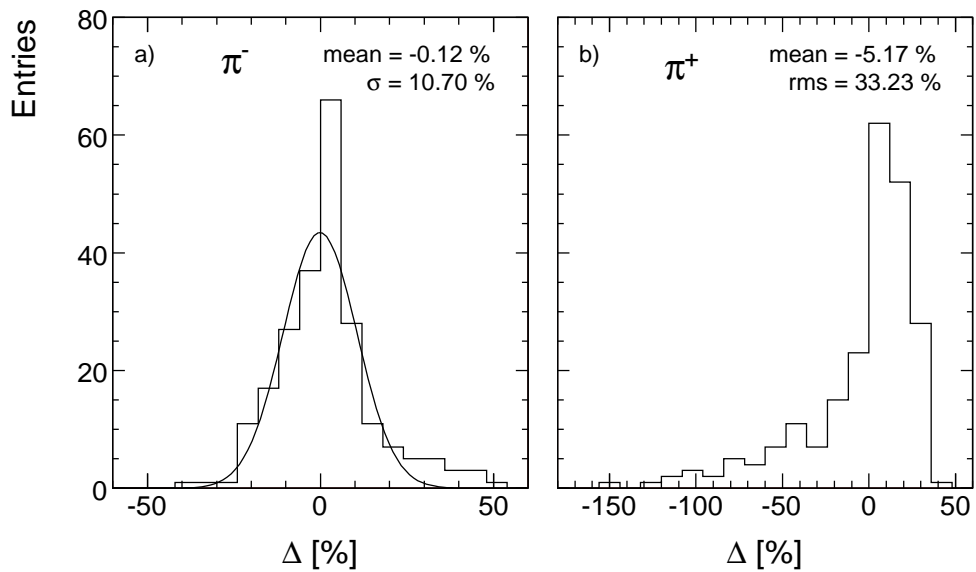}
      \caption{Histograms of the percent differences for all
               angles and beam momenta between \cite{harp-cdp} and \cite{harp}. Panel a) $\pi^-$, panel b) $\pi^+$}
      \label{fig:harpdiffdist}
  \end{center}
\end{figure}

These results cover a range from 3 to 12~GeV/c beam momentum
at $\Theta_{\textrm{lab}}$ between 25 and 117~degrees and 0.125~$< p_{\textrm{lab}} <$~0.75~GeV/c.
They are thus directly comparable to the ones from
\cite{harp-cdp} which are part of the global data interpolation.
Their differences to this interpolation are presented in
Fig.~\ref{fig:harpdiffdist} for all beam momenta and the standard grid of $p_{\textrm{lab}}$
and $\Theta_{\textrm{lab}}$ values.

\begin{figure}[h]
  \begin{center}
    \includegraphics[width=15.5cm]{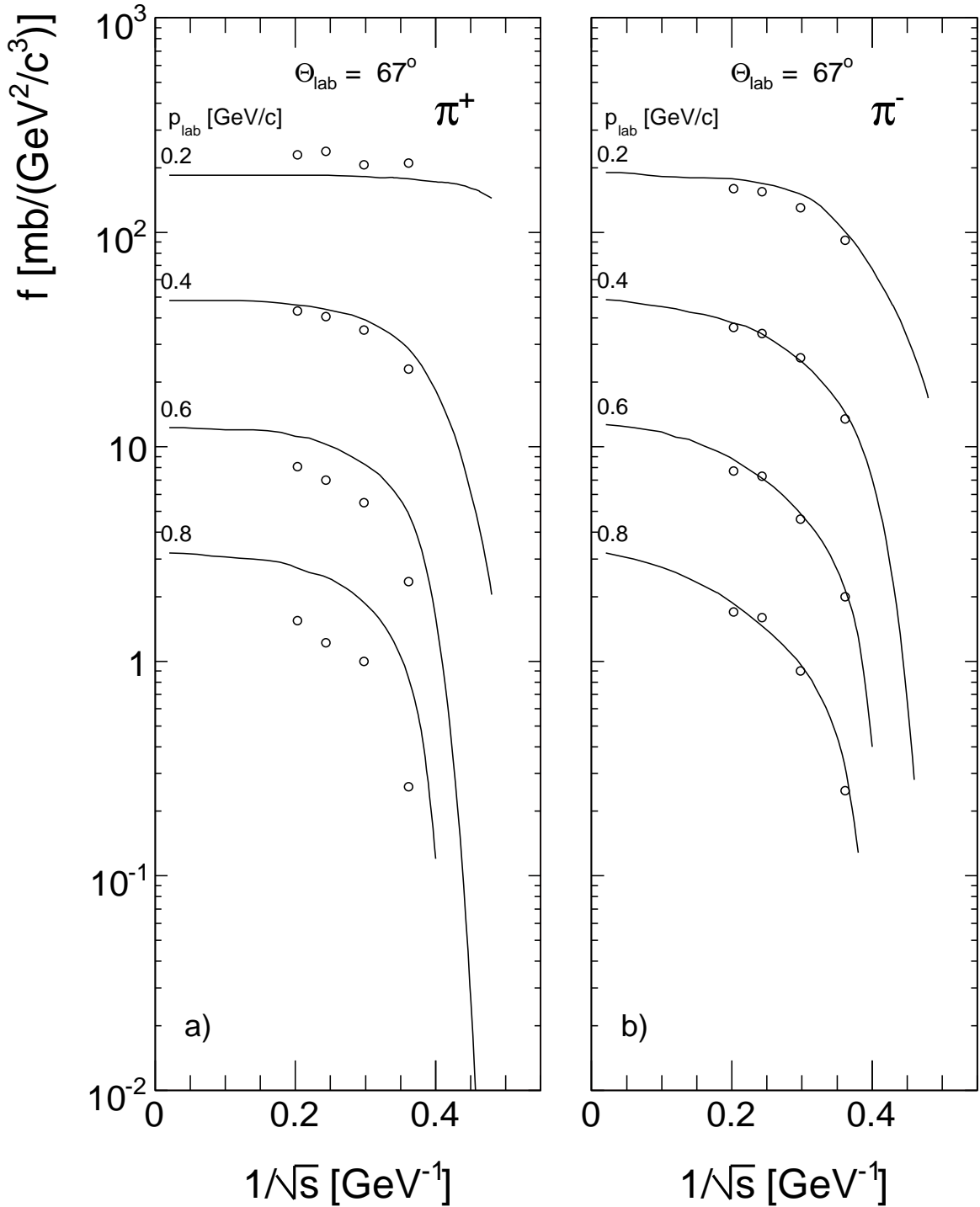}
    \caption{Invariant pion cross sections from \cite{harp} for $\Theta_{\textrm{lab}}$~=~67~degrees
             and four $p_{\textrm{lab}}$ values as a function of $1/\sqrt{s}$,
             (open circles) in comparison to the global data
             interpolation (full lines). Panel a) $\pi^+$, panel b) $\pi^-$}
     \label{fig:pion_sdep}
  \end{center}
\end{figure}

If the mean values of the differences are close to zero, 
their number distributions show wide spreads especially for
$\pi^+$. This is exemplified in Fig.~\ref{fig:pion_sdep} where a typical comparison
to the global interpolation (full lines) is given as a function
of $1/\sqrt{s}$ at $\Theta_{\textrm{lab}}$~=~67~degrees for four $p_{\textrm{lab}}$ values.

\begin{figure}[h]
  \begin{center}
    \includegraphics[width=11.cm]{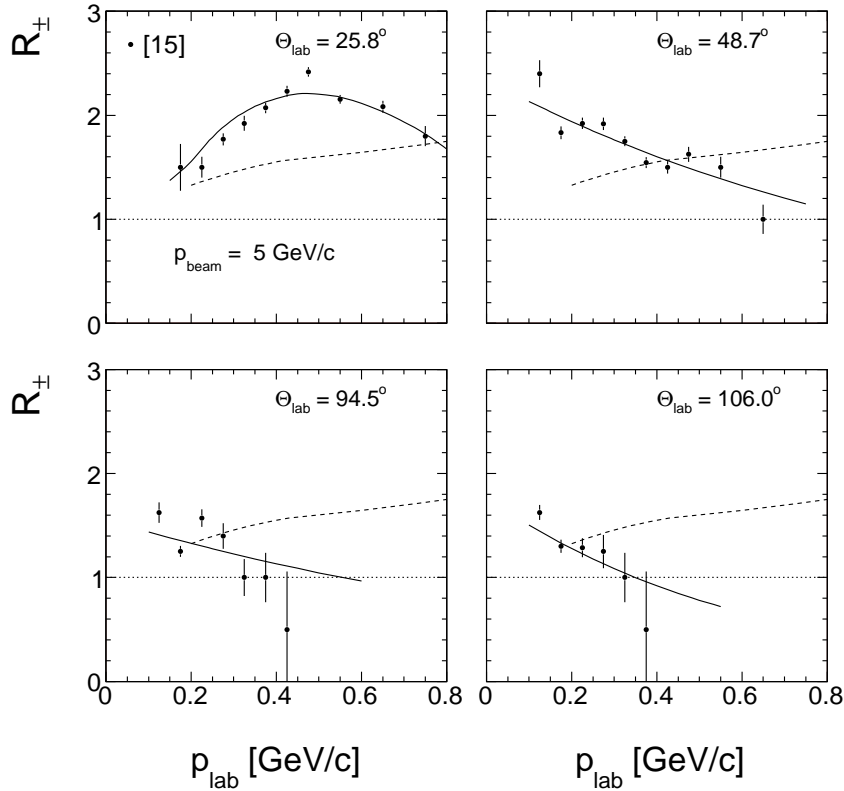}
      \caption{Charge ratio $R_\pm$ from \cite{harp} as a function of $p_{\textrm{lab}}$
               for four lab angles and a beam momentum of 5~GeV/c. The full
               lines give the results of the data interpolation from \cite{harp}, the
               full circles correspond to the ratios of the
               measured cross sections. The broken lines give the result
               of the global interpolation}
      \label{fig:harp_p2n_5}
  \end{center}
\end{figure}

A comparison of $\pi^+/\pi^-$ ratios is given in Fig.~\ref{fig:harp_p2n_5} as a function
of $p_{\textrm{lab}}$ for four values of $\Theta_{\textrm{lab}}$ at $p_{\textrm{beam}}$~=~5~GeV/c.

This figure demonstrates the importance of using, in addition to the invariant cross section proper,
the particle ratios which are strongly constrained by physical arguments, see Sect.~\ref{sec:p2n}.

%
%
\subsection{The pion data of ref. \cite{abgrall}}
\vspace{3mm}
\label{sec:shine}

These data have been obtained at a beam momentum of 31~GeV/c
in a $\Theta_{\textrm{lab}}$ range from 0.6 to 22.3~degrees and $p_{\textrm{lab}}$ from
0.2 to 18~GeV/c. If a large part of the given angular and
momentum coverage falls outside the backward region regarded
here, the low momentum range up to $p_{\textrm{lab}} \sim$~0.5~GeV/c for all
angles and the range 0.6~$< p_{\textrm{lab}} <$~1~GeV/c for angles above
about 9~degrees corresponds to negative $x_F$ and can therefore
be considered here.

\begin{figure}[h]
  \begin{center}
  	\includegraphics[width=9.cm]{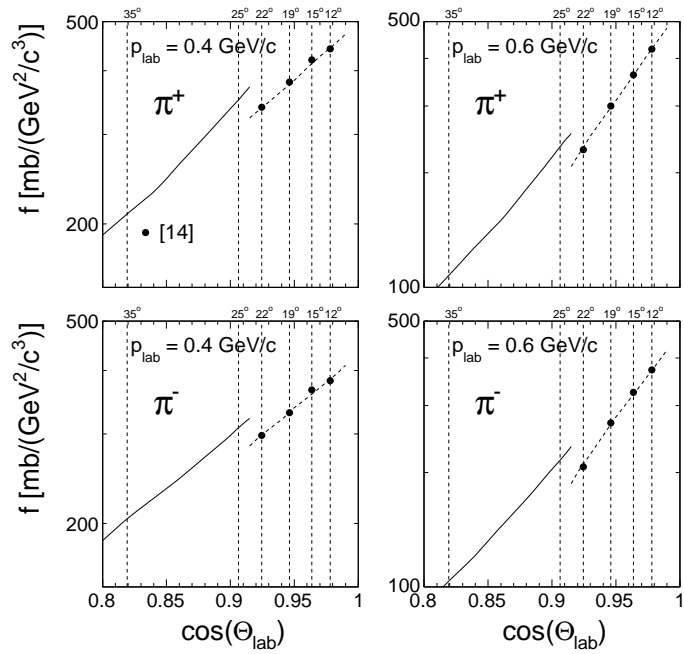}
      \caption{Invariant pion cross sections as a function of $\cos(\Theta_{\textrm{lab}})$ at
               $1/\sqrt{s}$~=~0.13~GeV$^{-1}$ in the $\Theta_{\textrm{lab}}$ region from 12 to
               35~degrees for $p_{\textrm{lab}}$~=~0.4 and 0.6~GeV/c. Full lines: global
               data interpolation. Full dots and broken lines: data from \cite{abgrall}}
  	  \label{fig:na61}
  \end{center}
\end{figure}

The complete $1/\sqrt{s}$ dependence established in the preceding
sections has a lower angular limit at 25~degrees corresponding
to the lowest value of the standard grid of angles. This angle is 
close to the highest angle of \cite{abgrall} at 22.3~degrees allowing
for a safe interpolation. This is shown in Fig.~\ref{fig:na61} where the 
global interpolation is compared to the data of \cite{abgrall} at their
highest angles between 12 and 22~degrees for two $p_{\textrm{lab}}$ values
both for $\pi^+$ and for $\pi^-$.

\begin{figure}[h]
  \begin{center}
  	\includegraphics[width=9.cm]{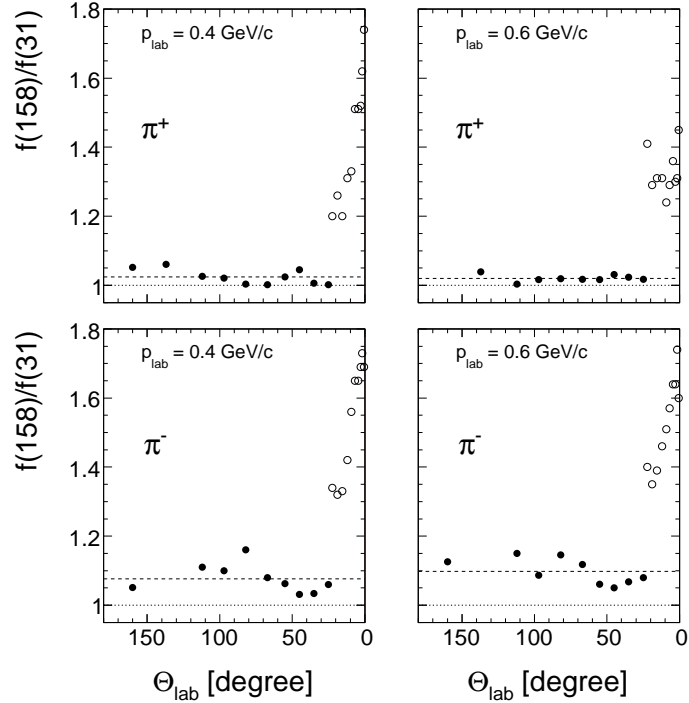}
      \caption{Ratio of invariant pion cross sections at 158/400 
               and 31~GeV/c as a function of $\Theta_{\textrm{lab}}$ for $p_{\textrm{lab}}$~=~0.4
               and 0.6~GeV/c. The full circles correspond to the
               global data interpolation, the open circles to the
               direct ratio \cite{pc_pion}/\cite{abgrall}. The broken lines give the
               mean ratio over the complete range of the global
               data interpolation}
  	  \label{fig:na61_ratio}
  \end{center}
\end{figure}

As the global interpolation is limited to $\Theta_{\textrm{lab}} >$~25~degrees,
another way of comparison is offered by the combined NA49 and
Fermilab results at 158/400~GeV/c where the former data cover
the complete angular range of \cite{abgrall}. The ratio of the available
high energy data to the results at 31~GeV/c beam momentum is 
shown, as a function of $\Theta_{\textrm{lab}}$, in Fig.~\ref{fig:na61_ratio}.

Apparently this cross section ratio is within errors angle independent over the full range
of the global data survey, with well defined averages below 1.05 for $\pi^+$ and 1.1 for $\pi^-$.
In contrast, the cross section ratio between NA49 and ref. \cite{abgrall} shows values
in the region of 1.4 increasing with decreasing $\Theta_{\textrm{lab}}$.

In conclusion to this Section it may be stated that the global data interpolation
between 15 different experiments attempted in this paper proves to be a useful tool
for the detection of deviating data sets. Further details concerning the above
comparisons can be found in an internal report on ref. \cite{site}.
%
%
\section{The separation of target fragmentation and intra-nuclear component for pion production at SPS energy}
\vspace{3mm}
\label{sec:separation}

Hadronic production in the backward direction of p+A collisions has
two components: the fragmentation of the target nucleons which have
been hit by the projectile proton, and the propagation of momentum
transfer into the nucleus by secondary nucleon-nucleon interaction
which follow, on a longer time scale, the initial excitation process.
Both processes are governed by the mean number of collisions $\langle \nu \rangle$ 
suffered by the projectile on his trajectory through the nucleus.

As only the sum of these two separate mechanisms is experimentally
accessible, a minimum assumption about the fragmentation of the target
nucleons is needed in order to allow the separation of the components 
in an otherwise model-independent fashion. This minimal assumption
consists in assuming that the fragmentation process of the hit nucleons
is equal to the basic nucleon-nucleon interaction, taking full account
of course of isospin symmetry. In addition and only valid for the
relatively small value of $\langle \nu \rangle$ in the Carbon nucleus, it will be
assumed that successive collisions result in hadronization at full
interaction energy of the corresponding elementary interactions.	

As far as the value of $\langle \nu \rangle$ is concerned, this has been determined
for pion production in some detail in \cite{pc_discus} using the forward
and the backward region at $x_F >$~-0.1 where no intra-nuclear cascading
is present, see below. This determination used three independent approaches:

\begin{itemize}
 \item A Monte-Carlo calculation using the measured nuclear density distributions.
 \item The relation between the inelastic cross sections of p+p and p+C interactions.
 \item The approach to $x_F$~=~-0.1 of the ratio of pion densities in p+C and p+p collisions.   
\end{itemize}

The two former methods have to make the assumption that the inelastic
interaction cross sections are independent of the number of subsequent
collisions $\nu$.

In \cite{pc_proton} a similar approach is used concerning the production of
protons and anti-protons, again in the regions where there is no
contribution from nuclear cascading as well as in the full backward
hemisphere. 

All methods mentioned above result in a consistent estimate of 
$\langle \nu \rangle$~=~1.6 in p+C collisions, with a relative systematic uncertainty 
of the order of a few percent.

In the following argumentation a prediction of the mean pion density
of target fragmentation in the backward hemisphere at $\sqrt{s}$~=~17.2~GeV 
will be used which is relying on the published pion data from NA49 \cite{pp_pion} 
and the estimated mean number of collisions, $\langle \nu \rangle$. The invariant pion 
cross sections are divided by the inelastic cross section to yield the 
quantity

\begin{equation}
   \langle f_{\textrm{pp}}(x_F,p_T) \rangle = 0.5(f^{\pi^+}_{\textrm{pp}}(x_F,p_T) + f^{\pi^-}_{\textrm{pp}}(x_F,p_T)) 
   \label{eq:mfpp}
\end{equation}
per inelastic event which establishes isospin symmetry, and

\begin{equation}
   f^{\textrm{pred}}(x_F,p_T) = 1.6 \langle f_{\textrm{pp}}(x_F,p_T) \rangle                       
   \label{eq:fpp}
\end{equation}

This prediction is transformed into the appropriate coordinates $p_{\textrm{lab}}$
and $\Theta_{\textrm{lab}}$ and divided by the measured invariant p+C cross sections
$f_{\textrm{pC}}(p_{\textrm{lab}},\Theta_{\textrm{lab}})$ per inelastic event yielding the ratio

\begin{equation}
   R^{\textrm{pred}}(p_{\textrm{lab}},\Theta_{\textrm{lab}}) = 
   \frac{f^{\textrm{pred}}(x_F,p_T)}{f_{\textrm{pC}}(p_{\textrm{lab}},\Theta_{\textrm{lab}})}
   \label{eq:pred}
\end{equation}

\begin{figure}[h]
  \begin{center}
  	\includegraphics[width=11.cm]{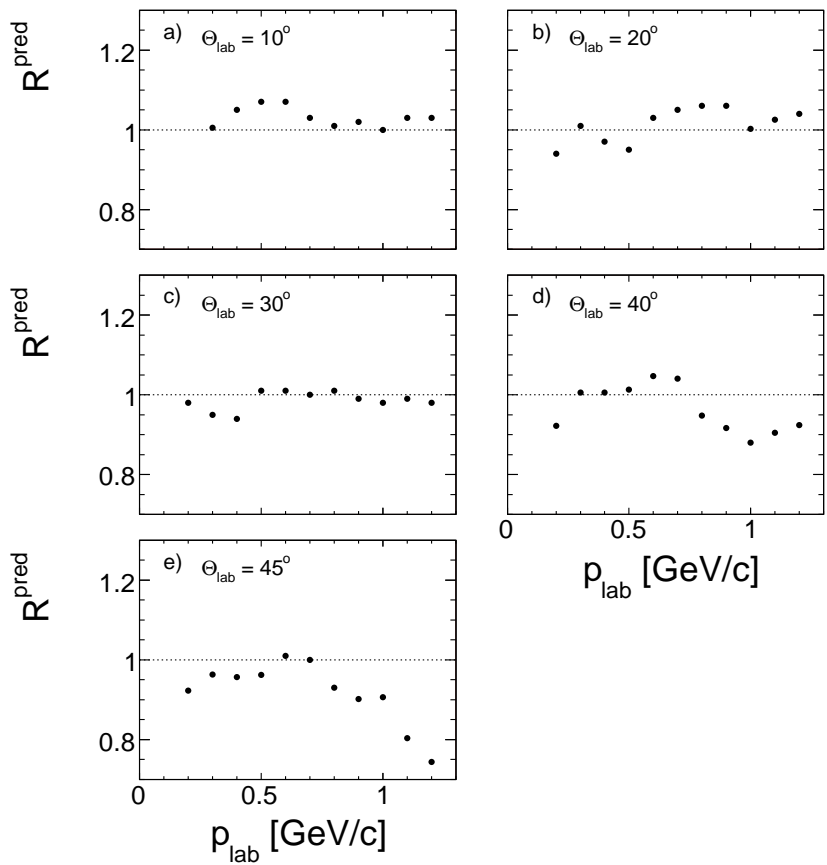}
 	 \caption{$R^{\textrm{pred}}(p_{\textrm{lab}},\Theta_{\textrm{lab}})$ as a function of
              $p_{\textrm{lab}}$ for the five angles 10, 20, 30, 40 and 45~degrees}
  	 \label{fig:rpred_plab}
  \end{center}
\end{figure}

This ratio is shown in Fig.~\ref{fig:rpred_plab} as a function of $p_{\textrm{lab}}$ for the lab angles
10, 20, 30, 40 and 45~degrees.

\begin{figure}[h]
  \begin{center}
  	\includegraphics[width=6.cm]{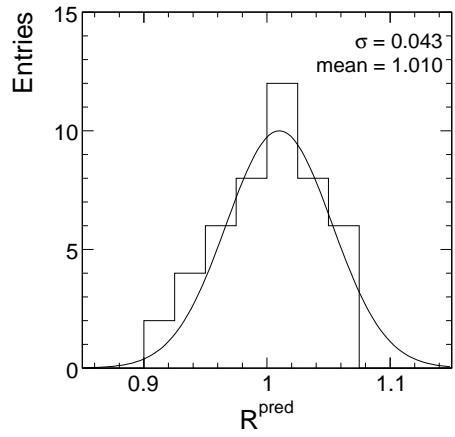}
 	 \caption{Distribution of $R^{\textrm{pred}}(p_{\textrm{lab}},\Theta_{\textrm{lab}})$ for the
              lab angles 10, 20, 30~degrees and 40 and 45~degrees at $p_{\textrm{lab}} <$~0.8~GeV/c. 
              Full line: Gauss fit to the distribution with a mean value
              at 1.01 and a relative rms of 4.3\%}
  	 \label{fig:rpred_hist}
  \end{center}
\end{figure}

It is evident that the ratio is close to one for the three lowest angles
at all $p_{\textrm{lab}}$ and for the region below 0.8~GeV/c for 40 and 45~degrees. 
This is quantified in Fig.~\ref{fig:rpred_hist} which gives the distribution of the ratio 
for the mentioned $p_{\textrm{lab}}$ ranges.

\begin{figure}[h]
  \begin{center}
  	\includegraphics[width=10.5cm]{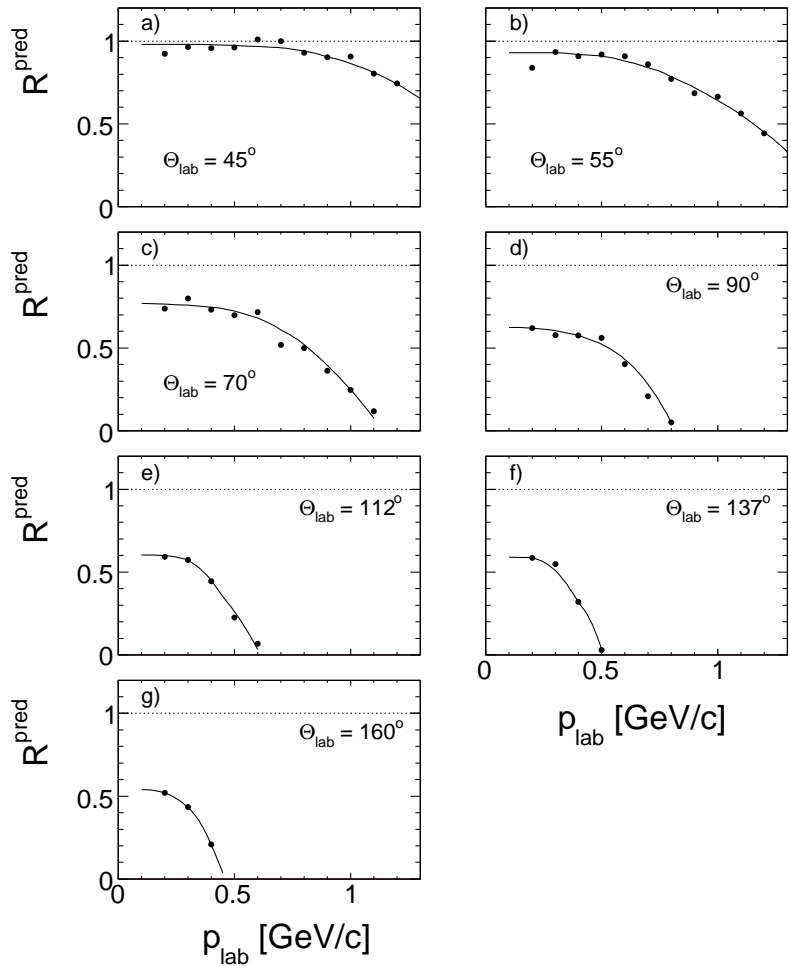}
 	 \caption{$R^{\textrm{pred}}(p_{\textrm{lab}},\Theta_{\textrm{lab}})$ as a function of $p_{\textrm{lab}}$
              for the angles of 45, 55, 70, 90, 112, 137 and 160~degrees. The full lines
              are local interpolations}
  	 \label{fig:rpred_plab_high}
  \end{center}
\end{figure}

The results show that indeed the measured pion cross sections correspond
for lab angles up to 45~degrees precisely to the prediction from elementary 
collisions. This indicates that there is no contribution from intra-nuclear
cascading in this region, in accordance with the results of \cite{pc_discus}.
A drop of the ratio becomes however visible in the higher $p_{\textrm{lab}}$ range
at 40 and 45~degrees. This marks the onset of a nuclear component which
becomes clearly visible in the ratios at larger angles shown in Fig.~\ref{fig:rpred_plab_high}.
    
\begin{figure}[h]
  \begin{center}
  	\includegraphics[width=14.cm]{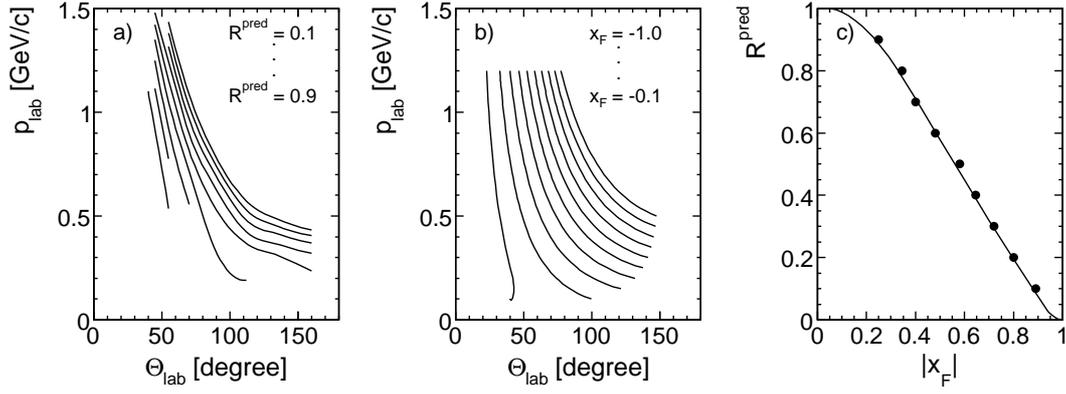}
 	 \caption{a) measured correlation between $p_{\textrm{lab}}$ and $\Theta_{\textrm{lab}}$
              for constant values of $R^{\textrm{pred}}(p_{\textrm{lab}},\Theta_{\textrm{lab}})$ between 0.1
              and 0.9, b) correlation between $p_{\textrm{lab}}$ and $\Theta_{\textrm{lab}}$ for fixed
              values of $x_F$ between -0.1 and -1.0 and c) $R^{\textrm{pred}}(p_{\textrm{lab}},\Theta_{\textrm{lab}})$
              as a function of $x_F$}
  	 \label{fig:rpred_kin}
  \end{center}
\end{figure}

It is interesting to note that the target fragmentation governs the pion
density up to the highest lab angles at low $p_{\textrm{lab}}$, with $R^{\textrm{pred}}$ values
of more than 50\%. The ratio decreases however steadily with increasing $p_{\textrm{lab}}$
and reaches zero at distinct momentum values indicating the approach to
$x_F$~=~-1 in the plots of Fig.~\ref{fig:kinematics}. This kinematic effect is more clearly brought
out in Fig.~\ref{fig:rpred_kin} showing that the fraction of target fragmentation is
essentially a function of $x_F$ and rather independent on lab angle.

The correlation between $p_{\textrm{lab}}$ and $\Theta_{\textrm{lab}}$ for fixed values of $R^{\textrm{pred}}$
shown in panel a traces rather exactly the kinematic correlation between
the same variables for fixed values of $x_F$, panel b. This allows
to establish a direct dependence of $R^{\textrm{pred}}$ on $x_F$ which is to first order
angle-independent, panel c.

\begin{figure}[h]
  \begin{center}
  	\includegraphics[width=11cm]{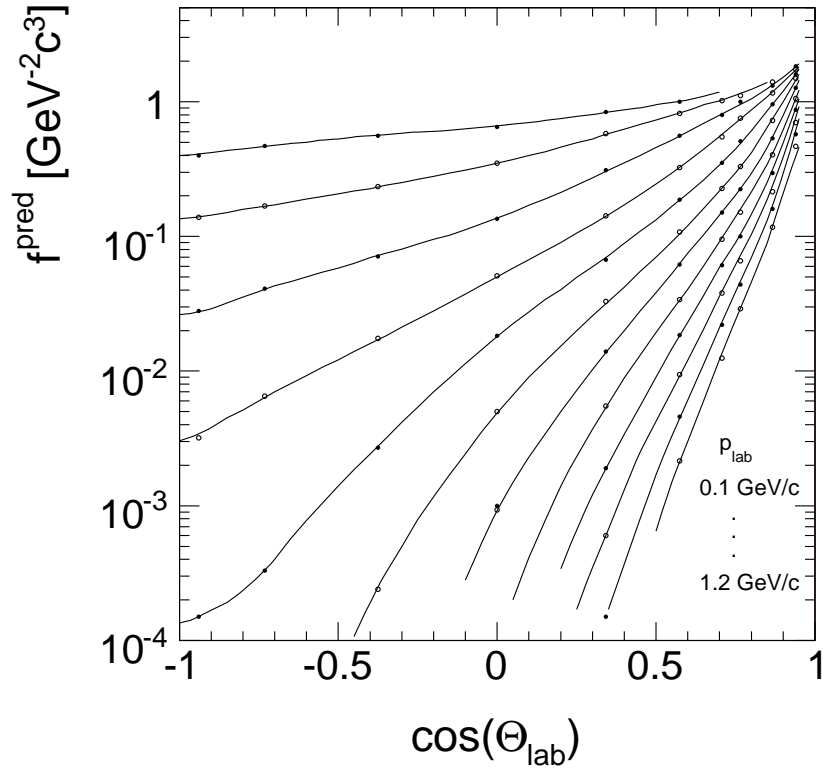}
 	 \caption{Predicted invariant density $f^{\textrm{pred}}(p_{\textrm{lab}},\Theta_{\textrm{lab}})$
              per inelastic event as a function of $cos(\Theta_{\textrm{lab}})$ for fixed values
              of $p_{\textrm{lab}}$ between 0.1 and 1.2~GeV/c. The full lines represent data
              interpolations}
  	 \label{fig:fpred_theta}
  \end{center}
\end{figure}

The invariant densities $f^{\textrm{pred}}(p_{\textrm{lab}},\Theta_{\textrm{lab}})$ per inelastic event
as predicted from the fragmentation of the participant target nucleons is
presented in Fig.~\ref{fig:fpred_theta}.

\begin{figure}[h]
  \begin{center}
  	\includegraphics[width=11cm]{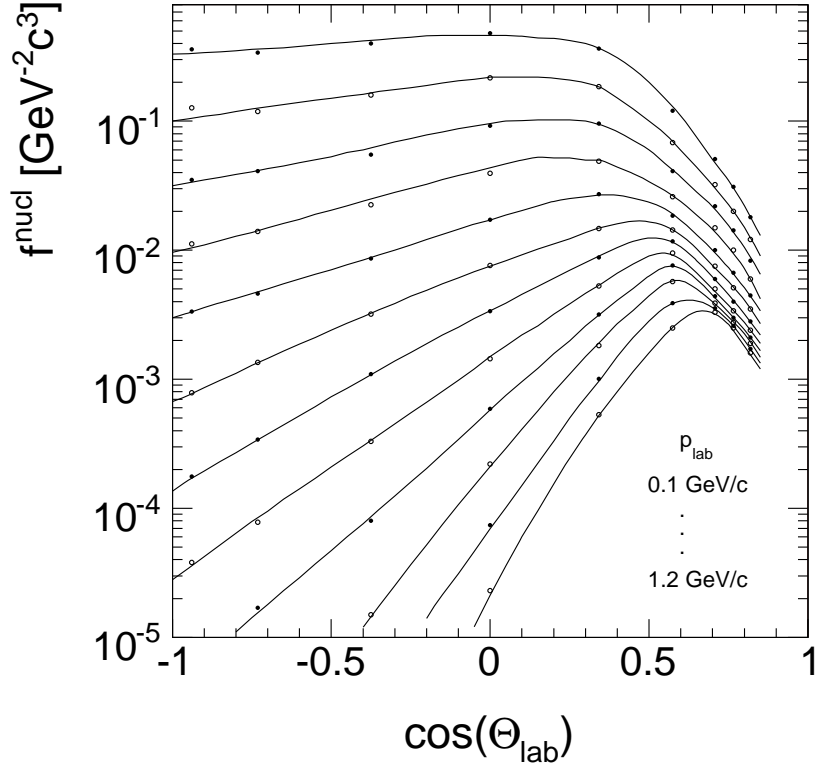}
 	 \caption{Invariant pion density $f^{\textrm{nucl}}(p_{\textrm{lab}},\Theta_{\textrm{lab}})$ 
              from intra-nuclear cascading as a function of $cos(\Theta_{\textrm{lab}})$ for
              fixed values of $p_{\textrm{lab}}$ between 0.1 and 1.2~GeV/c. The full lines
              represent data interpolations}
  	 \label{fig:nucl_theta}
  \end{center}
\end{figure}

This density may be subtracted from the pion density $f(p_{\textrm{lab}},\Theta_{\textrm{lab}})/\sigma^{\textrm{inel}}$ 
measured in p+C interactions which is within errors equal for 
$\pi^+$ and $\pi^-$, see Figs.~\ref{fig:pc_pip_theta} and \ref{fig:pc_pin_theta}. The resulting invariant density 
 
\begin{equation}
   f^{\textrm{nucl}}(p_{\textrm{lab}},\Theta_{\textrm{lab}}) = \frac{f(p_{\textrm{lab}},\Theta_{\textrm{lab}})}{\sigma^{\textrm{inel}}} -
   f^{\textrm{pred}}(p_{\textrm{lab}},\Theta_{\textrm{lab}})
   \label{eq:nucl}
\end{equation}
is shown in Fig.~\ref{fig:nucl_theta}.

This subtraction procedure becomes of course uncertain in the small angle
region where the nuclear component is on the few percent level and below
with respect to the target fragmentation, see Figs.~\ref{fig:rpred_plab} and \ref{fig:rpred_plab_high}.

The invariant angular distributions shown in Figs.~\ref{fig:fpred_theta} and \ref{fig:nucl_theta}
may be converted into number distributions following:

\begin{equation}
  \frac{d^2n^{\textrm{pred}}(p_{\textrm{lab}},\Theta_{\textrm{lab}})}{dp_{\textrm{lab}}d\Theta_{\textrm{lab}}} 
  = 2\pi \frac{p_{\textrm{lab}}^2}{E_{\textrm{lab}}} f^{\textrm{pred}}(p_{\textrm{lab}},\Theta_{\textrm{lab}})
\end{equation}
and

\begin{equation}
  \frac{d^2n^{\textrm{nucl}}(p_{\textrm{lab}},\Theta_{\textrm{lab}})}{dp_{\textrm{lab}}d\Theta_{\textrm{lab}}} 
  = 2\pi \frac{p_{\textrm{lab}}^2}{E_{\textrm{lab}}} f^{\textrm{nucl}}(p_{\textrm{lab}},\Theta_{\textrm{lab}})
\end{equation}

\begin{figure}[h]
  \begin{center}
  	\includegraphics[width=12cm]{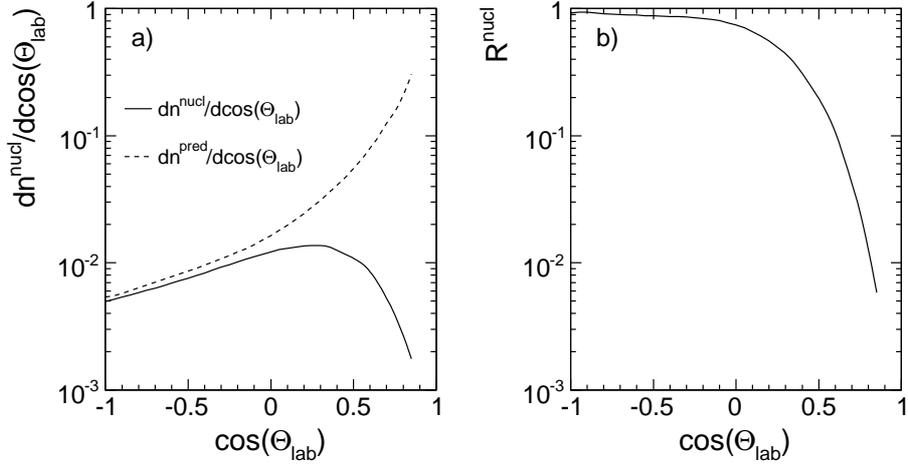}
 	 		\caption{a) $dn^{\textrm{nucl}}/d\cos(\Theta_{\textrm{lab}})$ as a function of
               $\cos(\Theta_{\textrm{lab}})$ (full line),  $dn^{\textrm{pred}}/d\cos(\Theta_{\textrm{lab}})$ as a function 
               of $\cos(\Theta_{\textrm{lab}})$ (broken line), b) the ratio $R^{\textrm{nucl}}$ as a 
               function of $\cos(\Theta_{\textrm{lab}})$}
  	 \label{fig:dndcostheta}
  \end{center}
\end{figure}

Integrating these distributions over $p_{\textrm{lab}}$, the number distributions

\begin{equation}
	\frac{dn^{\textrm{pred}}}{d\cos(\Theta_{\textrm{lab}})}   
\end{equation}
and

\begin{equation}
	\frac{dn^{\textrm{nucl}}}{d\cos(\Theta_{\textrm{lab}})} 
\end{equation}

are obtained which are shown in Fig.~\ref{fig:dndcostheta} together with the ratio

\begin{equation}
	R^{\textrm{nucl}}(\cos(\Theta_{\textrm{lab}})) = \frac{dn^{\textrm{nucl}}}{d\cos(\Theta_{\textrm{lab}})}\Big/\frac{dn^{\textrm{pred}}}{d\cos(\Theta_{\textrm{lab}})}
\end{equation}

Evidently the nuclear component of pion production stays comparable to
the target fragmentation in the full backward hemisphere of $\Theta_{\textrm{lab}}$.
It decreases rapidly for $\Theta_{\textrm{lab}}$ below about 60~degrees and vanishes
below $\Theta_{\textrm{lab}}$~25~degrees. 
 
Integration of $dn^{\textrm{nucl}}/d\cos(\Theta_{\textrm{lab}})$ over $\cos(\Theta_{\textrm{lab}})$ results in the 
total single pion yield from nuclear cascading 

\begin{equation}
	n^{\textrm{nucl}}_{\pi} = 0.105 
\end{equation}
per inelastic event. The predicted integrated yield from target fragmentation is

\begin{equation}
	n^{\textrm{pred}}_{\pi} = \frac{1.6 (n^{\textrm{pp}}_{\pi^+} + n^{\textrm{pp}}_{\pi^-})}{4} = 2.151 
\end{equation}
with

\begin{equation}
	n^{\textrm{pp}}_{\pi^+} = 3.018
\end{equation}
and 

\begin{equation}
	n^{\textrm{pp}}_{\pi^-} = 2.360
\end{equation}
from p+p interactions as measured by NA49, \cite{pp_pion}. This means that 
for p+C interactions the nuclear component of pion production amounts 
to 4.9\% of the pions originating from the fragmentation of the hit 
target nucleons. Applying isospin symmetry on the isoscalar C nucleus 
with

\begin{equation}
	n_{\pi^+} = n_{\pi^-} = n_{\pi^0}
\end{equation}
the total pion yields are 6.45 from target fragmentation and 0.315
from nuclear cascading.

\begin{figure}[h]
  \begin{center}
  	\includegraphics[width=12cm]{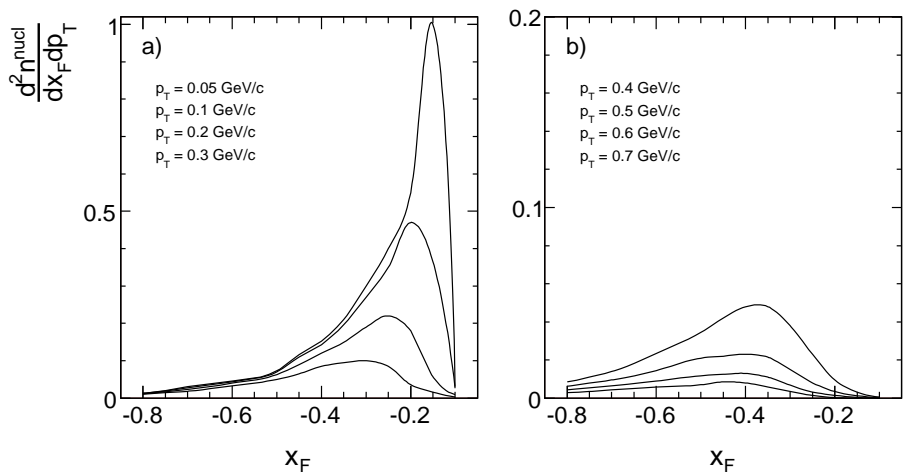}
 	 		\caption{Double differential pion density $d^2n^{\textrm{nucl}}/dx_Fdp_T$
               as a function of $x_F$ for a) 0.05~$< p_T <$~0.3~GeV/c and b)
               0.4~$< p_T <$~0.7~GeV/c}
  	 \label{fig:dndxfdpt}
  \end{center}
\end{figure}

Making use of the kinematic relation between the coordinate pairs
$p_{\textrm{lab}}$, $\Theta_{\textrm{lab}}$ and $x_F$, $p_T$, see Fig.~\ref{fig:kinematics}c, the double differential yields 
for the nuclear component as functions of $x_F$ and $p_T$

\begin{equation}
	\frac{d^2n^{\textrm{nucl}}}{dx_Fdp_T} = 2\pi p_{\textrm{max}} \frac{p_T}{E} f^{\textrm{nucl}}(x_F,p_T)
\end{equation}
may be obtained where $p_{\textrm{max}}$, (\ref{eq:xf}), and $E$ are cms quantities.
The resulting pion density distributions are shown in Fig.~\ref{fig:dndxfdpt} as
a function of $x_F$ for $p_T$ values from 0.05 to 0.7~GeV/c.

\begin{figure}[h]
  \begin{center}
  	\includegraphics[width=12cm]{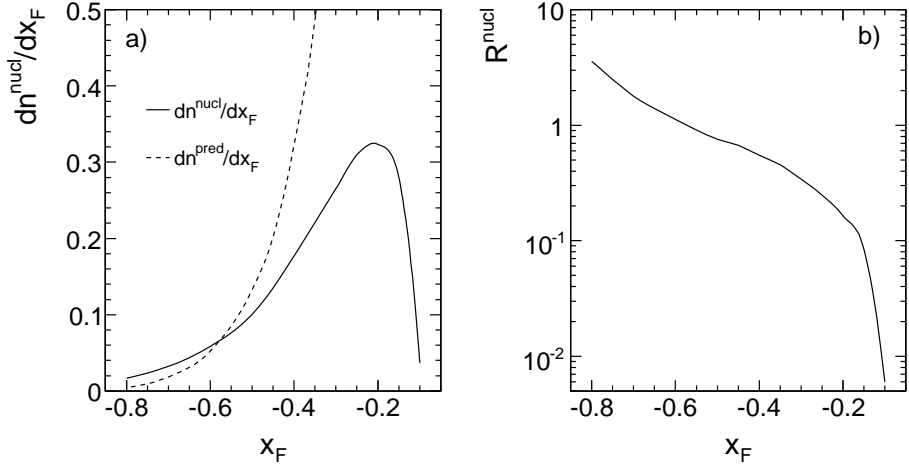}
 	 	\caption{a) Pion density $dn^{\textrm{nucl}}/dx_F$ as a function of $x_F$
                (full line). The predicted density distribution from target
                fragmentation $dn^{\textrm{pred}}/dx_F$ is shown as the broken
                line; b) Ratio $R^{\textrm{nucl}}(x_F) = (dn^{\textrm{nucl}}/dx_F)/(dn^{\textrm{pred}}/dx_F)$
                as a function of $x_F$}
  	 \label{fig:dndxf}
  \end{center}
\end{figure}

A peak at low $p_T$ and $x_F$~=~-0.15 is apparent which corresponds
to the location of pions with small lab momentum, see Fig.~\ref{fig:kinematics}. With 
increasing $p_T$ the maximum density decreases and shifts in $x_F$ to 
lower values which is again in accordance with the kinematic correlation
visible in Fig.~\ref{fig:kinematics}. Integration over $p_T$ results in the single differential
density $dn^{\textrm{nucl}}/dx_F(x_F)$ shown in Fig.~\ref{fig:dndxf} together with the predicted
density distribution $dn^{\textrm{pred}}/dx_F(x_F)$ from target fragmentation
and with the ratio of the two densities.

The $p_T$ integrated pion density $dn^{\textrm{nucl}}/dx_F(x_F)$ shows a peak
at $x_F \sim$~-0.2 and vanishes at $x_F \sim$~-0.08. As shown by the density ratio
with the predicted target fragmentation $dn^{\textrm{pred}}/dx_F(x_F)$ in Fig.~\ref{fig:dndxf}b,
the nuclear component reaches ~10\% of the target fragmentation at
$x_F$~=~-0.15 and exceeds this contribution for $x_F <$~-0.55.

The nuclear pion component extracted above is used in \cite{pc_proton} in conjunction 
with the complementary nuclear proton component to obtain the percentage of
cascading protons which are accompanied by pion emission.    

%
%
\section{Conclusion}
\vspace{3mm}
\label{sec:conclusion}

This paper presents a survey of available data concerning backward proton
and pion production in minimum bias p+C interactions, including new and 
extensive data sets obtained at the CERN PS and SPS. The backward direction 
being defined as the complete phase space at negative Feynman $x_F$, the data cover, 
for projectile momenta from 1 to 400~GeV/c, the ranges from
0.2 to 1.2~GeV/c in lab momentum $p_{\textrm{lab}}$ and from 10 to 180~degrees in lab angle $\Theta_{\textrm{lab}}$. 
The paper attempts an interconnection of the different data sets by 
a detailed three-dimensional interpolation scheme in the variables $1/\sqrt{s}$, $p_{\textrm{lab}}$, 
and $\cos(\Theta_{\textrm{lab}})$. This attempt allows a precise control of the internal data consistency 
as well as the study of the evolution of the invariant inclusive cross sections in all three variables.

A literature search has provided a set of 19 different experiments with a total of more than 
3500 data points. These measurements were obtained over
40 years of experimentation by collaborations employing widely different
experimental techniques. In this respect it may be stated as a first positive result 
that the majority of the data may be combined into a surprisingly
self-consistent ensemble. This global interpolation scheme results in
a considerable discriminative power against the systematic deviation
of particular data sets. Only 4 of the 19 quoted experiments show in fact
deviations which clearly mark them as systematically diverging. These
experiments are inspected in detail one by one in an attempt to clearly
bring out the discrepancies. In some of the cases, possible experimental
error sources are pointed out.

The underlying physics provides for additional constraints concerning
basic quantities like charge conservation and isospin symmetry as well
as the necessity of smoothness and continuity of the observed cross
sections. Whenever possible, contact to the complementary elementary
nucleon-nucleon interactions is established. This concerns in particular
the evocation of mesonic exchange processes for the description of $\pi^+/\pi^-$
ratios and the prediction of the target fragmentation from elementary
interactions and its separation from the component of nuclear cascading.

As far as the dependences of the invariant cross sections on the three
basic variables $p_{\textrm{lab}}$, $\Theta_{\textrm{lab}}$ and $1/\sqrt{s}$ is concerned, a well
constrained phenomenology emerges. The $p_{\textrm{lab}}$ dependences are exponential 
or close to exponential over a major part of the phase space with some exceptions 
mostly towards low interaction energies. This fact results
in an important constraint for the data interpolation. The $\cos(\Theta_{\textrm{lab}})$
dependences are not far from exponential and smooth and continuous
through all lab angles. In particular there is no indication of an
instability around 90~degrees for the proton yields. The $1/\sqrt{s}$
dependences converge, after strong variations close to production
threshold, smoothly to asymptotic behaviour in the SPS energy range.
This region is approached from above by the protons and from below
for the pions. This convergence is confirmed by the $\pi^+/\pi^-$ ratios which show, 
being governed by meson exchange at low $\sqrt{s}$ with large values marked 
by the projectile isospin, a smooth decline with energy towards unity
as expected from the underlying elementary exchange processes. 

\section*{Acknowledgements}
\vspace{3mm}
This work was supported by
the Polish State Committee for Scientific Research (P03B00630),
the Polish National Science Centre (on the basis of decision no. DEC-2011/03/B/ST2/02634)
the Bulgarian National Science Fund (Ph-09/05),
the EU FP6 HRM Marie Curie Intra-European Fellowship Program,
the Hungarian Scientific Research Fund OTKA (T68506) and
the Hungarian OTKA/NKTH A08-77719 and A08-77815 grants.

\end{document}